# Roadmap for Quantum Nanophotonics with Free Electrons

F. Javier García de Abajo,[1,2,*] Albert Polman,[3,*] Cruz I. Velasco,[1] Mathieu Kociak,[4,*] Luiz H. G. Tizei,[4,*] Odile Stéphan,[4] Sophie Meuret,[5] Takumi Sannomiya,[6,*] Keiichirou Akiba,[6,7] Yves Auad,[4] Armin Feist,[8,9,*] Claus Ropers,[8,9,*] Peter Baum,[10,*] John H. Gaida,[8,9] Murat Sivis,[8,9] Hugo Lourenço-Martins,[11] Luca Serafini,[12,13] Johan Verbeeck,[12,13] Andrea Konečná,[14,15,*] Nahid Talebi,[16,17,*] Beatrice Matilde Ferrari,[18,§] Cameron J. R. Duncan,[18,§] Maria Giulia Bravi,[18] Irene Ostroman,[18] Giovanni Maria Vanacore,[18,*] Ethan Nussinson,[19] Ron Ruimy,[19] Yuval Adiv,[19] Arthur Niedermayr,[19] Ido Kaminer,[19,*] Valerio Di Giulio,[8,9] Ofer Kfir,[20] Zhexin Zhao,[21] Roy Shiloh,[22] Yuya Morimoto,[23] Martin Kozák,[24,*] Peter Hommelhoff,[21,25,*] Francesco Barantani,[26] Fabrizio Carbone,[27,*] Fatemeh Chahshouri,[17,18] Wiebke Albrecht,[3,*] Sergio Rey,[3] Toon Coenen,[28,*] Erik Kieft,[29] Hoelen L. Lalandec Robert,[12,13,*] Frank de Jong,[29] and Magdalena Solà-Garcia[3]

[1]ICFO-Institut de Ciencies Fotoniques, The Barcelona Institute of Science and Technology, 08860 Castelldefels, Barcelona, Spain
[2]ICREA-Institució Catalana de Recerca i Estudis Avançats, Passeig Lluís Companys 23, 08010 Barcelona, Spain
[3]Center for Nanophotonics, NWO-Institute AMOLF, 1098 XG Amsterdam, The Netherlands
[4]Université Paris-Saclay, CNRS, Laboratoire de Physique des Solides, 91405 Orsay, France
[5]Centre d'Élaboration de Matériaux et d'Etudes Structurales (CEMES), Toulouse, France
[6]Department of Materials Science and Engineering, School of Materials and Chemical Technology, Institute of Science Tokyo, 4259 Nagatsuta, Midoriku, Yokohama, 226-8501, Japan
[7]Takasaki Institute for Advanced Quantum Science, National Institutes for Quantum Science and Technology (QST), 1233, Watanuki-machi, Takasaki, Gunma, 370-1292, Japan
[8]Department of Ultrafast Dynamics, Max Planck Institute for Multidisciplinary Sciences, Göttingen, Germany
[9]4th Physical Institute—Solids and Nanostructures, University of Göttingen, Göttingen, Germany
[10]Universität Konstanz, Fachbereich Physik, 78464 Konstanz, Germany
[11]CEMES, University of Toulouse and CNRS, 31055 Toulouse, France
[12]EMAT, University of Antwerp, Groenenborgerlaan 171, Antwerp, Belgium
[13]Nanolight Center of Excellence, University of Antwerp, Groenenborgerlaan 171, Antwerp, Belgium
[14]Central European Institute of Technology, Brno University of Technology, Brno 61200, Czech Republic
[15]Institute of Physical Engineering, Brno University of Technology, Brno 61669, Czech Republic
[16]Institute of Experimental and Applied Physics, Kiel University, 24098 Kiel, Germany
[17]Kiel Nano, Surface and Interface Science KiNSIS, Kiel University, 24118 Kiel, Germany
[18]Laboratory of Ultrafast Microscopy for Nanoscale Dynamics (LUMiNaD), Department of Materials Science, University of Milano-Bicocca, Milano, Italy
[19]Faculty of Electrical and Computer Engineering, Technion–Israel Institute of Technology, 3200003 Haifa, Israel
[20]School of Electrical Engineering, The Iby and Aladar Fleischman Faculty of Engineering, Tel Aviv University, Tel Aviv 69978, Israel
[21]Physics Department, Friedrich-Alexander-Universität Erlangen-Nürnberg (FAU), Erlangen, Germany
[22]Institute of Applied Physics, Hebrew University of Jerusalem (HUJI), Jerusalem, Israel
[23]RIKEN Cluster for Pioneering Research (CPR) and RIKEN Center for Advanced Photonics (RAP), RIKEN, Wako, Japan, and Department of Nuclear Engineering and Management, Graduate School of Engineering, The University of Tokyo, Bunkyo-ku, Tokyo, Japan
[24]Faculty of Mathematics and Physics, Charles University, Prague, Czech Republic
[25]Physics Department, Ludwig-Maximilians-Universität München (LMU), Munich, Germany
[26]Department of Physics, The University of Texas at Austin, Austin, 78712 TX, USA
[27]Institute of Physics, École Polytechnique Fédérale de Lausanne, Lausanne, 1015, Switzerland
[28]Delmic B.V., Oostsingel 209, 2612 HL Delft, The Netherlands
[29]Thermo Fisher Scientific, De Shakel 2 – 10, 5651 GG, Eindhoven, The Netherlands
[§]These authors contributed equally

**Corresponding authors:**
F. Javier García de Abajo: javier.garciadeabajo@nanophotonics.es; Albert Polman: a.polman@amolf.nl; Mathieu Kociak: mathieu.kociak@universite-paris-saclay.fr; Luiz H. G. Tizei: luiz.galvao-tizei@universite-paris-saclay.fr; Takumi Sannomiya: sannomiya.t.aa@m.titech.ac.jp; Armin Feist: armin.feist@mpinat.mpg.de; Claus Ropers: claus.ropers@mpinat.mpg.de; Peter Baum: peter.baum@uni-konstanz.de; Giovanni Maria Vanacore: giovanni.vanacore@unimib.it; Andrea Konečná: andrea.konecna@vutbr.cz; Nahid Talebi: talebi@physik.uni-kiel.de; Ido Kaminer: kaminer@technion.ac.il; Martin Kozák: m.kozak@matfyz.cuni.cz; Peter Hommelhoff: peter.hommelhoff@physik.uni-erlangen.de; Fabrizio Carbone: fabrizio.carbone@epfl.ch; Wiebke Albrecht: w.albrecht@amolf.nl; Toon Coenen: coenen@delmic.co; Hoelen L. Lalandec Robert: hoelen.lalandecrobert@uantwerpen.be

**ABSTRACT:** Over the past century, continuous advancements in electron microscopy have enabled the synthesis, control, and characterization of high-quality free-electron beams. These probes carry an evanescent electromagnetic field that can drive localized excitations and provide high-resolution information on material structures and their optical responses, currently reaching the sub-ångström and few-meV regime. Moreover, combining free electrons with pulsed light sources in ultrafast electron microscopy adds temporal resolution in the sub-femtosecond range while offering enhanced control of the electron wave function. Beyond their exceptional capabilities for time-resolved spectromicroscopy, free electrons are emerging as powerful tools in quantum nanophotonics, on par with photons in their ability to carry and transfer quantum information, create entanglement within and with a specimen, and reveal previously inaccessible details on nanoscale quantum phenomena. This Roadmap outlines the current state of this rapidly evolving field, highlights key challenges and opportunities, and discusses future directions through a collection of topical sections prepared by leading experts.

# 1. INTRODUCTION

**F. Javier García de Abajo* and Albert Polman**

Advancements in electron microscopy over the past century have enabled the generation and manipulation of free-electron beams (e-beams) with ever-increasing precision.[1] State-of-the-art transmission electron microscopes (TEMs) feature e-beams with a high degree of transverse coherence, enabling a spatial resolution limited by Abbe's diffraction to sub-ångström scales ($\sim\lambda_e/\mathrm{NA}$ in aberration-corrected instruments with numerical apertures $\mathrm{NA}\sim10^{-2}$ and electron kinetic energies of $30-200$ keV corresponding to electron wavelengths $\lambda_e \approx 7-2.5$ pm). Optical excitations with such spatial detail can be mapped using monochromatized e-beams combined with electron analyzers, achieving a final energy resolution in the few-meV range.[2,3] Localized optical excitations can thus be spectrally and spatially resolved by scanning the e-beam in a TEM while performing electron energy-loss spectroscopy[4,5] (EELS). In addition, the decay of these excitations into cathodoluminescence (CL) emission provides an alternative source of spectral and spatial information on such excitations when their radiative decay is significant.[6] Unlike EELS, CL does not require e-beam transmission through a specimen and can be performed in scanning electron microscopes[7] (SEMs). Furthermore, the simultaneous acquisition of EELS and CL in TEMs offers deeper insights, such as the strength and statistics of radiative decay channels.[7] Overall, this panorama highlights the primary role of e-beams in nanophotonics, serving as probes that enable spatial and spectral mapping of the optical response in nanostructures through the spontaneous conversion of electron kinetic energy into optical excitations in matter and outcoupled radiation.

In a visionary paper,[8] Archie Howie proposed combining light and free electrons to harness the best of both worlds in electron microscopy: the high spectral resolution of optical fields and the strong spatial focusing of electrons. Electron energy-gain spectroscopy (EEGS) was later proposed[9] and eventually demonstrated with deep sub-meV resolution.[10,11] In a separate development, inelastic electron–light scattering (IELS) was shown to produce multiple photon absorption and emission events by free electrons interacting with illuminated gases.[12] This principle was later applied in a pioneering study[13] utilizing spatially focused femtosecond electron pulses, facilitating strong interactions with ultrafast laser pulses mediated by scattering at a specimen. This breakthrough materialized in the realization of photon-induced near-field electron microscopy[13] (PINEM). Notably, when coherent laser light is used, the electron retains its coherence while interacting with optical fields. Consequently, once IELS occurs, free-space electron propagation reshapes the electron wave function and induces temporal compression, which has been demonstrated to reach the attosecond domain.[14,15] This advancement has direct application in achieving temporal resolution in the sub-femtosecond domain, as demonstrated by the sub-cycle reconstruction of optical field images.[16,17,18]

Beyond high-precision microscopy, free-electron–light interactions are gaining increasing attention as a powerful tool for expanding the capabilities of nanophotonic systems, introducing quantum degrees of freedom when the electron is post-selected (e.g., for generating quantum light[19,20,21]) and enabling entanglement between electrons and photonic states.[22] Free electrons are also sensitive to environmental fluctuations,[23,24,25] making them highly promising for quantum sensing and metrology.[26]

This Roadmap intends to capture key trends in the fundamentals and applications of free-electrons–light interactions through a selection of topical sections organized into four broad categories at the intersection of free electrons and optical fields (see Figure 1):

- *Recent Advances in Electron Microscopy*. This block explores the latest developments and future directions in electron spectroscopies for nanophotonics, including state-of-the-art EELS (Section 3) and CL (Sections 4 and 5), as well as enhanced spatial and spectral resolution through EEGS (Section 6). Theoretical insights and challenges are further examined in Section 2.
- *Ultrafast Electron–Light Interactions*. This category highlights recent progress and future objectives in ultrafast electron microscopy (Section 7), including the synthesis and exploitation of single attosecond electrons (Section 8) and quantum coherent aspects (Section 9), as well as electron–photon temporal coincidence spectroscopy (Section 10) and advancements in electron wave shaping to enable disruptive forms of microscopy (Sections 11 and 11).

- *Quantum Physics and New Concepts*. Free electrons are proposed as powerful probes for testing fundamental aspects of quantum electrodynamics (Section 13) and exploring quantum physics with potential applications, including enhanced microscopy and metrology (Section 14). Section 15 discusses recent achievements and prospects in optical low-energy electron acceleration, while Section 16 elaborates on the Kapitza-Dirac effect[27] (a form of stimulated IELS) and the scattering of optically shaped electrons. Finally, Section 17 proposes using shaped electrons to engineer many-body states in correlated materials.
- *Applications in Materials Science*. Electron microscopy plays a crucial role in analyzing the microscopic structural and optical properties of materials. Section 19 focuses on EELS for structural, compositional, and optical analysis, while Section 18 explores the use of EELS and CL for studying excitons. Additionally, the extraction of structural and near-field information through electron ptychography is discussed in Section 20. We conclude with an industrial perspective presented in Section 21, highlighting current trends in technology transfer for commercial instruments leveraging electron–light interactions.

As the field continues to evolve rapidly, new directions for research and applications will emerge. While the selection presented in this Roadmap is not exhaustive, we hope it is a useful resource for practitioners and inspires further advancements.

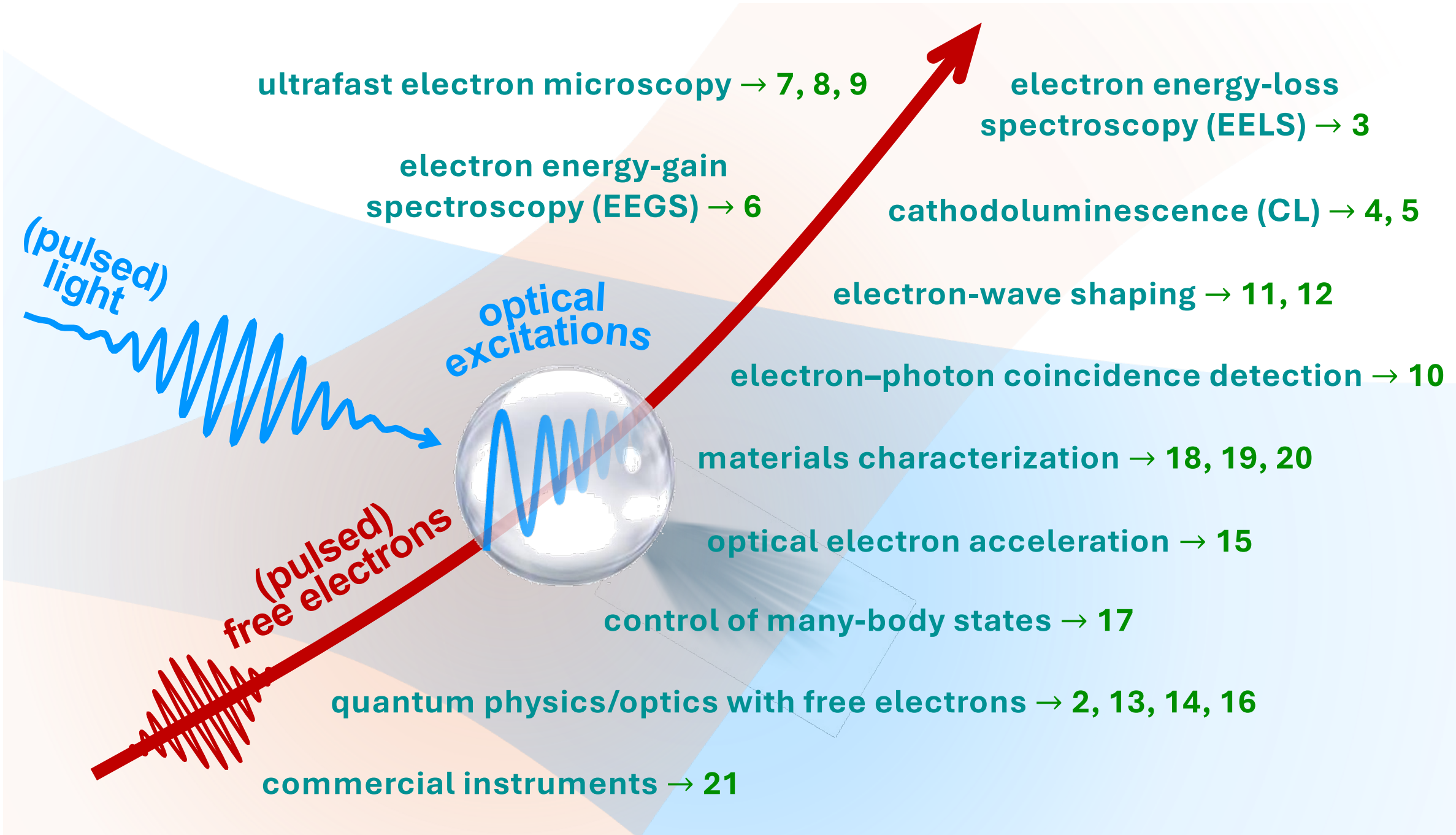

**Figure 1.** Quantum nanophotonics at the intersection of free electrons and optical fields. Disruptive forms of microscopy are emerging, offering an unprecedented combination of spectral, spatial, and temporal resolution. In addition, free electrons are increasingly recognized as powerful tools for exciting, characterizing, and manipulating nanoscale optical modes. This Roadmap highlights key trends in the field, organized into sections corresponding to the numbers in this graphic.

# ELECTRON MICROSCOPY: FUNDAMENTALS AND TECHNIQUES

## 2. THEORY AND CHALLENGES OF FREE ELECTRONS FOR QUANTUM NANOPHOTONICS

**F. Javier García de Abajo* and Cruz I. Velasco**

Electron microscopes enable exquisite control over free electrons and their interactions with material structures and optical fields.[28] The key elements involved in this scenario are summarized in Figure 2a. From the electron side, they include a source, electron optics to produce high-quality e-beams, and an analyzer with spectral and angular resolution capabilities. From the photonic side, a light source is needed to illuminate a specimen or directly interact with the electron, as well as angle-resolved light spectrometry, possibly combined with single-particle electron–photon coincidence detection schemes. Over the last few years, intense theoretical efforts have been devoted to describing these elements,[5,29] including quantum treatments of the electron and light,[30,31] that further enable the study of correlations among them.[28,32] In this section, we present a succinct summary of these elements and identify theoretical challenges and opportunities for future research in the context of quantum nanophotonics.

### 2.1 Primer on Free-Electron Interaction with Light and Optical Excitations in Matter

Inside an electron microscope, electrons are generally prepared in a state with a narrow distribution in kinetic energy and momentum relative to central values $\hbar\varepsilon_0 = m_e c^2(\gamma - 1)$ and $\hbar q_0 = m_e v\gamma$, respectively, where $v$ is the velocity, $\gamma = 1/\sqrt{1 - v^2/c^2}$ is the Lorentz factor, $c$ is the speed of light, and $m_e$ is the electron rest mass. For an e-beam oriented along the $z$ direction, it is convenient to write the electron wave function $e^{i(q_0 z - \varepsilon_0 t)}\phi(\mathbf{r}, t)$ as a function of spatial coordinates $\mathbf{r} = (x, y, z)$ and time $t$ in terms of a slowly evolving envelope $\phi(\mathbf{r}, t)$ subject to the Schrödinger equation[33,34]

$$i\hbar\left[\partial_t + v\partial_z - (i\hbar/2m_e\gamma)\left(\partial_{xx} + \partial_{yy} + \gamma^{-2}\partial_{zz}\right)\right]\phi = \mathcal{H}^{\rm int}\phi. \tag{2.1}$$

Here, $\mathcal{H}^{\rm int} = (ev/c)A_z + (e^2/2m_e c^2\gamma)(A_x^2 + A_y^2 + A_z^2/\gamma^2)$ describes the electron interaction with an electromagnetic field of vector potential $\mathbf{A}(\mathbf{r}, t)$ in the minimal coupling prescription, working in a gauge with vanishing scalar potential.

#### 2.1.1 *Stimulated Inelastic Electron–Light Scattering*

When the interaction only produces small changes in the kinetic energy of the electron compared with $\hbar\varepsilon_0$, the second derivatives in eq 2.1 can be dismissed and the wave function admits the analytical solution

$$\phi(\mathbf{r}, t) = \phi^{\rm inc}(\mathbf{r}, t)\exp\left\{-(i/\hbar v)\int dz'\, \mathcal{H}^{\rm int}(x, y, z', t + (z' - z)/v)\right\}, \tag{2.2}$$

where $\phi^{\rm inc}(\mathbf{r}, t)$ is the electron wave function before interaction. This expression assumes classical external illumination (e.g., that provided by a laser). For monochromatic light of frequency $\omega$, writing the vector potential $\mathbf{A}(\mathbf{r}, t) = (2c/\omega)\,{\rm Im}\{\mathbf{E}(\mathbf{r})e^{-i\omega t}\}$ in terms of the optical electric field $\mathbf{E}(\mathbf{r})$ and dismissing $A^2$ terms in $\mathcal{H}^{\rm int}$, eq 2.2 becomes[30,35]

$$\phi(\mathbf{r}, t) = \phi^{\rm inc}(\mathbf{r}, t)\sum_l J_l(2|\beta(\mathbf{R})|)\, e^{i\,l\,\arg\{-\beta(\mathbf{R})\}}e^{i\,l\omega(z - vt)/v}, \tag{2.3}$$

where $\mathbf{R} = (x, y)$ denotes transverse coordinates, $l$ runs over the number of photons absorbed ($l > 0$) or emitted ($l < 0$) by the electron, $J_l$ is a Bessel functions of order $l$, and

$$\beta(\mathbf{R}) = \frac{e}{\hbar\omega}\int dz\; E_z(\mathbf{r})e^{-i\omega z/v} \tag{2.4}$$

is a dimensionless coupling coefficient proportional to the spatial Fourier transform of the optical field. In EEGS, for small $\beta$ values, the gain probability (fraction of electrons that have gained energy) reads $\Gamma_{\rm EEGS} = |\beta|^2$, whereas in PINEM, the probability of sideband $l$ is given by the squared Bessel function $J_l^2(2|\beta|)$. The integral in eq 2.4 imposes the phase-matching condition

$$\omega = \mathbf{k}\cdot\mathbf{v} \tag{2.5}$$

for the optical-field wave vectors $\mathbf{k}$ that can couple to the electron [i.e., in the $k$-decomposition of $E_z(\mathbf{r})$]. This condition is represented as a red-shaded region in the energy–momentum dispersion diagram of Figure 2b. We conclude that only field components outside the light cone can couple to the electron, such as those associated with the scattering of light by nanostructures, possibly involving the excitation of polaritons in a specimen. Importantly, eq 2.3 shows that the wave function is transformed into an energy comb with sidebands separated by multiples of $\hbar\omega$ relative to the incident electron energy.

In free space, only ponderomotive $A^2$ terms contribute to the electron–light interaction, imprinting a phase on the electron (e.g., in the Kapitza–Dirac effect;[27,36] see also Sections 11, 11, and 16), but also yielding a similar modulation as in eq 2.3 with $\omega = \omega_1 - \omega_2 = \mathbf{k}_1 - \mathbf{k}_2$ when bi-chromatic light fields of frequencies $\omega_{1,2}$ and wave vectors $\mathbf{k}_{1,2}$ are employed (see Section 16, ref 37, and Figure 2b).

The solution in eq 2.1 becomes more complicated when the external optical field is not in a coherent state (e.g., thermal light), requiring a quantum treatment of the optical field and leaving distinct signatures on the electron (e.g., enabling the measurement of the statistical properties of the employed illumination[31]). This scenario is further discussed in Sections 13 and 14.

2.1.2 *Spontaneous Free-Electron Transitions*

In the absence of external illumination, there is still a vector potential associated with the evanescent field of the electron and its scattering by material structures. The scattered field acts back on the electron and produces energy losses that can be resolved through EELS. The probability that an electron is inelastically scattered per unit of transferred energy $\hbar\omega$ reads[30] $\Gamma_{\rm EELS}(\omega) = (1/\pi){\rm Re}\{\beta(\mathbf{R})\}$, where $\beta(\mathbf{R})$ is given by eq 2.4 with $E_z(\boldsymbol{r})$ substituted by the corresponding frequency component of the self-induced electric field $E_z^{\rm ind}(\mathbf{r},\omega) = \int dt\, E_z^{\rm ind}(\mathbf{r},t)e^{i\omega t}$, and the lateral position $\mathbf{R}$ is taken at the location of the e-beam. Numerous analytical studies have been devoted to obtaining analytical solutions for $\Gamma_{EELS}(\omega)$ in different geometries.[5] In addition, the CL light emission spectrum can be calculated from the far field produced by the electron treated as a classical point charge.[5,30]

2.1.3 *Electron Reshaping During Free-Propagation*

The second-derivative terms in eq 2.1 account for recoil effects to the lowest order and become relevant to describe free electron-wave propagation over large distances. In particular, $\partial_{xx}$ and $\partial_{yy}$ produce lateral e-beam spreading along paraxial propagation, while $\partial_{zz}$ causes the mixing of electron components moving with different energies and, therefore, different velocities. This term introduces a correction in eq 2.2 consisting of an $l$-dependent phase factor $e^{-2\pi i l^2 z/z_T}$, where $z_T = 4\pi m_e v^3\gamma^3/\hbar\omega^2$ is the so-called Talbot distance. After the electron propagates over a distance $z$, the initial wave packet is transformed into a train of temporal pulses whose degree of compression depends on the interaction coefficient $\beta$. This effect was first identified in ref 14 as the result of the preservation of quantum coherence in the electron state after interaction with laser light.

A recent work considers the electron interaction with spectrally broad optical pulses to achieve temporal compression down to the zeptosecond regime by invoking the concept of temporal lensing.[38] With the illumination acting as a lens, the incident electron wave packet as an object, and the compressed electron as an image, by placing the *object* at the *lens* plane, this scheme is robust against jitter in the time of arrival of the electron, thus producing temporal compression even when the incident electron pulses possess a limited degree of temporal coherence.

2.1.4 *Electron Post-Selection*

Remarkably, in the nonrecoil approximation, the EELS and CL probabilities are independent of the incident electron wave function along the e-beam direction.[30] In particular, for a point-like electron, which, in virtue of the uncertainty principle, spans an infinite momentum range, any finite-energy exchanges with optical fields or material excitations do not change the point-like character of the probe, and therefore, those excitations remain fully coherent with the classical electromagnetic field associated with the moving electron. In contrast, when the electron is prepared as a plane wave of well-defined momentum $\hbar\mathbf{q}_0$, every excitation changes the electron state in a distinguishable manner, so the interaction produces entanglement between the final electron states $|\mathbf{q}\rangle$ and the sampled excited states $|j\rangle$ of the specimen and the radiation field. In the most general case, the final state of the electron–specimen–radiation system has the form $\sum_j|\mathbf{q}_j\rangle\otimes|j\rangle$. When an electron is detected with a given energy and scattering direction (i.e., a given value of $\mathbf{q}$), the rest of the system is projected onto the states $|j\rangle$ that share that value of the final electron wave vector $\mathbf{q}_j=\mathbf{q}$. Electron post-selection thus adds quantumness to the system by projecting it onto nonclassical states. This type of projection can be used, for instance, to generate photon-number states[19,21,39,40] and entangled electron–excitation states.[22] In this context, we have recently designed an efficient electron–photon coupler that should enable the realization of quantum sensing and metrology based on the measurement of electron currents alone, promising orders-of-magnitude improvement relative to existing optics-based schemes.[26]

## 2.2. Challenges and Opportunities

The preceding section describes free electrons and their interaction with optical fields, possibly mediated by the presence of material scatterers. Several elements in this description are still poorly understood or completely undeveloped, although they could be useful to gain further control of free electrons and their interactions with quantum nanophotonic fields. We discuss several of these elements in what follows, emphasizing the theoretical challenges and identifying some opportunities.

### 2.2.1 *Electron Recoil*

Besides the effects described by the second derivatives in eq 2.1, producing electron reshaping upon propagation over large distances, recoil during the interaction with optical fields can be leveraged to widen the range of accessible electron-compressed states[37] and to deterministically generate quantum light.[41] Further insight could be obtained by systematically exploring analytical solutions of eq 2.1 in different geometries. Numerical solutions have been presented for low-energy electrons and classical light,[42] while an extension to quantum light has revealed additional insight into the effects associated with few-photon exchanges.[43] Recoil in the presence of elastic diffraction by crystal surfaces has recently been shown to boost the strength of inelastic electron–light interaction.[44] This is a promising direction that deserves further exploration of more general scenarios involving diffraction by other types of structures in combination with structured light fields.

### 2.2.2 *Non-Beam and Multi-Beam Electrons*

Free-form electrons (without a preferential direction of motion in contrast to e-beams) constitute an ultimate form of electron waves in combination with nanophotonic environments. We envision the evolution of such waves in a designed nanoscale potential landscape to go beyond the current paradigm of e-beam-based electron microscopy. This could be supplemented by nanophotonic fields, nanoscale emitters, and small-footprint electron detectors. All of these elements need to be developed by not only extending available free-electron theories but also proposing disruptive approaches to engineer nanoscale electron optics, electron–light interactions, localized electron emitters, and practical spectrally resolved electron detection schemes. From the theoretical front, theory could be elaborated on eigenstates of the employed potentials. In macroscopic designs, non-beam and multi-beam free electrons could be prepared by resorting to beam splitters and mixers, going boldly away from paraxial conditions, which should be feasible at low kinetic energies. The lateral degrees of freedom could be incorporated in the theory through transmission and reflection matrices similar to those employed to describe light diffraction by gratings and beam splitters. As a concrete example, although still in the paraxial approximation, multi-path electron splitting and recombination by means of gratings has been proposed to achieve spectrometer-free electron spectromicroscopy (SFES), promising a similar level of spectral resolution as EEGS by also resorting to fine tuning of the light with which the electron interacts, but without requiring electron monochromation or spectrometry;[45] this method uses two gratings in conjugated planes to block electron transmission, so that, when an illuminated specimen is introduced in the region between them, PINEM-like sidebands disrupt the grating imaging, thus enabling electron transmission in proportion to the strength of the optical near field.

### 2.2.3 *Electron Decoherence*

The interaction between free electrons and extended structures has been recently shown to produce decoherence even when macroscopic distances are involved.[25] Decoherence is a valuable source of information on the environment, highly dependent on fluctuations in the vacuum field and material polarization modes. This phenomenon manifests in the electron density matrix by generally reducing off-diagonal elements, whose measurement (see Section 14) would open a window into previously inaccessible magnitudes (e.g., the statistics and strength of polariton and photonic fields). In this context, exciting experimental and theoretical results were obtained in inelastic electron holography, which targeted well-defined extended plasmonic systems.[23,24] Further theoretical efforts are needed to materialize this potential, extend available free-electron theories with an accurate description of decoherence, and explore new phenomena associated with this ubiquitous phenomenon.

2.2.4 *Strong Electron–Polariton Coupling*

The probability that an electron excites a given optical mode is generally orders of magnitude lower than unity. However, the window opened by free electrons into quantum nanophotonic interactions critically depends on the ability to realize order-unity coupling, as explored in recent works.[19,44,41] Strong electron–photon interaction at the single-electron/single-photon level could be achieved under aloof, phase-matched interaction with waveguides[19,41] (in particularly when realizing quasi-parabolic electron trajectories by reflection of the electron at a finite distance from the waveguide, assisted by a repulsive perpendicular DC field[26]), and also by going to low electron energies and exploiting lattice resonances[44] or interacting with highly polarizable Rydberg atoms.[46] This area is ready for the development of disruptive schemes that can materialize this important ingredient to enhance the role of free electrons in quantum nanophotonics.

2.2.5 *Multi-Electron Interactions*

Multi-e-beams have been recently explored and shown to reveal Coulomb-mediated energy/time correlations.[47,48,49] An interesting prediction has also been formulated on the strong multielectron correlation arising from interaction with quantum light.[50] Several questions arise in this context, including the degree of entanglement among electrons and the realization of deterministically shaped multi-electron pulses, which require further theoretical developments, including a careful account of the early stages of electron generation in a photoemission source and their possible use for electron–electron pump–probe analysis.

2.2.6 *Combined Quantum Description of Free Electrons and Material Excitations*

A general formalism exists for describing the interaction of free electrons and bosonic modes such as plasmon- and phonon-polaritons in terms of electromagnetic Green functions.[32] While insight into non-bosonic excitations in atoms, defects, and few-level systems can be obtained by representing them through effective Hamiltonians,[51] the complexity of real many-body systems requires more advanced formulations at the level of density function theory or beyond. The integration of these systems with free electrons and light still demands further development of the theory to account for quantum-optical, free-electron, and material-excitation states at a more fundamental level.

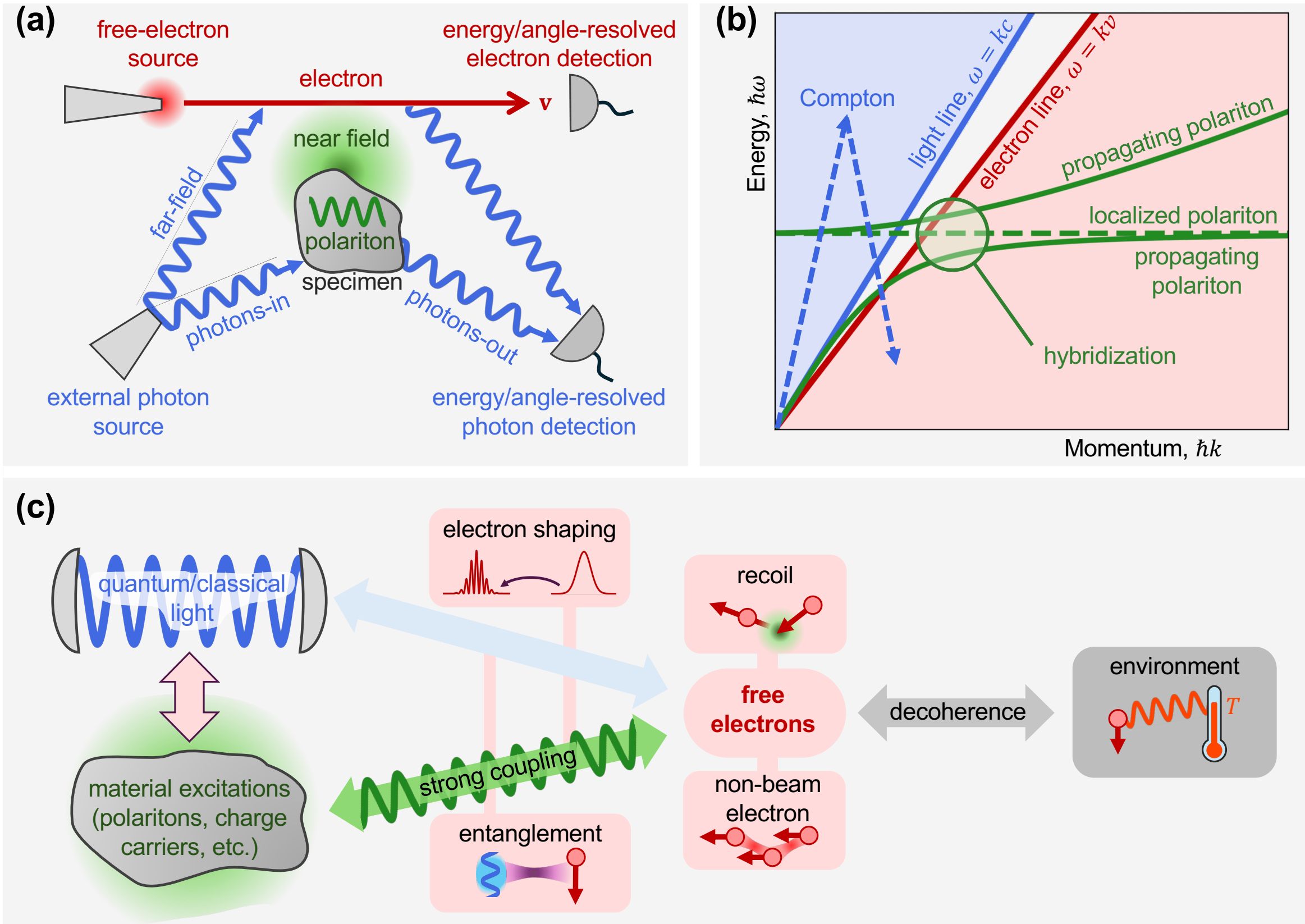

**Figure 2.** Theoretical elements involved in the description of free electrons for quantum nanophotonics. (a) The e-beam is a central element and requires a source, electron optics, and energy/angle-resolved detection. Far- and near-field light can interact with the electron, modifying its energy and momentum distribution and, thus, providing information on polaritons and other types of excitations supported by a scattering specimen. Light emitted by excitation of the latter upon passage of the electron is also a valuable source of information. (b) In a dispersion diagram (energy vs momentum), the range of kinematically accessible excitations produced by a recoilless electron (constant velocity) does not overlap the light cone, but it can intersect propagating and localized optical modes in material structures. Inelastic Compton scattering (upward and downward dashed arrows standing for photon absorption and emission) can also reach the electron excitation region with a zero net exchange of photons. (c) Recoil effects can extend the range of allowed electron excitations, while free-form (non-beam) electrons should unfold additional possibilities for spectromicroscopy, quantum sensing, and quantum metrology. Additional elements in need of further development are the interaction between electrons and quantum light; the realization of strong coupling between electrons and polaritons at the few- or single-quantum level; leveraging the interaction with the environment, which produces decoherence in the electron state, and therefore imprints information on such environment; and better understanding recoil effects, particularly during the interaction of electrons with optical fields.

## 3. SPATIALLY RESOLVED ELECTRON ENERGY-LOSS SPECTROSCOPY (EELS): NOVEL EXPERIMENTS ENABLED BY HIGHLY SPATIALLY COHERENT AND MONOCHROMATED MICROSCOPES

**Mathieu Kociak,* Luiz H. G. Tizei, and Odile Stéphan**

### 3.1 Introduction

EELS in electron microscopes is known historically for its usage in material science. In an EELS experiment, a fast electron (~30~300 keV) interacts with a thin (~100 nm) material or a nanostructure. The electron transfers energy to the material, creating excitations at various energy ranges, from the far-infrared (< 80 meV) to the hard X-ray (>10 keV) regimes passing by the visible range (see Figure 3a). Therefore, EELS is instrumental in exploring vibrations, phonons, plasmons, excitons, band gaps, photonic modes, and core-loss excitations. Given the spatial resolution of TEMs, this technique can access vibrational, optical, chemical, and electronic properties at nanometer scales, if not down to the single-atom or single-atomic-column scale.[4] Initially devoted to the study of fundamental excitations such as bulk or surface plasmons by physicists, EELS has been mostly concerned by the study of core-loss excitations were most of the material science applications lie. Together with the spread of scanning TEM (STEM), the spectral-image mode (SI), and the aberration correction introduced in the early 2000s, atomically resolved chemical mapping of materials is now routine.[52] In parallel, in the mid-2000s monochromators and deconvolution techniques reached ~100 meV resolution. Since this resolution is of the order of typical core losses (in the X-ray regime) and plasmons (in the visible regime) linewidths, it permitted performing full spatial and spectral characterization of plasmonic,[53] chemical, and electronic properties of various materials.[54] Monochromation then developed[2] in the mid-2010s to a point where few-meV energy resolution is reached, and the far-infrared regime accessed. The recent EELS advances presented here have been enabled by using high-brightness guns, which guarantee minimal loss of intensity while monochromating and/or focusing the electron probe to sub-ängström sizes.[2] Likewise, it permits extreme momentum resolution,[55] easily reaching the $\mu$m range, as well as better spatial coherence, which is of utmost importance for phase-dependent applications.[56] The overall increased stability of the monochromated (S)TEM and spectrometers have been needed to cope with measuring sub-10 meV peaks. EELS has also been revolutionized by the arrival of Poisson-noise limited, optimized detector quantum efficiency (DQE) in direct electron detectors, and also event-based electron detectors reaching in addition nanosecond temporal resolution.[57] Although not described here, it is worth noting that EELS theory (see Section 2) is now robustly established for almost any energy range.[58] In this section, we shall review the impact of related technological advances for the investigations of various excitations.

### 3.2 Applications in the X-Ray Range

Core-loss studies were not radically impacted by the recent rise of ultrahigh monochromation, given their relatively large linewidths. They have been probably more influenced by the increased signal-to-noise ratio brought by the combination of ultrahigh monochromation, high-brightness guns, and improved DQE. Also, the fact that multiple different signals from different energy ranges, including linewidth-limited core losses, can be acquired almost in parallel (multimodal approach), proved invaluable in deciphering complex samples, such as biomaterials[59]. Beyond monochromation, core-loss studies have benefited from recent development in electron phase manipulation. In particular, magnetic dichroic signals can be now mapped in one dimension at the atomic resolution using advanced electron magnetic circular dichroism (EMCD) techniques[60]. Phase shaping beyond EMCD, in particular using phase engineering with phase plates,[56] has been the promise of two-dimensional (2D) circular dichroism mapping but is still to be demonstrated.

### 3.3 Applications to Optical Excitations

EELS is now a major means for studying surface-plasmon physics[61], especially thanks to the very high dynamical range in space (from the nanometer to microns) and energy (from a few tens of meV to several eV) needed to study plasmons in general. This is particularly useful given the numerous applications of surface plasmons in fields such as sensing, nonlinear optics, and metamaterials. Such combination of dynamical ranges is difficult to access from pure optical techniques, but has been made possible by the recent advances in EELS. Beyond conventional metallic surface plasmons, exotic plasmons such as found in doped oxides can be studied even in sub-10 nm nanoparticles.[62] Together with the increased brightness of the guns that allows for relatively high intensities with good reciprocal (momentum) resolution, sub-20 meV spectral resolution permits us to measure dispersion relations in plasmonic crystals.[55] Photonic modes, such as whispering gallery modes (WGM) in spheres[63] or modes in photonic band gap materials[64] have much higher quality factors ($Q$) but could be tackled already by former ~100 meV monochromator technologies. The use of ~10 meV monochromation helped in accessing higher quality WGMs (~100), but could not really compete with CL in that respect. It also helped mapping cavity modes in the infrared (IR) regime in photonic band gaps,[65] although their enormous $Q$ (~$10^6$) requires a ~μeV spectral resolution to be resolved, which advocates for the development of new sorts of spectroscopies (see Section 6). The measurement of band gaps by EELS is of major interest, in particular for applications, but still did not clearly benefit from recent technical advances, probably because the real issue is the band-gap signal itself being smooth and not a sharp feature such as, for example, a plasmon. This is different for excitons, even if they feature weak signals such as in transition-metal-dichalcogenide monolayers (see Section 18). In that case, the vanishing of the zero-loss-peak (ZLP) tail intensity in the visible range thanks to monochromation, the use of Poisson-noise-limited detectors, and the fact that excitons have a peaky spectral signature make an even tiny ($10^{-5}$ of the ZLP) signal detectable. In addition, the spectral resolution of tens of meV is now approaching the natural linewidths of some excitons, so that altogether EELS is now a useful tool to map excitonic properties at the nanometer scale.[66] Finally, the study of coupling mechanisms between optical excitations is at the heart of nano-optics. While plasmon–plasmon coupling has been extensively studied for a long time by EELS due to its reasonably large coupling (induced peak splitting typically larger than 100 meV), it is only since 2019 that more exotic couplings have been unveiled using this technique, in particular exciton–plasmon coupling,[67] phonon–plasmon coupling,[68] or Fano-like exciton-to-continuum coupling, putting EELS on par with other nano-optical approaches.[66] Finally, the first attempt to measure the local response to the optical polarization[69] in the visible range in EELS using phase plates has been performed on plasmonic nanoantennas.[70]

### 3.4 Applications to Vibrational Excitations

The most spectacular impact of the recent advances in EELS is undoubtedly the possibility to access vibrational modes and phonons with extreme spatial resolution.[71] Leveraging their strong interaction with electrons, optical-phonon modes have been extensively measured in two[72] and even three[73] dimensions. Due to their low energy, phonons can be thermally excited, resulting in the emergence of spontaneous gain peaks at room temperature and above, which can be used to measure temperature locally.[74] As it is sustained only on fundamental principles, the measurement of the temperature is absolute, forming the basis of a very robust nanothermometry technique.[75] Despite still having an energy resolution that does not compete with the best IR spectroscopy (in particular Fourier Transform IR, FTIR), vibrational EELS starts to tackle the same use cases, such as isotopic labelling[76] or chemical fingerprints in e-beam sensitive (or electron-dose sensitive) materials.[59] Taking advantage of the high coherence of high-brightness guns, the necessary compromise between spatial and momentum resolution can be optimized. Indeed, the phonon density of states and its modifications can be probed at the atomic level[77] (see Figure 3c), while phonon dispersions with high momentum resolution can be obtained.[78]

### 3.5 Current Challenges in EELS

Despite impressive advances, the energy resolution in EELS is now stagnating at a few meV,[2] which is still lagging behind state-of-the-art spectroscopy techniques such as FTIR for phonons and orders-of-magnitude insufficient for tackling some of the quantum-relevant structures discussed in Sections 13 and 14. Solutions to this problem may arise from developing new techniques such as the EEGS (see Section 6).

Event-based-driven direct electron detectors, now reaching close to 1 ns temporal resolution[57], are redefining how and why EELS experiments are performed. Together with the relevant scanning engine, they permit realizing sparse spectral imaging in the same way as it is done for  analysis, with potential impact for dose-sensitive materials. They also open the way to nanosecond-resolved coincidence techniques, where EELS events are associated with EDX or CL detection events (see Section 10). One important point to note is that the nanosecond resolution is obtained at the detection level (i.e., after the e-beam interaction with the sample). This means that all the properties of the microscope are preserved, in particular the spatial and spectral resolutions. This is in contrast with technical approaches in which the temporal resolution is obtained with pulsed guns (either under laser illumination or thanks to the use of fast deflectors). Of importance for this section, this means that the EELS spectral resolution remains essentially the same when combined with nanosecond (or the more common) millisecond-to-second time resolution. Even more, the sole fundamental limit is the Heisenberg energy–time relation. Therefore, meV resolution can be accessed even at the nanosecond. Practical applications may be for example the monitoring of the phonon spectrum and, therefore, the local temperature[75] within the nanosecond and nanometer scales, as recently demonstrated.[79]

Finally, the access to very low energy ranges permits one to probe new sorts of excitations such as magnons.[80] Together with stable He-cooling technologies, it advances the limits for investigating condensed-matter-physics effects at the atomic level.

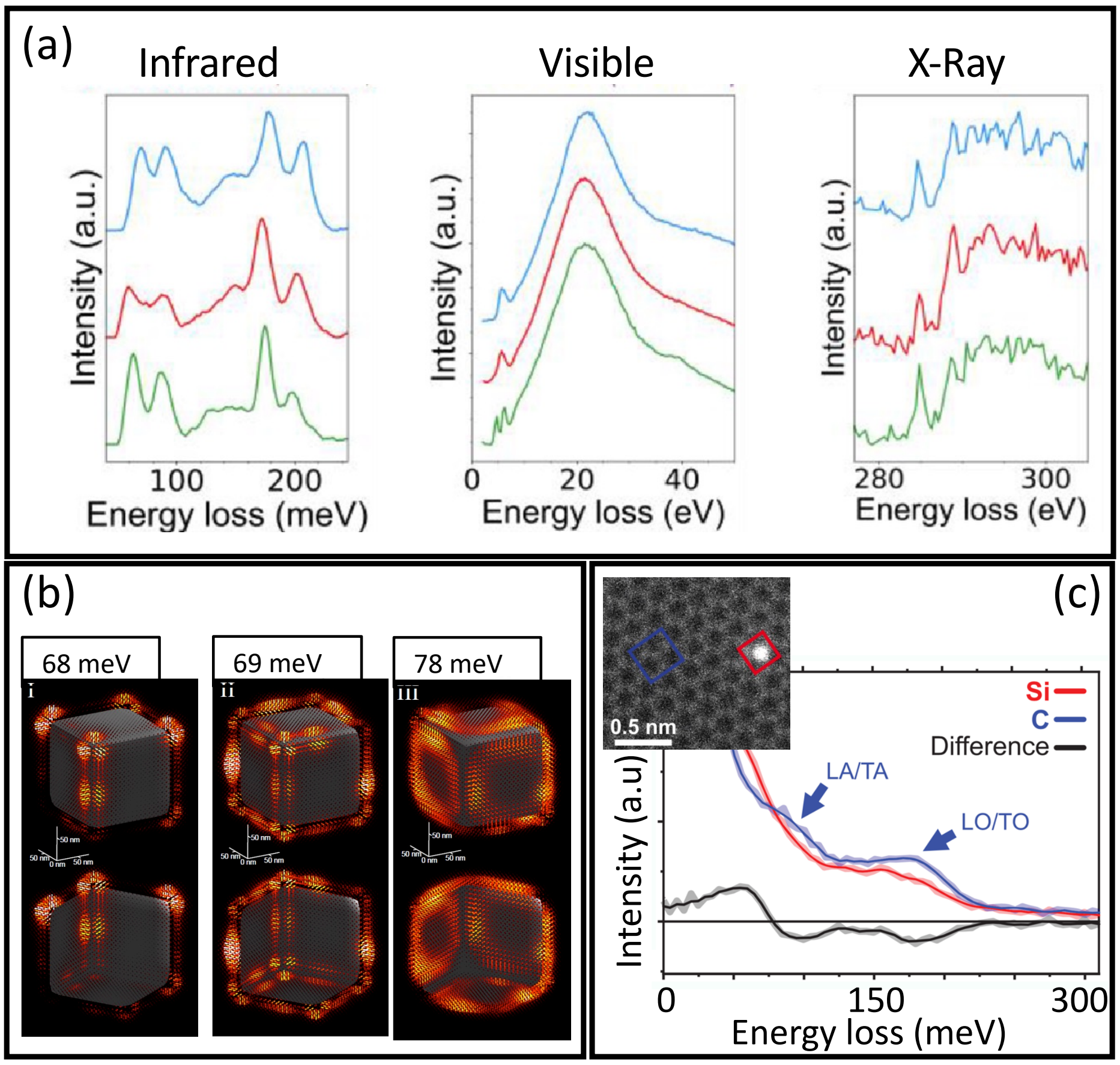

**Figure 3.** Highly monochromated EELS (a) The different energy ranges accessible with recently developed monochromators are illustrated here on three different types of metal-organic framework nanoparticles. Adapted from ref 59 (Copyright 2023 American Chemical Society). (b) Nanoscale 3D and vectorial mapping of plasmon polaritons confined at the surface of a MgO cube. Adapted from ref 73 (Copyright 2021 American Association for the Advancement of Science). (c) Measurement of the local phonon density of states in a graphene monolayer as modified locally by a Si substitute atom. Adapted from ref 77 (Copyright 2020 American Association for the Advancement of Science).

## 4. SPACE, TIME, AND PHASE RESOLUTION IN CATHODOLUMINESCENCE (CL) MICROSCOPY: STATUS AND OPPORTUNITIES

**Albert Polman[*] and Sophie Meuret**

### 4.1 Cathodoluminescence for Materials Analysis: Incoherent Emission

An exciting new research direction in CL spectroscopy is the development of time-resolved techniques to probe ultrafast phenomena at high spatial resolution. Ultrafast electrostatic beam modulators, placed in the electron column, can now create electron pulses in the SEM as short as 30 ps, possible using an electrical pulse generator[81] and 100 fs using photoconductive switching.[82] Sub-picosecond electron pulses can also be created in the SEM/STEM by photoemission of the electron cathode using femtosecond pulsed laser excitation. In a further advanced configuration, the laser pulse drives both the electron cathode and the sample, enabling pump-probe CL spectroscopy, in which the laser or electron pulse first excites a material, that is then probed after a well-defined time delay with the electrons or the laser pulse, as was first demonstrated for the state conversion of nitrogen-vacancy (NV) centers in diamond[83] (Figure 4a).

Time-resolved SEM-CL is now routinely used to map the carrier lifetime in semiconductors.[84,85] Time-resolved CL has also been demonstrated in the TEM, [86] and enables correlation of radiative emission with materials structure and composition at atomic resolution. The e-beam excitation of semiconductors leads to incoherent CL emission; there is no fixed phase relation between the incident electron and the emitted light, due to the femtosecond materials excitation-relaxation sequence upon electron excitation.

In photoemission, the number of electrons that is generated per laser pulse can be tuned in the range from 1 to 1000, enabling spectroscopies where the excitation density must be controlled. The excitation of semiconductors by a single electron creates many materials excitations and, thus, many photons, resulting in strong photon bunching in these CL experiments. [87,88] Measurements of photon bunching give insights into both carrier lifetimes and electron excitation probabilities. The correlation of CL and low-loss EELS using single-photon and single-electron detectors enables lifetime measurements of single-photon emitters and enhances the sensitivity of CL.[89]

### 4.2 Cathodoluminescence for Near-Field Imaging: Coherent Interactions

An exciting research area is the use of high-energy electrons as fundamental electrodynamic sources of coherent excitation of materials. When a swift electron passes through or near a material, its evanescent field drives materials polarizations that in turn create localized near fields that act back on the electron.[90] This interaction results in coherent CL emission: there is a well-defined phase relation between the electron impact and the emitted light.[91] Coherent CL has given many insights into the plasmonic resonances of single noble metal nanoparticles. Their CL spectra are characteristic of their size and shape, with the linewidth being a marker of their radiative and nonradiative dissipation and the coupling to their dielectric environment.[6] Similarly, e-beams can excite Mie modes in small dielectric particles and resonant whispering gallery-type modes in optical microcavities,[92] with the CL maps and angular profiles representing the multipolar field distributions.

Angle-resolved CL enables momentum spectroscopy to measure the local photonic band structure of periodic and aperiodic structures. In this way, cavity modes and interfaces in photonic and topological crystals and waveguides have been identified at deep subwavelength spatial resolution. Correlated measurements of CL and EELS in the coherent mode have been carried out where the energy loss heralds the generation of single (or more) photons.[19,21]

Polarization-resolved CL measurements enable the identification of the degree of linearly and circularly polarized light emitted from a sample. Electron excitation of specially structured surfaces has created CL with unique vectorial properties such as vortex e-beams.[93] Angle- and polarization-resolved CL has also created insights into the control of directional emission using plasmonic antennas that increase the performance of (nano-)lasers, light-emitting diodes, and solar cells. It has also inspired the development of optical metamaterials with unique properties enabling applications in imaging and integrated optics. An overview of earlier experiments in these fields is given in ref 7.

The ultrashort oscillation in time of the electric field carried by the electron corresponds to a spectral bandwidth from 0 eV to several 10s of eV.[94] As discovered by Smith and Purcell, an array of electron-excited dielectric scatterers can be used to effectively collect such light in the visible spectral range. [95] With the advance of metamaterials design, focusing these broadband CL in space and time pulses creates unique opportunities to perform materials spectroscopies with a time resolution down to the femtosecond time domain and a spectral range from deep ultraviolet to far IR[96,97] (Figure 4c).

As described in Section 2 of this Roadmap, the electron effectively probes the strength of a single Fourier component of the spatial distribution of the near field along its trajectory at a spatial frequency given by the electron velocity. In this way, CL is a unique metrology technique that characterizes 3D electromagnetic field distributions at the true nanoscale[98,99] (Figure 4d). As these field distributions are strongly linked to the materials' shape and composition, CL effectively also probes the 3D materials geometry at the nanoscale. The coherent phase relation with the incident e-beam enables holography and interferometry using the electron-generated CL signals, which creates further opportunities for 3D materials metrology with very high spatial resolution.[100] It also enables detailed studies of polaritonic excitations, for example, in excitonic 2D semiconductors, where far-field CL interference can probe the polariton dispersion relation, as described in Section 18.[101]

As it turns out, the strength of the electron-near-field interaction increases for lower electron energies,[98,99] and SEMs, operating in the 1–30 keV energy range, are ideal to carry out these studies. Electron excitation at even lower energies, in the 10–1000 eV range, is also appearing as a promising new field of research, taking advantage of the high interaction strength, and does not require the complex infrastructure of a complete electron microscope.

### 4.3 New Multidisciplinary CL Research Areas

Several new cross-disciplinary research opportunities are emerging. One upcoming application of resonant nanostructures is in light-driven sustainable chemistry, where light drives catalytic reactions at locally heated plasmonic nanoparticles, either in the gas or liquid phase. The CL emission may then provide an *in situ* fingerprint of time-varying reactions at very high spatial resolution, by probing the resonant properties of the plasmonic catalyst. Spectral shifts in CL can serve as a means to perform local thermometry. As a first step toward this goal, CL nanothermometry was demonstrated on semiconductor nanowires, where a band-gap energy shift serves as a sensitive probe of local temperatures (Figure 4b). The use of time-modulated excitation enabled the determination of the thermal conductivity at the nanoscale.[102]

Finally, we note the new development of PINEM that is reviewed in several sections of this Roadmap. Here, strong electron-near-field interaction creates a quantum-mechanical superposition state of the electron wave packet.[14] Entangling electron wave packets that are shaped in time and space with materials excitations may enable entirely new forms of ultrafast CL spectroscopy. Exploiting the entanglement of the electron states and the emitted CL photons may also provide a new way to perform quantum metrology. A further analysis of these opportunities is provided in the Section 5.

### 4.4 Applications

Advances in CL technology can drive the development of sustainable technologies such as energy-efficient lighting, high-efficiency photovoltaics, quantum technologies, and much more. Many new research directions in these fields lie ahead, and include plasmon-induced chemistry, optimization of photovoltaic materials, semiconductor metrology, and quantum CL spectroscopy, to mention just a few.

The CL spectroscopy modalities that we presented here offer several advantages for these application fields compared to conventional optical spectroscopy. First of all, the very high spatial resolution of SEM-CL and TEM-CL enables identification of defects, band gap variations, carrier lifetimes, temperature, and materials composition in semiconductors at much higher precision, enabling nanoscale studies in, for example, photovoltaics and solid-state lighting. Moreover, the high-energy electron cascade enables excitation of materials with very wide band gaps for which lasers are not easily available, which is relevant for solid-state lighting and high-power electronics. Similarly, in plasmonic materials, the broadband excitation spectrum carried by a high-energy electron can efficiently couple to deep-UV plasmons in materials such as gallium and indium. By varying the electron energy, and hence penetration depth, depth-resolved information can be achieved that cannot always be acquired with conventional optical spectroscopy.

In light-induced nanochemistry, operando CL may provide spectroscopic fingerprints that can be correlated to TEM tomography, providing atomic-scale information and chemical reaction pathways that cannot be probed by conventional optical techniques. With CL interferometry, nanoscale dimensions can be retrieved that cannot be easily acquired by conventional optical scattering or SEM imaging techniques.

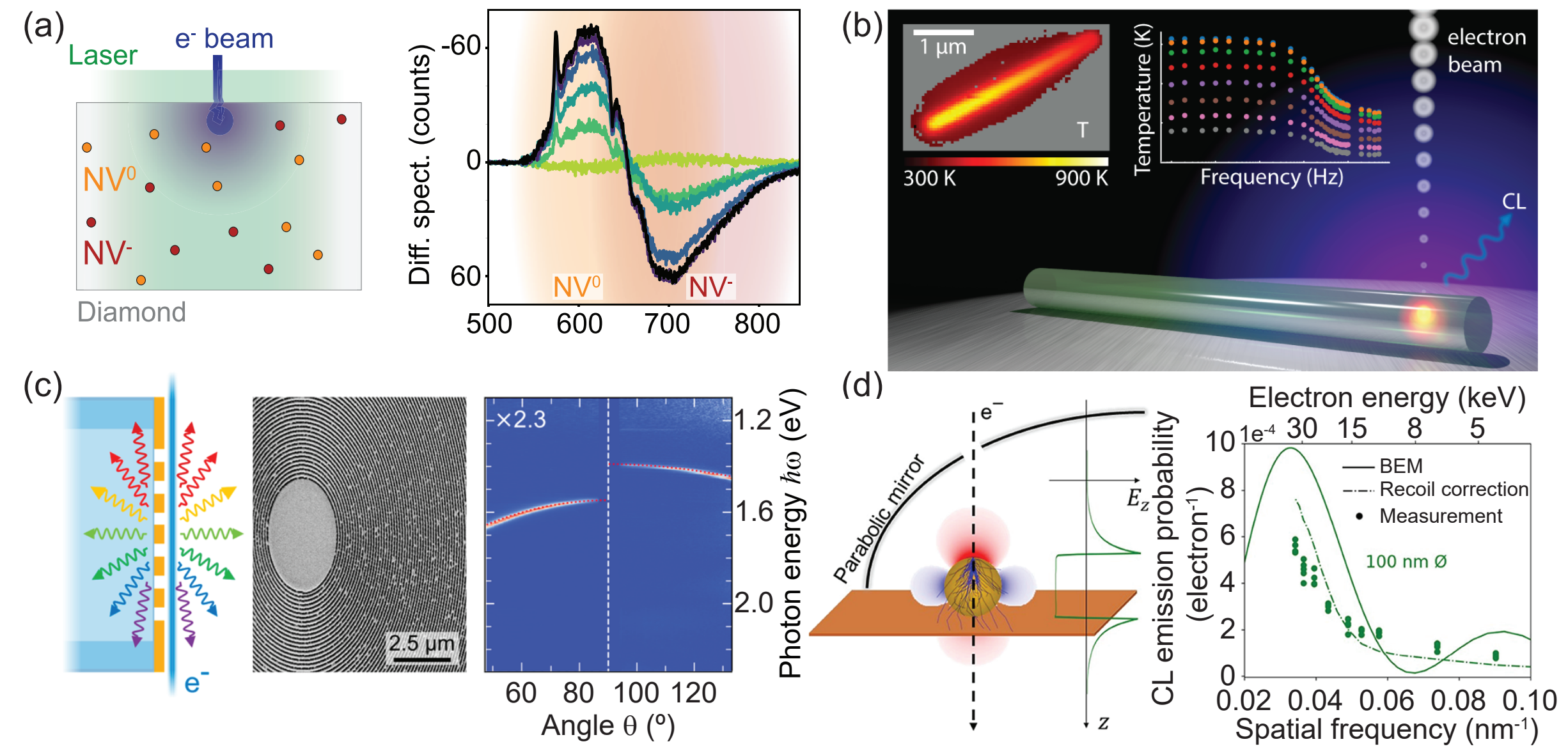

**Figure 4.** Recent advances in incoherent (a,b) and coherent (c,d) CL spectroscopy. (a) Pump-probe CL spectroscopy of electron-induced state transfer of nitrogen-vacancy (NV) centers in diamond. Adapted with permission from ref 83 (Copyright 2019 American Chemical Society). (b) Nanothermometry from semiconductor nanowires. Adapted with permission from ref 102 (Copyright 2021 American Chemical Society). (c) Collecting Smith–Purcell radiation from plasmonic bullseye-covered silica fibers. Adapted with permission from ref 97 (Copyright 2024 American Chemical Society). (d) Electron-beam near-field coupling strength dependence on electron velocity as a probe of characteristic spatial frequencies. Adapted with permission from ref 98 (Copyright 2024 American Chemical Society).

## 5. QUANTUM CATHODOLUMINESCENCE: POSITION- AND MOMENTUM-RESOLVED DETECTION

**Takumi Sannomiya* and Keiichirou Akiba**

### 5.1 State of the Art

Cathodoluminescence in SEM and STEM provides superresolution optical imaging surpassing the diffraction limit of light alongside structural imaging. For CL measurements, parabolic mirrors are commonly used to collect light. By placing a pinhole in the angular plane or by imaging it, angular or in-plane momentum information can be obtained (Figure 5a).[103,104,105] This momentum-resolved CL approach has been utilized to investigate dispersion relations,[104,106] discrimination of coherent and incoherent CL,[91] optical multipoles,[107] etc. Optical momentum-resolved measurements are also available in EELS,[108] which is complementary to the CL method since EELS provides information of non-radiating modes.[109] However, EELS suffers a tradeoff between spatial and momentum resolution: for momentum-resolved detection, the spatial resolution must be reduced to the micrometer akin to the purely optical measurement. CL bypasses this limitation by resolving space with electrons and momentum with photons,[109,110] which becomes useful for investigating quantum features.

Although the excitation position in CL is precisely controlled by the e-beam, the photon emission position has been mostly overlooked despite its importance for investigating spatially separate emission modes. Emission-position-resolved CL measurements have recently been demonstrated by using a parabolic mirror as an optical lens to optically image the emission position (Figure 5b),[111] or by placing an optical objective lens below the sample.[112] The former parabolic-mirror-based method offers additional flexibility by angle selection, allowing selection of the projection plane of the emission position imaging in 3D space including the $z$ axis (parallel to the e-beam). Emission-position imaging is particularly important for certain quantum CL measurements where the photon states from different positions (e.g., $|\psi_1\rangle$ and $|\psi_2\rangle$ corresponding to positions 1 and 2) should be detected as distinguishable spatial modes (Figure 5b).

In quantum domains, the CL photon intensity (second-order) correlation using Hanbury Brown and Twiss (HBT) interferometry (Figure 5c) has been performed quite intensively over the past decade. One of the pioneering studies revealed photon bunching,[87] which has been applied to the measurement of emission lifetimes, excitation efficiencies, etc.[113,114] The origin of the photon bunching effect was attributed to the inclusion of vacuum states between the photon states excited by single free electrons (Figure 5c).[115] By excluding the vacuum, it has been shown that the true CL photon statistics is different for coherent CL —essential for quantum photonics using free electrons—and incoherent CL involving multiple cascade excitation processes.[116]

### 5.2 Challenges and Future Goals

The coherent CL photon generation, when appropriately designed, satisfies energy and momentum conservation, leading to quantum entanglement between the primary electrons and emitted photons.[31,117] Using this electron–photon entanglement through a parametric scattering process, nonclassical light can be generated. For example, by selecting the energy of the scattered primary electron, specific photon number states can be extracted.[20] This electron-heralded light source substantially differs from existing quantum light sources, offering, for example, wavelength and bandwidth selectivity that is not available in current nonlinear crystal-based methods. A key advantage of this CL photon generation scheme is that the photon source is engineerable by the photonic structure, such as a photonic chip waveguide, undulator, etc.[21,118] Additionally, the polarization and angular momentum of photons can also be controlled or selected.[110] The potential applications of quantum CL extend beyond light sources: it could revolutionize microscopy, leveraging entangled electron–photon pairs generated in well-designed platforms. Ultimately, an electron–photon analog of a nonlinear optical crystal entangler might become available as an entangled electron–photon generator. Such a particle source is not only useful as a toy system for quantum physics experiments but also holds promise for advanced imaging techniques, enabling, for example, ultrasensitive electron microscopy, electron–photon ghost imaging, or CL measurement with complete phase information. This approach using entangled electron–photon pairs could potentially overcome some of the technological and fundamental challenges of quantum electron microscopy.[119]

### 5.3 Suggested Directions to Meet These Goals

While the entanglement of free electrons and photons has been extensively discussed and theoretically applied to various systems, its experimental verification remains unachieved. In contrast, simply extracting the interacted electron–photon pairs in coherent CL has recently been demonstrated by correlating energy-filtered electrons and emitted photons.[119,120] While energy selection offers one avenue, the abovementioned momentum selection (Figure 5a) provides an additional degree of freedom, for example, in the measurement basis conversion. Momentum-resolved measurement is also essential when handling recoils.[39] To assess such electron–photon correlations including entanglement, a quantitative measure of the correlation strength has been recently proposed.[121] Interference measurement based on the delayed-choice principle, commonly known as a quantum eraser, is also a way to verify the entanglement.[122] In quantum eraser experiments, photons emitted from two distinct sources must be detected as spatially separate modes, which is readily addressed using emission-position-resolved CL techniques (Figure 5b).[111]

In the intensity correlation measurement, apart from the need to extract the true photon state in single-excitation events (Figure 5c),[116] time resolution is a limiting factor because the lifetime of a coherent CL mode ($\lesssim 100$ fs) is typically far shorter than the resolution of typical detector systems ($\sim$1 ns), making, for example, single-photon state observation difficult. Techniques to convert the time dimension to a space dimension, such as Michelson interferometry, which enabled observing photon bunching in blackbody radiation, could potentially enhance time resolution down to femtosecond ranges.[123] First-order interference, such as homodyne measurements, would also become essential to assess phase information (i.e., off-diagonal elements or coherence terms) of the density matrix, which is not accessible by intensity correlations. Such interference systems incorporating reference sources could be integrated within a chip or nano- and micro-structures.[124] Finally, it is worth emphasizing that techniques or methodologies associated with the quantum CL approach hold significant technological and scientific values for advancing CL measurement itself and foster classical CL analysis.

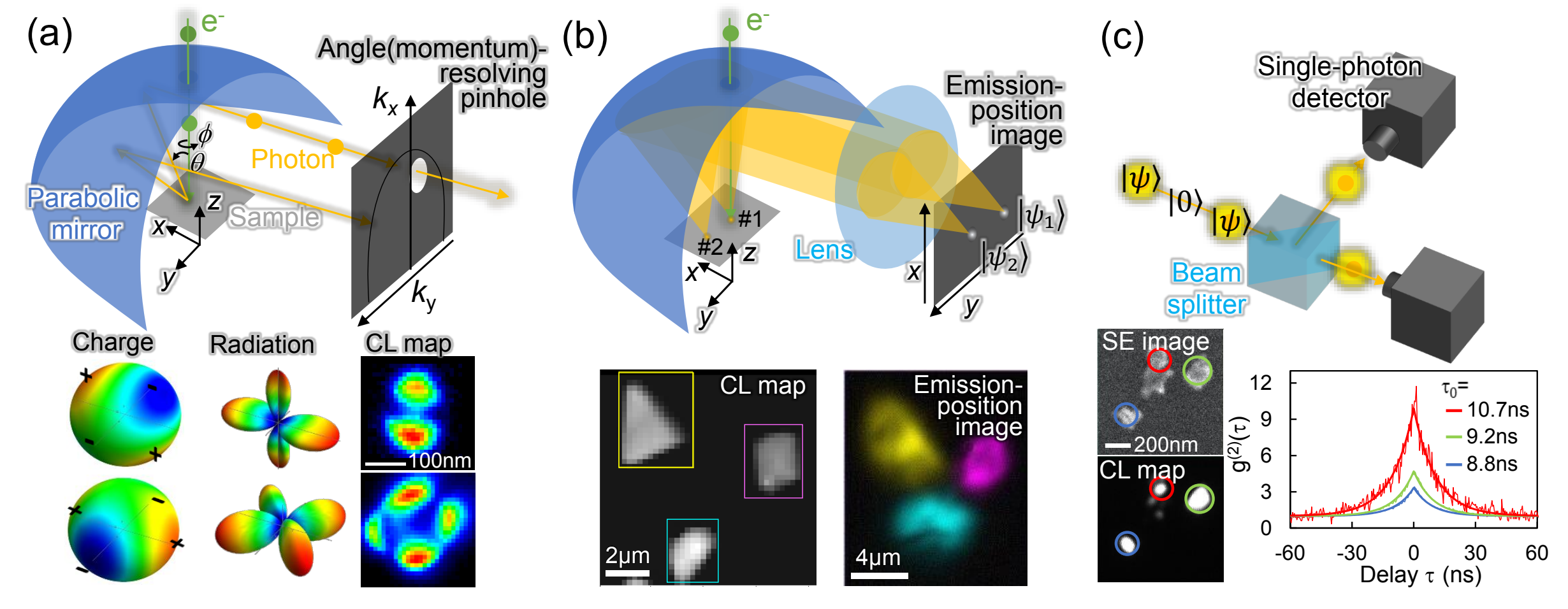

**Figure 5.** Illustrations of (a) momentum-resolved, (b) emission-position-resolved, and (c) Hanbury Brown and Twiss CL measurement setups with examples shown in the bottom insets. Adapted from refs 107, 111, and 114 (Copyright 2018 and 2022 American Chemical Society, and 2021 American Physical Society).

## 6. ELECTRON ENERGY-GAIN SPECTROSCOPY (EEGS) FOR MICROELECTRONVOLT/SUBNANOMETER ENERGY/SPACE RESOLUTION

**Mathieu Kociak,* Yves Auad, Armin Feist, Claus Ropers, and F. Javier García de Abajo**

### 6.1 Introduction

One of the most exciting aspects of e-beam science is the close connection between fundamental electron–light–matter interactions and their practical application in a broad range of fields, from condensed matter physics and materials science to biology. Facilitated by picometer-scale electron wavelengths, the spatial resolution of electron microscopes allows us to characterize material structures at all relevant distances, including atomic bonds. Beyond atomic-scale structures, e-beam spectroscopy grants us access to a wide variety of material excitations. For instance, in the spectral range of X-ray transitions, line broadenings typically exceed 150 meV[125] due to the very short core–hole lifetimes. In the visible range, even the sharpest plasmonic excitations exhibit linewidths of tens-to-hundreds of meV. Thanks to the remarkable technical advancements discussed in Sections 3 and 7, modern monochromatized (and even conventional) microscopes can directly map such excitations via EELS,[126] complemented by CL spectroscopy.[6]

However, the situation becomes more challenging for excitations with narrower linewidths and a complex mode structure, which have remained largely elusive for established e-beam techniques. Specifically, probing phonons, excitons, and high-quality-factor (high-$Q$) photonic modes require significantly higher spectral resolution, far below 1 meV. Specifically, phonon studies[3] necessitate resolutions of hundreds of μeV, while excitons are often barely resolved using EELS[127] or CL.[128] In addition, high-$Q$ photonic modes,[64] relevant in optomechanics and quantum nanophotonics, display linewidths as narrow as a few μeV and below (see Section 7).

A technique is therefore needed to reach a high spatial resolution alongside the selective probing of μeV spectral features. Unfortunately, the advancement spectral resolution in EELS has remained at a plateau around an (already impressive) spectral resolution of 3 meV for several years.[2] In addition, although CL may seem like an obvious alternative, as it leverages well-established optical technologies where spectral resolution can easily reach the μeV range, it suffers from relatively weak signals. Indeed, besides the challenges of adapting optical technologies to an electron microscope, CL relies on spontaneous emission processes, where a very narrow linewidth (i.e., a very high $Q$) implies a weak coupling to the far field, rendering measurements of ultrasharp spectral signatures challenging.

In a visionary discussion,[8] Archie Howie proposed using a laser within an electron microscope to study the optical properties of defects, mentioning in passing the possibility of observing energy gain. In 2008, two of us proposed using this physical phenomenon to combine the spectral resolution of a laser with the spatial resolution of an electron microscope—a technique coined EEGS.[9] Following Barwick *et al.*'s demonstration of inelastic scattering by photons in 2009,[13] it took another thirteen years before Henke *et al.*[10] measured an EEGS spectrum far surpassing the resolution attainable in EELS. This section briefly describes the principles of EEGS, the initial steps made to demonstrate it, and some prospects for future developments.

### 6.2 The EEGS Principle

In brief, the principle of EEGS can be explained as follows[129] (see Figure 6a): if a laser source irradiates a nanostructure, the scattered field can couple to the electron, thus producing inelastically scattered electron components; in particular, electrons that gain energy (in quanta of the photon energy) can be resolved and provide a measure of the strength of the optical field at the applied optical frequency; the areas of such gain features integrated over energy gain and plotted as a function of photon energy $\hbar\omega$ permit us to build a spectrum of the optically active excitation modes in the specimen, with a spectral resolution depending on our ability to monochromatize the laser, which is independent of the electron energy resolution (determined by the incident electron energy width and the used electron spectrometer), provided the photon energy exceeds such an instrument-intrinsic electron resolution. We note that the technique associated with mapping the gain signal in space at one specific energy is commonly referred to as PINEM. More precisely, an electron traversing the specimen with constant velocity is exposed to light components with spatial frequencies placed outside the light cone, thus breaking the photon–electron mismatch in free space (see Section 2). As a consequence of this interaction, the electron develops a series of energy sidebands indexed by integers $\ell$ and having probabilities $P_\ell = J_\ell^2(2|\beta|)$ (see eq 2.3), where $\beta$ is the electron–light coupling coefficient defined in eq 2.4.[29,35] By measuring electron spectra (and thus $|\beta|$) as a function of the laser energy, one can then deduce the spectral response of the specimen at each spatial point. In general, the coupling parameter can be retrieved by fitting the modulated electron energy spectrum[14] and integrating the area under the gain peak in the linear limit,[11] as noted above and initially proposed[9] and explained in Figure 6a. As this form of excitation spectroscopy is largely independent of the EELS resolution (see above), the technique is essentially limited in spectral resolution by the laser bandwidth (see the difference in resolving mode peaks along the electron spectrum axis and the laser wavelength axis in Figure 6b). Using excitation by laser pulses, the achievable spectral resolution $\sigma_E$ (standard deviation) is given by the optical bandwidth, constrained by the uncertainty principle $\sigma_E\sigma_t \geq \hbar/2 \approx 0.329$ eV fs, where $\sigma_t$ is the optical pulse duration. Using electron pulses, only the optical bandwidth of the fields overlapping with the electron pulse becomes relevant.[17] As discussed below, this can be used to enhance spectral resolution by stretching laser pulses to a duration exceeding the electron pulse. Depending on $\sigma_t$, EEGS and related techniques first allowed surpassing the limits imposed by the EELS spectral resolution of the microscopes on which PINEM experiments were conducted and eventually led to record-high combinations of spectral and spatial resolution (see below). It should be noted that, in contrast to PINEM, EEGS does not require multiple sidebands, as it relies on the determination of the optical near-field strength through the measurement of the (optionally energy-integrated) gain side of the electron spectrum.

### 6.3 Overcoming the Spectral Resolution Limit Imposed by EELS

During the development of PINEM,[13] which initially relied on laser pulses with durations of hundreds of femtoseconds, the energy resolution of EEGS and its derivatives was limited to a few meV (by the uncertainty principle in the laser pulses). Although not surpassing the capabilities of high-resolution EELS (Section 3), this was a remarkable advance when comparing the spectral resolution achieved with EEGS to that of the microscopes in which the experiments were conducted (ranging from 600 to over 1000 meV). The first demonstration was performed on plasmonic modes of a nanoantenna, with peak widths typically in the range of tens of meV.[130] In these experiments, due to the Boersch effect[131] in the electron pulses, the spectral resolution in EELS was degraded (~ 6 eV) to the point that the PINEM replicas were no longer visible and EEGS spectra were reconstructed from variations in the zero-loss peak. Remarkably, this version of EEGS achieved a resolution of 20 meV under these conditions. Subsequently, a related technique was used to retrieve the band structure of a photonic crystal by measuring electron energy spectra as a function of the angle and wavelength of the incident laser[132] (Figure 6c). Again, a femtosecond laser was used, limiting the resolution to $\sim 10-20$ meV. A similar technique was developed using dispersively stretched broadband optical pulses to encode narrowband spectral information in time.[133,134] The delay between a very short electron pulse (200 fs) and a picosecond laser pulse was controlled so that the electrons saw a different wavelength for each delay time, thus achieving a resolution of 10 meV on a silica microsphere[133] and the dielectric modes of thin transition-metal-dichalcogenide films,[134] limited by the lifetime of the excitations.

### 6.4 Sub-meV to μeV EEGS Spectroscopy

To achieve a resolution exceeding that of highly monochromatized electron microscopes, several obstacles had to be overcome. In this direction, two works demonstrated the possibility of performing PINEM (not EEGS yet) experiments on plasmonic nanostructures with sub-meV laser linewidth, using either nanosecond pulsed lasers[135] or continuous-wave (CW) lasers.[136] This is relevant because the effective cross-section of PINEM—and thus, the signal-to-noise ratio of EEGS—depends on the instantaneous power of the laser field (i.e., the near field $E_z$): for a given average laser power, higher peak field amplitudes are possible for shorter pulse durations. However, the main challenge in substantially improving the spectral resolution suffers from a drawback analogous to CL (see above): it is challenging to efficiently couple far-field light to a nanostructure with a very narrow linewidth (i.e., a very high $Q$). The aforementioned approaches were separately explored by the authors. In one of them, a frequency-stabilized CW laser was coupled to a waveguide, which was in turn coupled the optical near field of a high-$Q$ photonic resonator (a ring microresonator with $Q\sim 0.7\times10^6$, see Figure 6d). This approach led to optimal light coupling into the resonator, resulting in a high signal-to-noise PINEM signal for a modest injected power. This permitted the measurement of a 3.2 μeV linewidth, with line shape features reaching the nanoelectronvolt range. Moreover, the efficiency of near-field coupling was later exploited to study nonlinear effects.[137] However, this method required the development of custom sample holders and samples. Conversely, a far-field coupling approach was developed, where a nanosecond laser beam was focused on a spot a few optical wavelengths in size, positioned with subwavelength precision close to a spherical microresonator. Optimal light-to-sphere whispering-gallery-mode (WGM) coupling was achieved with the technical challenge now shifted from sample and sample-holder fabrication (near-field coupling) to using a high-numerical-aperture, high-precision light injection system (far-field coupling) inserted in a highly stable, monochromatized electron microscope.[11] Performed under these conditions, EEGS revealed its superior spectral resolution and signal-to-noise ratio compared to EELS and CL (Figure 6b): under the same geometrical conditions, the EEGS signal was found to be 20 times larger in EEGS compared to CL, the latter being below the noise level.[11] Spectroscopy of WGMs with $Q$'s as high as $10^4$ (linewidth < 200 μeV) in nanospheres was performed in this way. Ultimately, only the laser linewidth and stability determine the achievable spectral resolution (e.g., $\sim$40 peV for at 10 kHz frequency-stabilized laser). Of course, a practical and meaningful limit is the intrinsic linewidth of the probed excitation.

### 6.5 Current and Future Challenges in EEGS

EEGS provides a spectral resolution exceeding by orders of magnitude the one that can be achieved with EELS, while producing spectra with a better signal-to-noise ratio than CL. In the first demonstrations of EEGS, due to the large volume of the investigated modes, the spatial resolution of the e-beam was only partially harnessed. However, a combination of spatial and spectral resolution like that in EEGS is required to investigate low-mode-volume resonators[65] or image more complex spatial modes. Additionally, extending EEGS to the IR domain is highly desirable, as that kind of spectral resolution is far from attainable with EELS. Although EEGS works best for bright modes that feature a large coupling to light (i.e., in-coupling of externally supplied light), one could exploit the interaction of dark modes with optical nanoantennas acting as intermediate coupling elements, thus suggesting a form of EEGS assisted by additional material structures that mediate the interaction between the external light and optically dark modes in a specimen (e.g., nondipolar excitons and acoustic vibrations[67,68]).

As an interesting possibility for future developments in EEGS, one could leverage the fact that energy resolution is provided by one particle (the photon) while time resolution can be imprinted in another particle (the electron, for example, via PINEM modulation). Specifically, one could achieve a combination of temporal resolution $\Delta t$ and spectral resolution $\sigma_E$ below the uncertainty limit[30] (i.e., such that $\sigma_E\sigma_t < \hbar/2$, which would be impossible if these quantities were associated with the same particle). We thus envision using spatiotemporally pre-shaped electrons in the form of energy combs with a wide energy spacing (e.g., $\sim$2 eV) in their sidebands, combined with the analysis of inelastically scattered electrons similar to EEGS but using a lower, scanned photon energy (< 2 eV) as a way to reconcile sub-fs time resolution and sub-meV energy resolution.

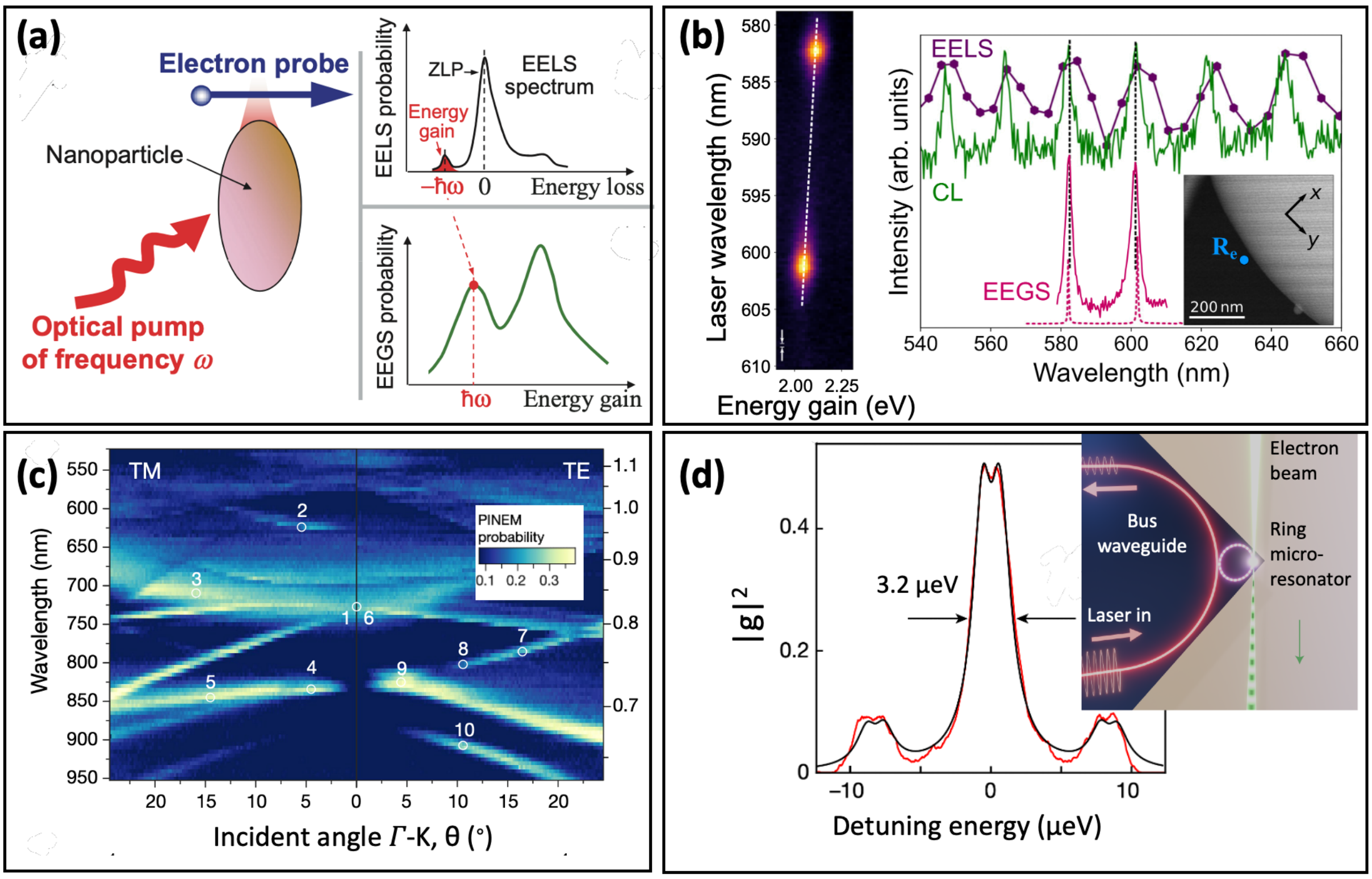

**Figure 6**. Electron energy-gain spectroscopy (EEGS) (a). EEGS principle: a nanostructure is irradiated with a laser beam of a given energy/wavelength. The electron probes the induced near-field, and its energy spectrum presents an energy-gain peak at the photon energy. In its simplest realization, the EEG spectrum is reconstructed by scanning the laser wavelength, measuring the area under the gain peak, and plotting the latter as a function of light wavelength. Adapted from ref 9 (Copyright 2008 Institute of Physics). (b) Experimental realization of EEGS. Left: electron spectra (gain side) as a function of laser wavelength, taken on a ~ 4 μm silica sphere illuminated from the far field; two WGMs are resolved. Right: comparison of EELS, CL, and EEGS signals measured at the same e-beam position, showing the superiority of EEGS in terms of spectral resolution (cf. EEGS and EELS) and signal-to-noise ratio (cf. EEGS and CL). Adapted from ref 11 (Copyright 2023 Springer Nature). (c) Band structure of a photonic crystal (see inset) revealed by spectrally resolved PINEM. Adapted from ref 132 (Copyright 2020 Springer Nature) (d) Record spectral resolution obtained on a ring microresonator coupled to a CW laser in the near field. Adapted from ref 10 (Copyright 2021 Springer Nature).

ULTRAFAST ELECTRON–LIGHT INTERACTIONS

## 7. ULTRAFAST ELECTRON MICROSCOPY

**Armin Feist,*  F. Javier García de Abajo, and Claus Ropers**

### 7.1 Introduction, Background, and State of the Art

Electron microscopes excel in analyzing the static atomic-scale structure and electronic properties of complex materials. However, understanding nonequilibrium behavior and resulting functionalities requires the study of dynamical processes in response to external stimuli. Acquisition speeds in conventional transmission electron microscopes using continuous e-beams are limited by the employed detectors, typically reaching milliseconds for full-frame recording or microseconds for fast spectroscopy. Unfortunately, many relevant nanoscale phenomena, including electronic excitation and relaxation, energy transfer, and structural transformations, occur on picosecond-to-femtosecond timescales, or even faster. This requires a measurement technology that enables faster observation of transient states of matter following tailored excitation.

A growing set of methodologies comprising ultrafast electron microscopy (UEM) accomplishes this goal by probing dynamics in a specimen with a time-structured e-beam at temporal scales much faster than the framerate of the employed detector. Inspired by ultrafast optical pump-probe spectroscopy, a pulsed (usually optical) stimulus (the *pump*) excites an investigated specimen, and the resulting dynamical changes are tracked using a pulsed e-beam (the *probe*) after a variable time delay $\Delta t$ (see Figure 7, left).

Early implementations of ultrafast electron imaging using picosecond stroboscopic e-beams in SEM or nanosecond electron pulses in TEM date back several decades.[138] Combining these approaches and propelled by the availability of high-quality femtosecond lasers, ultrafast TEM reached sub-picosecond and nanometer resolutions by using photoemission of low-charge electron pulses from planar photocathodes.[139,140] The exceptional coherence of field-emission sources was translated to the ultrafast domain in TEM[14,141,142,143,144] and SEM,[81,145,146] allowing for the full capabilities of modern electron microscopes to be harnessed at high temporal resolution. While most dynamical measurements have been carried out using photoemission electron sources, ultrafast beam blanking is being pursued in parallel.[81,147,148,149]

Enabled by these technological advances and the unique capabilities of the simultaneous nanometer-femtosecond spatiotemporal resolution, a growing community of TEM/SEM researchers is exploring a broad range of ultrafast physical phenomena (see Figure 7, right). Examples include the nanoscale study of ultrafast phase transitions,[150,151] optically driven phononic systems,[152,153,154,155] ultrafast magnetism,[156,157] and carrier dynamics in semiconductors.[84,145] Furthermore, inelastic electron–light scattering in optical near fields facilitates the study of optical excitations such as surface-plasmon-polaritons,[17,140] propagating phonon-polaritons,[158] and cavity modes.[132,137] (see Section 9).

In the following, we provide a perspective on anticipated instrumental advances, new techniques, and promising applications in the field, emphasizing selected long-term challenges and opportunities.

### 7.2 A Platform for Nanoscale Light–Matter Interaction

Ultrafast electron microscopy presents us with unique tools to address fundamental questions in a broad range of subjects, from nanoscale energy transfer and transformations in correlated materials for spintronics and ultrafast electronics to free-electron quantum optics and photonics (see Sections 13 and 14). Equipped with a great flexibility of possible excitations and a vast range of complementary observables (see Figure 7, center), ultrafast electron microscopes are able to capture energy conversion processes as well as couplings among different microscopic degrees of freedom in materials via their time-domain evolution far from thermal equilibrium.

Femtosecond laser pulses can be used for tailored electronic and vibrational excitation as well as localized heating and the generation of thermal stress. Nonlinear field-driven processes are accessible from the visible to the mid-IR and terahertz ranges. Excitations can be supplied by free-space radiation, waveguides, antennas, or nanofabricated optically triggered current switches. Conceptually, reversible dynamics are observed in a stroboscopic pump-probe fashion, while ultrafast quenching promotes the study of laser-induced long-lived metastable states.[159] Current research focuses on extending these excitations into a broader frequency range and designated nanoscale sample environments, including high-frequency currents[160] and strongly localized optical excitations.[150] Here, future sample designs will harness localized secondary excitations, such as laser-induced sound waves, free charge carriers, and propagating optical modes. All of these phenomena can then be probed with nanoscale resolution by ultrafast free-electron pulses.

A particular strength of electron microscopes is the broad selection of external control parameters, commonly used for *in situ* experiments, which allow for the investigation of the response of materials to external perturbations, based on a well-adjusted thermal equilibrium state. This includes static electromagnetic fields, base temperatures, static compression, and the gas/liquid environment (e.g., as needed for studying nanoparticle catalysis). The availability of magnetic field-free electron lenses and cryogenic sample stages further extends these capabilities.

The induced ultrafast dynamics is routinely sampled using the versatile imaging, diffraction, and spectroscopy capabilities of state-of-the-art electron microscopes. These consist of direct imaging of atomic positions, lattice deformations, and structural phase changes in high-resolution bright- and dark-field imaging. Slowly varying strain, electromagnetic fields, and local magnetization can be imaged by phase-sensitive techniques. Using scanning probing of a focused e-beam grants us quantitative access to local structures, electromagnetic potentials, electronic occupations, optical near fields, and more.

### 7.3 Novel Measurement Schemes

Harnessing the particular coherence of the e-beam, pulsed-field emitters enable elaborate techniques like ultrafast Lorentz microscopy,[156] ultrafast darkfield imaging,[150] and nanoscale diffractive probing[152,153] with femtosecond time resolution. With more coherent pulsed electron sources, ultrafast (STEM) holography and atomic resolution ptychography come within reach. Regarding spectroscopy, future developments in time-resolved electron microscopy will aim to approach the time-bandwidth limit in ultrafast probing, using EEGS and μrad-meV angle-resolved phonon spectroscopy (see Sections 3 and 6).

A promising approach not traditionally available in *in situ* electron microscopy involves laser quenching for the preparation of metastable states in magnetism[157,159] and structural biology.[161] Another unique capability stems from the recent preparation of Coulomb-correlated few-electron states in a TEM[47] that enable shot-noise-reduced imaging and the probing of materials with a well-defined sequence/number of electrons. This may be particularly useful for studying delocalized material excitations and resonant sample responses, in which the momentum and time correlations of probing electrons are harnessed to access intrinsically correlated excitations. Such correlation-enhanced probing techniques rely on event-based electron detection, as also discussed in Section 10. Finally, optical phase modulation and coherence transfer can result in new contrast mechanisms and imaging modalities, accessing the optical phase-coherent sample response and the state of individual quantum systems (see Sections 13 and 14).

### 7.4 Opportunities from Functional Nanostructures to Biology

Many experiments in UEM are inspired by open scientific questions in ultrafast science that remain unresolved by spatially integrating techniques such as ultrafast optical spectroscopy or ultrafast electron diffraction. Electron probing techniques are suited explicitly to probe lattice degrees of freedom due to their strong diffraction by atomic nuclei. In addition, complementary coherent imaging and inelastic interactions with optical near fields give direct access to electrical and magnetic fields as well as photonic modes. This yields unique capabilities to simultaneously study electronic, lattice, and spin excitations during nonequilibrium evolution, rendering ultrafast transmission electron microscopy an ideal technique to probe energy conversion and dissipation processes in nanostructured materials.

Regarding future prospects, there is a strong case for studying correlated materials characterized by strong couplings between microscopic degrees of freedom. Cryogenic (liquid-helium) specimen holders and resonant IR or terahertz driving promise access to low-energy excitations. Integrating such advanced ultrafast light sources poses experimental challenges, but will enable the investigation of nanoscale light-matter interaction in quantum materials and assist in designing new tailored functionalities.[162] A complementary approach to study these materials in thermal equilibrium is direct probing with meV-resolution using monochromatized e-beams (see Section 3). This will bridge the field of UEM with ultrafast Terahertz science,[163] high-harmonic-generation spectroscopy,[164] and free electron lasers.[165]

Quantum metrology remains largely unexplored in this field. Recent work has proposed a method to boost sensitivity and resolution in the determination of optical phases by measuring electron currents alone after strong electron interaction with waveguided photons.[26] In addition, high-frequency measurement schemes in transmission (TEM) or reflection (SEM and low-energy electron microscopy, LEEM) geometries may provide enhanced sensitivity and new contrast mechanisms for precision measurements and single-defect characterizations in 2D and bulk semiconductor structures. Further possibilities span the imaging of functional devices, including operating micro- and nano-electromechanical systems (MEMS/NEMS), magnetic storage, logical circuits, and potentially superconducting qubit structures with a stroboscopic e-beam at megahertz-to-gigahertz sampling rates.[147,160]

Beyond the proof-of-principle demonstration of high-resolution imaging using a pulsed e-beam, ultrafast atomic-resolution imaging of laser-excited samples remains an outstanding challenge. Thermal drifts will require strategies for long-time exposure and high-coherence ultrafast electron probes to implement dose-efficient techniques such as ptychography. Similar constraints apply to the real-space imaging of coherent optical phonons. While highly monochromatized electron microscopy can map the phonon density of states and thermal occupations (see Section 3), dynamical studies using ultrafast electron pulses are presently restricted to momentum-space observations using thermal diffuse scattering. Future studies at higher contrast and resolution may trace combined real- and reciprocal-space ultrafast phonon scattering and dissipation cascades.

Nonequilibrium dynamics in biological systems is another interesting field that can benefit from cryo-TEM and its power to resolve the structure of proteins via single-particle ensemble tomography. An ambitious goal is to add ultrafast time resolution to study transient structures or even folding dynamics, as well as energy transfer and photoactivated processes. Similarly, environmental gas- and liquid-phase experiments may be complemented by optical excitation to investigate (photo-)catalytic reactions at the atomic scale, although stochastic processes and irreversible dynamics will present a major challenge.

### 7.5 Future Enabling Technology

A central challenge in ultrafast electron probing is the limited time-averaged brightness of the pulsed e-beam for coherent and high-resolution imaging applications. Continuous electron guns optimize the transverse beam coherence using field emitters (see Section 20). In UEM, Schottky-[141,145] and cold-field emitters[142,143] are in use, and in particular, linear single-photon photoemission yields highly tunable electron sources.[141] Further improvement of transverse beam brightness may be achieved by novel source concepts, including carbon-nanotube or $LaB_6$ needles, low-emittance planar photocathodes, and radiofrequency or electric e-beam chopping/blanking.[81,147,148,149] Another flavor of time-resolved electron microscopy uses high-charge electron pulses, particularly useful in low-contrast diffraction and for single-shot imaging. Here, the main challenge is to overcome the Coulomb-induced pulse degradation. Promising strategies for mitigation are tailored electron guns with high-acceleration fields, MeV beam energies, and time-dependent chromatic aberration correction.

Regarding longitudinal phase space, attosecond-bunched and optically phase-structured e-beams will drive the evolution of attosecond microscopy.[16,17,18] Such modulated e-beams also hold promise for studying coherent material excitations using CL (see Sections 4, 5, and 18).

A fundamental phase-space limit is imposed by the uncertainty principle for the pulse duration and the energy width, $\sigma_E\sigma_t \geq \hbar/2$, which translates into a bandwidth-limited 3.65-fs pulse duration (full width at half maximum (FWHM), $\sqrt{8\ln 2}\,\sigma_t$) for a 500-meV (also FWHM) spread in beam energy. Current electron guns provide pulses close to two orders of magnitude longer in duration, even in the single-electron limit. Underlying technical and fundamental reasons include a propagation-induced dispersive broadening, the bandwidth of the photoemission process itself, and fluctuations in high tension. Some of these issues can be overcome by more stable microscopes and active electron pulse compression schemes (see Section 8), such that, ultimately, nanoscale imaging and spectroscopy in the few-fs/few-meV range may come into reach.

Existing aberration correction will enable higher current densities and faster acquisition times for nanoscale local probing and spectroscopy (STEM-EELS/CBED/4D-STEM), further enhanced by low-noise, high-detective-quantum-efficiency direct electron detectors for shot-noise-limited electron imaging. Regarding tailored electron optics, complementing recently established optical phase plates (see Sections 11 and 16), light-driven e-beam shaping promises beam splitters and aberration correctors with femtosecond switching capability, also harnessing new contrast mechanisms (see Section 11).

Current UEMs are based on TEM or SEM instruments, but the approach can be translated to other platforms. This includes transmission SEM (STEM-in-SEM), which promises flexible sample geometries, or state-of-the-art LEEM instruments for ultrafast surface-sensitive imaging and diffraction. Besides the use of free e-beams, other techniques such as photoemission electron microscopy (PEEM), scanning near-field microscopy (SNOM), and scanning tunneling microscopy (STM) provide complementary information in ultrafast nanoscale imaging.

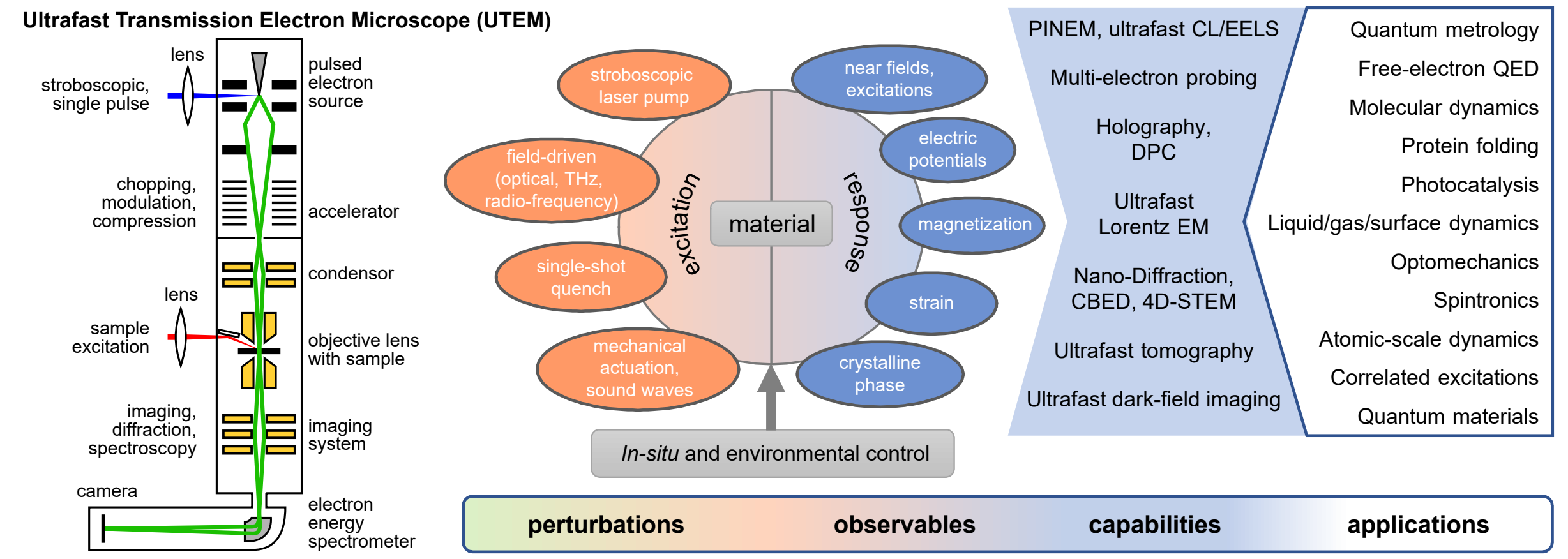

**Figure 7.** . Ultrafast electron microscopy for the study of nanoscale dynamics: instrument, concept, and future prospects. Left: A conventional electron microscope is integrated with a pulsed electron gun, realizing a laser-pump–electron-probe measurement scheme. Center: Various sample stimuli can excite a material out of equilibrium, and electrons yield versatile information about the transient sample state. Right: The vast toolbox of available ultrafast techniques promises novel applications extending beyond ultrafast condensed matter physics.

## 8. SINGLE ELECTRONS AND ATTOSECOND ELECTRON MICROSCOPY

**Peter Baum**[*]

### 8.1 Introduction and Overview

From a fundamental perspective, the foundations of nanophotonics are electrical and magnetic fields that oscillate in space and time on dimensions much smaller than the wavelength of light. While optical spectroscopy or related methods can reveal the overall response of a macroscopic material, a fundamental insight requires access to electric and magnetic fields with a resolution that resolves the optical cycles of light in space and time.[16] Ultrafast electron microscopy (see Section 7) with single electrons[166,167] is one of the most established and promising methods for accessing such a regime because high-energy electrons have a de Broglie wavelength in the picometer range that allows focusing a beam down to the size of an atom or less. Also, it is possible to generate ultimately short pulses in the attosecond[14,15,168,169] and sub-attosecond regime.[170,171] In addition, electrons are sensitive to dynamical electric and magnetic fields due to their elementary charge.[172] Therefore, many researchers contribute to advancing electron microscopy to ultimate time resolution and sensitivity for measuring electrical and magnetic dynamics in and around nanostructured materials. This section concentrates on the contributions from our laboratories; see the other sections and the references in the cited papers for a more detailed and balanced overview.

If an e-beam shall be focused tightly in space and time to nanometer and attosecond dimensions, it cannot contain much more than one or a few electrons per pulse[167]. In pulses with much more than one electron per pulse, space charge effects broaden the spectrum and prevent compression in the time domain.[173] Ideally, only one electron is generated at the source and later never filtered away.[167,174] Its properties are then determined by the physics of the photoemission process and unaffected by space charge effects.[175] However, even the most modern TEMs (see Sections 3 and 7) currently still generate hundreds of electrons per pulse of which only a tiny fraction, typically 0.01-5 electrons per pulse, are later selected by apertures for experiments. The phase space volume of the single electrons then expands by intra-pulse heating effects, and the post-filtered electron pulses are less coherent than they could be.[175] However, the generated single-electron or few-electron state can be further manipulated and controlled in the time–energy domain by microwaves,[176] terahertz pulses,[177] or by the optical cycles of laser light[14,15,168,169] to produce electron pulses that are shorter than one optical cycle of terahertz or near-IR light. This compression usually only works with an additional structure at the interaction point of the electron and photon beam as a modulator element, because the interaction of photons with electrons is mostly forbidden in free space. The third-body element can be a subwavelength resonator,[177] a nanometer needle tip,[14] or a free-standing thin-film membrane that is transparent for both photons and electrons.[178,179] Pulses shorter than one femtosecond can be created in this way,[14,15,168,169] enabling attosecond electron microscopy.[16,17,18] It is also possible to form single electrons into a chiral coil of mass and charge.[180]

In principle, single electrons can be compressed in the time domain as much as desired as long as the product of pulse duration and energy bandwidth remains within the limits of the uncertainty principle.[168,175] However, the laser damage threshold of the modulator element limits the field strength of the optical cycles and thereby the achievable electron pulse duration.[175] This limit can be circumvented by using two photons to control one single electron in free space without any further material.[170,171] Indirect spectroscopic evidence has recently been reported on the generation of 5-as electron pulses in the form of a sequence or pulse train.[171] Isolated electron pulses can be created by single-cycle filtering[181] or ponderomotive control.[182]

Using these technologies, it recently became possible to use an attosecond electron pulse train in a TEM to image the optical response of a nanophotonic material on the level of the cycles of light[16] (see Figure 8). We create attosecond electron pulses[15,169] and let them pass through or around an optically excited nanostructure. These electron pulses are then accelerated or decelerated in the time-frozen electromagnetic fields and acquire a position- and time-dependent energy gain or loss. Using a stroboscopic technique and an imaging energy filter then allows one to produce a movie of the electric fields in and around the material.[16] Alternatively, advanced holographic techniques with phase-modulated e-beams provide similar information without the need for attosecond electron densities.[17,18] The ability to see the optical electric fields in and around nanostructures or metamaterials with a spatial resolution smaller than one wavelength and a temporal resolution better than half an optical cycle period provides probably the most direct and fundamental insight into the response, functionality, and quantum properties of a nanophotonic material.

**8.2 Challenges and Future Directions**

A highly coherent and efficient production of single-electron pulses is probably the most central prerequisite for all ultrafast electron microscopy and attosecond imaging of nanophotonic materials. So far, the product of energy and time is far away from the fundamental limits of a matter wave. Even with the best available needle emitter tips (see Section 7), the electron pulse duration is hundreds of femtoseconds at an energy bandwidth of hundreds of meV. The product is ~100 times worse than theoretically allowed by the uncertainty principle. An ongoing challenge in quantum nanophotonics is therefore the production of single-electron pulses, or few-electron emission events, at the best possible product of pulse duration and energy spread. Ideally, one genuine single electron, not the typical $0.01-0.1$ electrons per pulse, is shaped into a beam of picometer diameter and pulses of attosecond length. We expect that smaller needle tips, cycle-driven field emission,[183] or emitter materials with narrowband structures can be helpful for this goal. If such a beam can be achieved, it would not only be relevant for ultrafast microscopy and attosecond nanophotonics but also useful for ongoing endeavors on the quantum physics of the electrons themselves (see Sections 13 and 14), for example, the generation of qubits.[184,185]

In attosecond electron microscopy, one of the desirable demonstrations is a measurement of optical nonlinear effects and single-cycle response. We expect that isolated sub-femtosecond electron pulses[181] or a ponderomotive selection of one of multiple spikes[182] might be a valuable way. In ultrafast electron diffraction, researchers have already seen the dynamics of electric and magnetic fields in nanostructures[186] but not yet the motion of valence electrons in crystalline materials.[187] The direct signal from such dynamics is very weak[187] and beyond the capabilities of modern instruments,[188] but we expect that a clever heterodyne detection[17,18] may provide access. We hope that many researchers take up on these challenges and join us in using non-filtered single electrons[167] under the control of laser light[168] for attosecond imaging,[16] to create novel and exciting ways for future investigations of quantum phenomena on nanometer, atomic, and subatomic scales.[189]

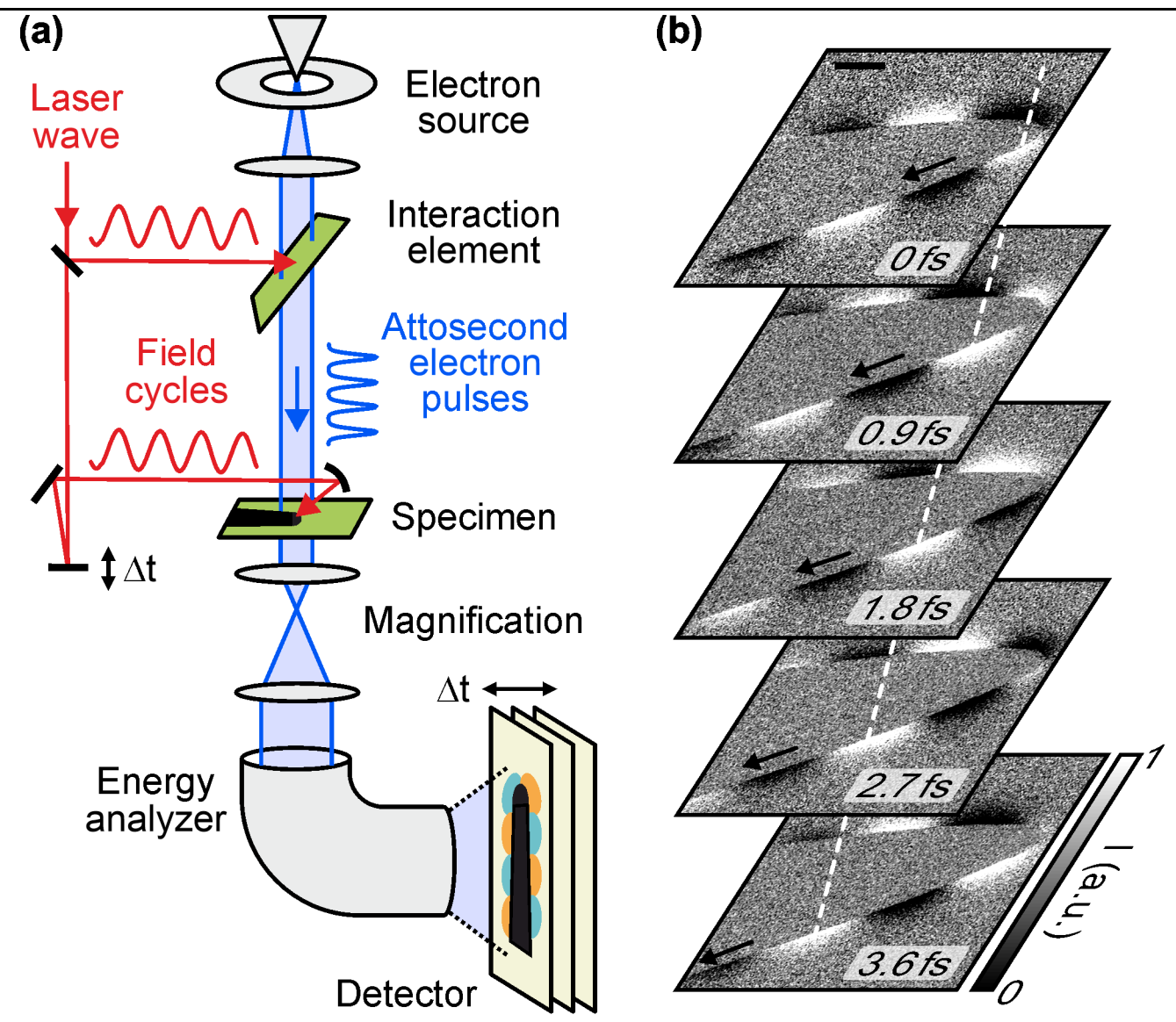

**Figure 8.** Attosecond electron microscopy with single electrons. (a) Concept and experiment. An e-beam (blue) is modulated by the optical cycles of laser light (red) into attosecond pulses that pass through a laser-excited specimen. The time-frozen near fields cause time-dependent energy changes that can be measured with an electron energy analyzer. (b) Attosecond–nanometer movie of the longitudinal electric fields of a light wave that travels around a nanostructured needle tip.[16]

## 9. QUANTUM-COHERENT PHOTON-INDUCED NEAR-FIELD ELECTRON MICROSCOPY

**John H. Gaida, Armin Feist, Murat Sivis, Hugo Lourenço-Martins, and Claus Ropers***

### 9.1 Introduction

The imaging of optical near fields is essential for understanding nanoscale light-matter interactions, and will promote advances in nanophotonics, plasmonics, and quantum optics. Various experimental techniques yield subwavelength field distributions, including scanning near-field optical microscopy (SNOM) and photoemission electron microscopy (PEEM), as well as their interferometric variants. Electron microscopy presents a particularly powerful methodology for studying electromagnetic fields and excitations without influencing the near field with the probing tip as in SNOM or requiring pristine surfaces as in PEEM. Scanning techniques using EELS (Section 3) and CL (Sections 4 and 5) can visualize confined optical modes corresponding to local absorption and scattering, respectively. These methods address excitations across a broad spectral range, employing spontaneous inelastic scattering, with typically low probability per mode.

External excitation that selectively populates specific modes can lead to drastically enhanced stimulated interactions that involve all electrons passing through the optical near field. The advent of ultrafast transmission electron microscopy (see Section 7) has enabled the use of femtosecond laser pulses for strong optical pumping and the creation of intense near fields for the short duration of electron probe pulses. In turn, this has enabled the implementation of PINEM by Barwick *et al.*,[13] in which stimulated interactions create discrete spectral sidebands in the electron-energy spectrum, spaced by the photon energy $\hbar\omega$ of the laser used (see Figure 9a).[5,33] In its early implementations,[13,140,152] PINEM used energy-filtered full-field imaging to obtain spatial representations of the optical near field based on the total number of inelastically scattered electrons. However, in this approach, the near-field contrast was generally nonlinear, saturated at higher field strengths, and also did not yield phase information.

### 9.2 Recent Developments in Quantum Coherent and Phase-Resolved Near-Field Imaging

In Figure 9b-d, we display a set of further developments using full-field imaging (Figure 9c) and a focused probe (Figure 9b,d) made in our laboratory, which have led to the electron-based quantitative and optically phase-resolved metrology of near-field distributions. These developments harness the fact that stimulated inelastic scattering induces a quantum-coherent optical phase modulation of the electron wave function.[14] Experimentally, this theoretically predicted scenario[5,33] can be observed if the probing electron pulses are shorter in duration than the envelope of the optical excitation. Under such conditions, all parts of the longitudinal wave function are homogeneously modulated by the same amplitude, revealing multilevel Rabi oscillations of the free-electron states, which represent the physics of a continuous-time quantum walk.[14] As a further consequence, it was theoretically predicted[14] and experimentally shown[169] that this phase modulation allows for a coherent shaping of the electron wave function and a temporal compression into a train of attosecond pulses. The quantum-mechanical phase-space distribution of this attosecond-shaped state was retrieved using a tailored quantum-state tomography scheme,[169] resulting in a reconstruction of the free-electron density matrix.

The strength of the electron-light coupling is encoded in the electron spectrum with a single coupling parameter, as also discussed in Section 2. Measuring a complete spectrum for every position using scanning-PINEM[14,190,191] allows for a quantitative determination of the optical near field (Figure 9b). However, this approach does not yet exploit the quantum coherence of the phase modulation of the electron wave function to extract the optical near-field phase. Figure 9c,d display two complementary approaches in full-field and scanning-probing to resolve also the optical phase, utilizing the coherent phase modulation in the transverse (Figure 9c) and longitudinal (Figure 9d) directions. In the transverse plane, stimulated scattering can be employed for wavefront shaping,[192,193] the preparation of vortex beams,[194] and the demonstration of quantized electron-light momentum transfer[195] (see also Section 11).

In ref 196, we imaged the spatial variations imprinted onto the electron wavefront by defocus phase-contrast imaging, which is sensitive to spatial phase gradients. Translating Fresnel-mode Lorentz microscopy from the mapping of static or magnetic fields to the domain of optical fields, in this approach, phase-retrieval techniques on energy-filtered defocus images yield the spatially varying near-field phase. This full-field implementation of energy-filtered phase-contrast imaging relies on the high spatial coherence of electron pulses generated by field-emitter tips.[14]

An alternative approach for retrieving phase-resolved sample responses employs phase-locked sequential interactions, as in free-electron quantum-state reconstruction[169] and Ramsey-type phase control.[197] To coherently map nanoscale responses, however, the modulation of the electron wave function needs to be sampled at every position. In free-electron homodyne detection[17] (FREHD), this is accomplished by applying a controlled and phase-varying reference interaction with the electron wave function at every position when scanning across an excited sample. In this way, both the amplitude and phase of the sample response can be retrieved. This scheme provides a quantitative alternative to energy-filtered imaging using sequential interactions with or without attosecond density modulation[15,16,18,169,192,198,199] (see also Section 8). Importantly, although inelastic scattering at optical fields yields phase modulation of the electron wave function, amplitude modulations stemming from, for example, time-periodic changes in the structure factor, can also be probed.

### 9.3 Future Perspectives

Over the last decade, nanoscale electron imaging of optical near fields has been an invaluable resource for studying nanophotonic systems. Facilitated by recent advances, numerous opportunities for even broader applications have come into reach.

Specifically, a full nanoscale reconstruction of the optical quantum state, beyond coherent-state excitations, is desirable in future experiments. Amplitude and phase information can be retrieved simultaneously by reconstruction algorithms, including *spectral quantum-interference for the regularized reconstruction of free-electron states* (SQUIRRELS),[169,200] also accounting for amplitude modulations and phase-space shearing arising from dispersive propagation.[16,17,18] STEM-type imaging offers local structural and optical information on a deep subwavelength scale. This calls for adopting different phase-resolving techniques, including off-axis or STEM holography, which have not yet been explored for ultrafast near-field imaging. Furthermore, while PINEM typically samples only a single spatial frequency along the e-beam direction, combining tunable-frequency light, white-light probing, and variable e-beam energy will enable studies of the complete broadband optical response in three spatial dimensions as well as time and frequency. In particular, this will be essential for addressing local optical nonlinearities. A further challenge is to achieve the necessary time resolution for probing partially coherent optical states, their scattering at interfaces, and dissipation. This will be enabled by generating few-femtosecond electron pulses, optical gating methods, or broadband optical spectroscopies, each requiring different approaches for the quantitative reconstruction of the optical state.

Another frontier in applying PINEM is access to continuous e-beams[10,201] and nonresonant structures. Here, a crucial tradeoff is given by a structure's optical field enhancement and quality factor. Practical limitations involve laser-damage thresholds, which may require temporal gating either using femto-to-picosecond optical pulses or nanosecond gating of the electron beam.[11] In this context, bandwidth- and pulse-duration tunable lasers and probing at an optimized duty cycle may significantly increase e-beam currents and improve the signal-to-noise ratio (SNR) in typical imaging applications.

A further information channel in nanoscale near-field probing is the local transverse momentum transfer. Momentum-resolved spectroscopy employing ω–q-type mapping (cf. refs 132, 196, and 195) promises access to in-plane mode polarization and band structures. This may be complemented by a full tomographic near-field reconstruction using sample rotation or beam tilting. An extension of PINEM-type imaging to very low electron energies or reflection geometries has the potential to access low momentum, large coupling efficiency, and slow light excitations in tailored optical structures. Considering higher kinetic energies, MeV e-beams may grant us access to thick samples and buried internal interfaces, with possible challenges in electron-light coupling strength.

Besides these technological improvements, future work will address an even broader range of materials excitations and geometries. This may include excitation at soft X-ray and extreme ultraviolet wavelengths, tuned to transitions exhibiting elemental contrast. At lower photon energies, infrared and terahertz excitations will yield information on low-energy and correlated excitations, benefiting from low sample temperatures. This may complement the capabilities of meV-resolved EELS instruments. Generally, correlative spectroscopy approaches (e.g., combining EELS/CL/PINEM;[191] cf. Sections 4, 5, and 10) can yield further insights into quantum photonic systems. Finally, a particular strength of ultrafast TEM is the possibility of simultaneously accessing electronic, spin, and lattice degrees of freedom (see Section 7). Future experiments will follow a combined approach to studying nanoscale ultrafast dynamics by tracing energy conversion, transfer, and dissipation in inhomogeneous systems.

These approaches will provide versatile multimodal access to the study of various functional systems and devices, ranging from nanoscale heterosystems used for light harvesting and optically driven catalysis to energy transfer in biological systems, which remains largely unexplored in ultrafast TEM. Both single-particle excitations and correlated modes can be traced, while the study of single defects and individual quantum systems remains a challenge.

### 9.4 Conclusions

Nanoscale optical and structural imaging contributes to the development of novel materials and devices with tailored electronic and optical properties. The coherent reading of sample-induced changes to the quantum state of electrons adds a new dimension of measurements to electron microscopy. Beyond the phase-resolved probing of electromagnetic fields, arbitrary dynamical changes of nanoscale specimens will become accessible. Ultimately, in this way, electron microscopes hold the promise to become universal devices for probing attosecond dynamics and local quantum states.

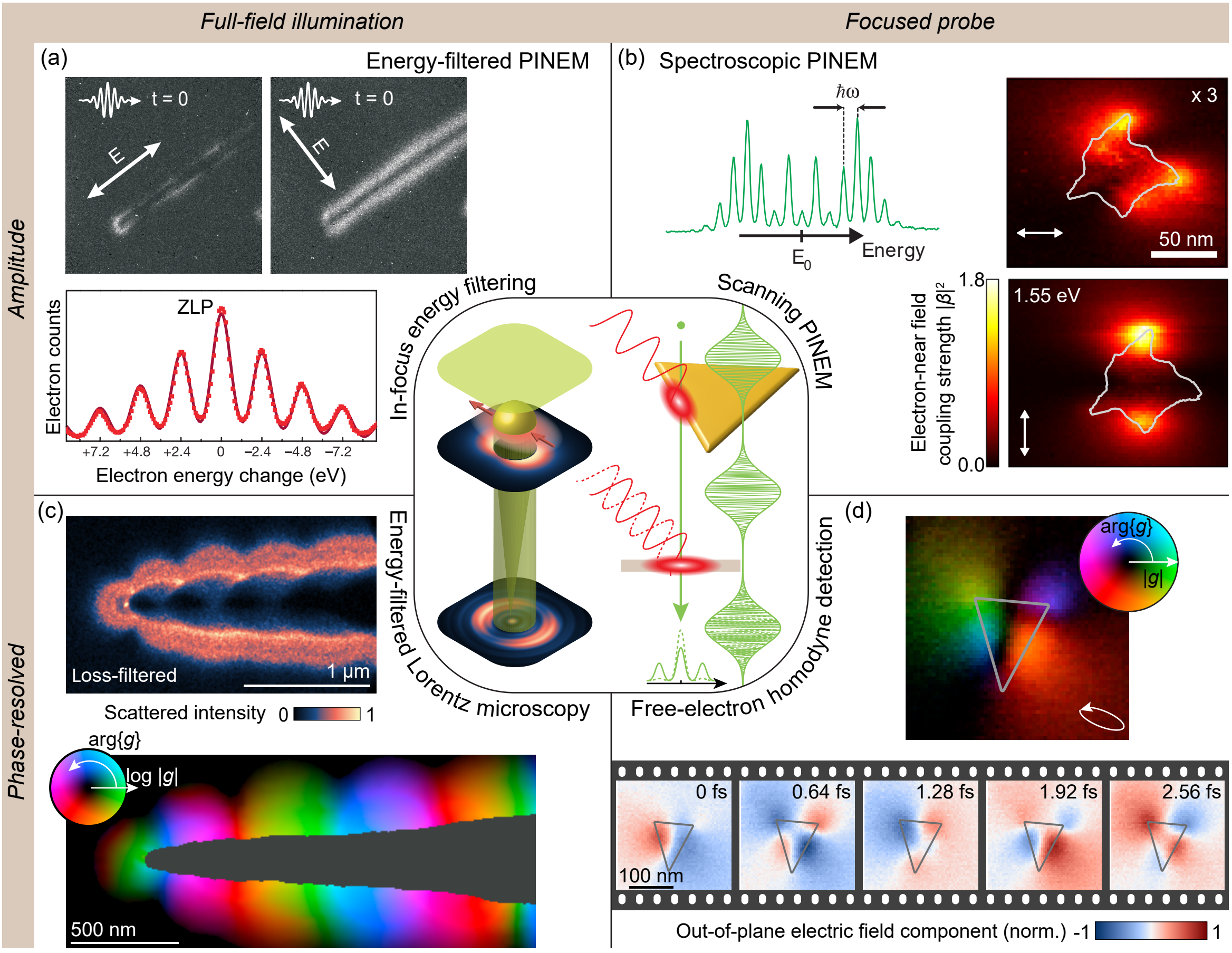

**Figure 9.** Quantum-coherent photon-induced near-field electron microscopy for the phase-resolved imaging of nanoscale optical fields. The measurement techniques can be categorized into full-field illumination (left) and focused probe (right) techniques. The bottom row shows approaches resolving the optical phase. (a) Energy-filtered TEM micrographs of a carbon nanotube. Contrast arises from filtering gain-scattered electrons out of the broadened energy spectrum shown below. Adapted from ref 13 (Copyright 2009 Springer Nature). (b) Coherent scattering results in quantum inference and multilevel Rabi oscillations. Measuring the spectrum with a focused probe by raster-scanning across the sample allows us to quantitatively image the electric field amplitude. Left: adapted from ref 14 (Copyright 2015 Springer Nature). Right: adapted from ref 191 (Copyright 2021 Springer Nature). (c) Defocused imaging in Lorentz microscopy converts phase profiles imprinted from the optical field onto the electron sidebands into measurable intensity contrast. The optical phase profile can be reconstructed from loss and gain energy-filtered micrographs. Adapted from ref 196 (Copyright 2023 Springer Nature). (d) A phase-controlled reference interaction enables free-electron homodyne detection (FREHD). Adapted from ref 17 (Copyright 2024 Springer Nature).

## 10. FREE ELECTRON AND PHOTON TEMPORAL COINCIDENCE SPECTROSCOPY

**Luiz H. G. Tizei,* Yves Auad, Luca Serafini, Johan Verbeeck, Armin Feist, and Claus Ropers**

### 10.1 Introduction

Temporal coincidence spectroscopy effectively distinguishes specific scattering mechanisms in experiments in which many channels are available. A byproduct of this selectivity is background suppression. Specifically, for electron spectroscopies, the coincidence between the scattering of a primary electron and the generation of secondary electrons,[202] X-ray,[203] visible photons,[204] and Auger electrons[205,206] have been explored. These coincidence experiments have evidenced, for example, that secondary electrons arise due to a cascade of events starting at the primary losses, which is dominated by bulk plasmon excitations[202] and specific deexcitation paths leading to Auger electron generation.[206] Some of these early experiments occurred in electron microscopes, benefiting from the available nanometric spatial resolution. However, spatially resolved measurements were limited by the available detector quantum efficiencies, temporal resolution and noise, the intrinsic low signal in coincidence experiments, and the temporal stability of available hardware.

Here, we focus on recent experiments describing the temporal coincidence between electron energy losses (measured through EELS) by an electron traversing a thin material[4] and the emission of one or more X-ray (energy-dispersive spectroscopy, EDS) or IR-visible-ultraviolet (CL) photons,[207] which benefit from a new class of event-based electron detectors (Timepix3). In Section 10.3, some perspectives on how these experiments can be improved are discussed.

### 10.2 Electron–Photon Coincidences

The coincident detection of the energy lost by electrons and the emission of X-rays has been considered an effective way to decrease background both in EELS (tails from bulk plasmon and absorption edges) and EDS (bremsstrahlung).[208,209] A clear application of this suppression is the improvement of the detection limits for elemental traces and chemical quantification. This method is currently limited by the low time resolution of column-mounted silicon drift detectors (SDDs), which have evolved to allow for high acquisition rates over a large collection solid angle. The latter comes at the cost of loss of temporal resolution of the detected X-rays due to varying drift times of the extrinsic charge carriers across the large surface of the SDD.

Considering IR-visible-ultraviolet photons, it has been shown that EELS–CL temporal coincidence allows for contrast-enhanced photonic imaging using electron–photon pairs,[21] heralding nonclassical light[210] (Figure 10a-d), the distinction of de-excitation pathways following electron excitation[120] (Figure 10e-f), and the measurement of excitation lifetimes.[89,121]

Post-selection of coincident electron–photon pairs reveals information that is obscured when examining average electron scattering and photon emission spectra. For instance, electron scattering in an optical cavity produces photons in multiple optical modes. Due to the limited spectral resolution of EELS (typically above a few meV), it is challenging to observe the spatial distribution of scattering at each individual optical mode, particularly if the modal density is high.[21] However, post-selection of photon-electron pairs containing photons of a specific energy or mode can address this limitation. Similarly, post-selection of electron–photon pairs involving photons emitted by defects in a semiconductor can elucidate the excitation pathways that contribute to CL, including bulk plasmons and core-hole excitations.[120]

### 10.3 Perspectives in Electron–Photon Coincidences

In recent years, advancements in temporal coincidence experiments have been driven by the introduction of nanosecond-resolved, event-based electron detectors relying on the Timepix3 chip. These detectors have effectively replaced the less versatile delay-line detectors, which, while offering lower time resolution, suffered from limited spatial resolution, being beam-sensitive and having lower detector quantum efficiency. As previously mentioned, in EELS–EDS coincidence experiments, the bottleneck typically arises on the X-ray detection side. Manufacturers of commercial EDS detectors tend to focus on larger SDDs to enhance collection efficiency for fast elemental mapping, where high temporal resolution is not considered.

To address this limitation, researchers are exploring custom X-ray detector designs. One promising approach involves using a small silicon PIN photodiode, equipped with a preamplifier and mounted directly on the TEM holder. The compact size of the photodiode helps maintain low capacitance, enabling faster acquisition times and reducing the drift effect of signal charge carriers that impairs time resolution. Additionally, placing the sample right next to the diode ensures high collection efficiency, making this design well-suited for coincidence experiments requiring high temporal resolution.

For IR-visible-ultraviolet electron–photon temporal coincidence experiments, electron detectors are becoming the bottleneck, given their restricted temporal resolution and limited event rate for studying weak-scattering processes. While precise zero-loss filtering can enhance the fraction of coincident electrons, reaching the picosecond time resolution of typical photon detectors will be enabled by integrating pulsed electron sources or fast blankers. A further challenge is the efficiency of detecting photons and their spectral analysis, which may benefit from high-numerical-aperture free-space light collection or fiber-coupled samples. Also, experiments at lower temperatures closer to liquid helium will increase the range of phenomena and materials accessible. Beyond the direct study of materials excitations, parametric photon generation at waveguides facilitates heralded single-electron sources, promising shot-noise-reduced electron imaging and spectroscopy.

Finally, multi-modal event-based electron spectroscopy could combine electron energy losses and correlated photon generation with momentum resolution and other detection channels, including the emission of secondary or Auger electrons. In practice, one would need the timed detection of all required signals. For example, electron–photon–photon timing for energy-resolved and momentum-resolved (EELS) second-order correlation function measurements (two CL detectors for Hanbury Brown and Twiss interferometry). Or angle-resolved electron (EELS) and photon timed detection, as recently demonstrated for electron-photon entanglement detection.[211,212] A difficult point in these multi-modal coincidence techniques is the requirement of increased exposure time due to the decrease in the number of coincidences. This will allow for new insights into ultrafast energy transfer pathways in complex materials using nanoscale e-beams.

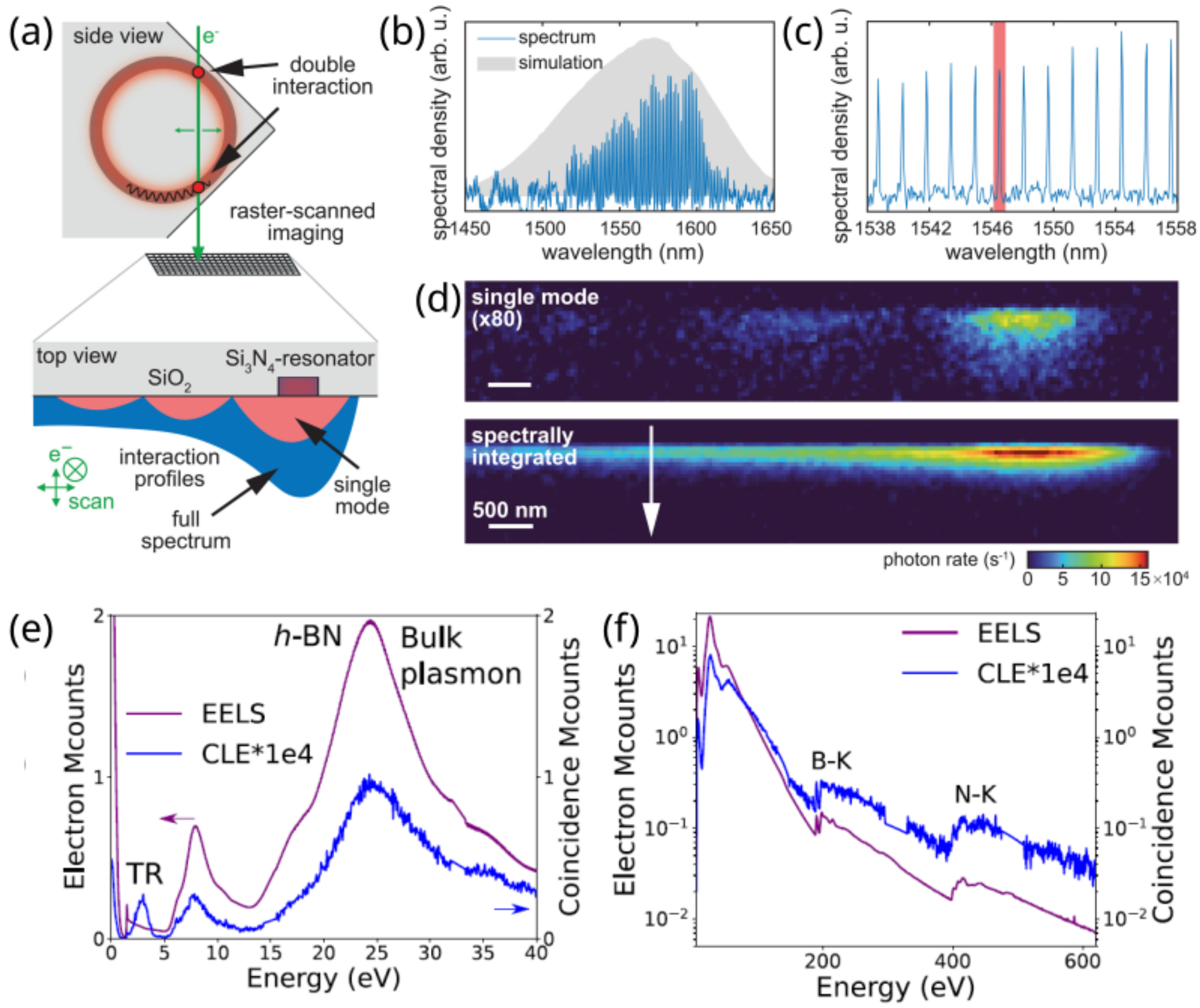

**Figure 10.** Electron–photon temporal coincidence experiments (a-d). Generation of electron–photon pairs mediated by an optical cavity. Post-selection of electron–photon pairs with photons of specific energy allows mapping the electron scattering probability at one optical mode (d).[21] (e-f) Temporal coincident electron–photon pairs elucidate the excitation paths leading to photon emission from defects in hexagonal boron nitride (hBN).[120] Panels (a-d) and (e-f) are reproduced with permission from refs 21 and 120, respectively (Copyright 2022 American Association for the Advancement of Science).

## 11. ELECTRON-BEAM SHAPING WITH LIGHT

**Andrea Konečná,* Nahid Talebi, and F. Javier García de Abajo**

### 11.1 Introduction

The capability of on-demand modulation and control of the spatiotemporal profile of an e-beam is vital for many standard and emerging techniques in electron microscopy. For example, the spatial control of e-beams is required for phase-contrast imaging,[213] single-pixel imaging,[214,215] mode-selective EELS,[70] or adaptive imaging and spectroscopy.[216] In addition, temporarily shaped and, in particular, ultra-compressed electron pulses are exploited to reach higher temporal resolution in ultrafast electron microscope setups.[15,16,169] Among the physical mechanisms allowing for the spatiotemporal control of free e-beams, the interaction with photons defines a research frontier due to the currently available excellent capabilities in preparing ultrashort light fields suitable for achieving the desired modification of the electron wave function (see also Section 12).

Electrons can interact with photons in two scenarios, corresponding, respectively, to the linear and quadratic terms (in the field amplitude) of the electron–light interaction Hamiltonian (see Section 2), and depending on the nature of the fields: (1) electrons can efficiently absorb or emit near-field photons confined in the neighborhood of a material, which is the key mechanism in EELS[5] and PINEM,[29,35] both relying on the interaction with optical components outside the light cone (i.e., evanescent or scattered light, typically associated with near fields or reflected light waves), which is mediated by the linear interaction term; in addition, (2) although energy-momentum mismatch between freely propagating electrons and light prevents photon absorption or emission by the electron, inelastic photon scattering can take place (i.e., Compton scattering with a zero net photon exchange), described by the quadratic term in the interaction Hamiltonian. Indeed, near-field and free-space electron–photon interactions have both been successfully used to generate spatially and temporarily shaped e-beams.

A first example of spatial modulation of free electrons due to the interaction with light dates back to 1933, when Kapitza and Dirac[27] predicted that electron waves should diffract from standing light waves. The Kapitza–Dirac effect was experimentally demonstrated seven decades later by collecting diffraction peaks in electrons traversing such standing waves.[36] However, free-space electron–photon interactions are not restricted to light fields forming standing waves. Recent experiments have shown that quasi-monochromatic freely propagating focused light pulses[217] can imprint an on-demand phase on the electron wavefront. With spatially structured optical fields, free-space ponderomotive interaction was suggested[34] and experimentally demonstrated[193] to enable a high degree of control over the transverse (in the plane perpendicular to the beam axis) electron wave function.

In contrast to the free-space interaction of electrons and light, near-field mediated processes can result in a net absorption or emission of multiple photons by the electron, producing a coherent electron energy comb that evolves by forming trains of attosecond electron pulses upon propagation over relatively large distances.[15,169] Exploiting this effect, together with the photon phase imprinted on the lateral profile of the inelastically scattered electron wave function, it has been shown that an on-demand spatial modulation of the electron wave function in the plane perpendicular to the e-beam axis can be obtained by interaction with a shaped optical near field,[218] including components corresponding to different numbers of photon exchanges to compensate aberrations and generally shape the lateral electron distribution.

Different levels of theoretical description of the electron–photon interaction have been applied, ranging from classical to fully quantum-mechanical approaches. However, except in a few exceptions,[31,219] light has been treated classically, which is well justified when resorting to intense coherent laser fields as those used in experiments. The interaction between electrons and photons often requires a quantum-mechanical description to account for quantum interference or diffraction effects.[220] To that end, the Dirac equation can be simplified under the assumptions of reasonable field strengths and electron energies, so that a relativistically corrected Schrödinger equation is produced[34] (see eq 2.1 in Section 2). In addition, some aspects of the free-space interaction can be addressed with semiclassical equations of motion for an electron evolving in the presence of spatiotemporally varying electromagnetic fields. Furthermore, when the interaction extends over many cycles of the optical field, one can average over the fast carrier oscillations and derive an effective ponderomotive potential due to the slower evolution of the light field envelope.[221] However, these fast oscillations emerge when directly simulating electron wave packets interacting with light in a complete Hamiltonian that includes both linear and quadratic terms in the optical field amplitude, a description that becomes important for slow electrons.[42]

### 11.2 Applications of Ponderomotively Shaped Electron Beams in Electron Microscopy

The Kapitza–Dirac effect has recently been utilized to create a Zernike phase plate for electrons,[213] which could enable the extraction of otherwise undetectable phase contrast emerging, for example, when electrons interact with weakly scattering specimens of organic compounds. The experiment was performed with a continuous e-beam, which discarded the possibility of using intense femtosecond laser pulses. To operate in the CW regime, the input optical power from a continuous laser source was enhanced 4000 times using precisely placed mirrors forming a Fabry–Pérot cavity. The performance of the experimental setup was demonstrated by recording diffractograms and also by detecting the enhanced phase contrast on a thin carbon sample.

Recent theoretical[34] and experimental[193] works demonstrated nearly arbitrarily tailored e-beams achieved via ponderomotive interactions with pulsed-shaped light. The idea behind this development relies on the modulation of a pulsed laser beam by a spatial light modulator (SLM). This type of pixelized device modifies the phase of light wavefronts, which translates into a transversely modulated optical intensity after focusing the transmitted laser light onto the region of interaction with the electrons, as schematically depicted in Figure 11a. Under the conditions in ref 193, the transverse profile of the in-focus laser intensity is directly imprinted onto the transverse phase profile of the transmitted electrons. Clear measurable modulations of the resulting electron intensity were observed by operating the electron microscope in the pulsed-beam regime and utilizing a precise synchronization of the electron and laser pulses in the interaction region inside the microscope. The achieved exotic e-beam shapes could find application in novel imaging and spectroscopic techniques. For example, the possibility of having a quick and versatile modulation of the e-beam shape available enables the implementation of adaptive measurement schemes to enhance image contrast.[216]

On-demand transversely tailored electron phase profiles could also find application in the design of alternative electron-optics elements. For example, the ponderomotive interaction with certain Hermite–Gaussian laser modes has been suggested to produce lensing effects.[193,222,223] The first available experimental results already showed that ponderomotive lenses could achieve focal distances comparable to those of conventional electron lenses, and they could produce both converging and diverging action on the e-beam.[193] It has also been proposed theoretically that the interaction of electrons with a vortex laser beam could compensate for spherical aberration and serve as an aberration corrector of standard objective lenses.[34]

When multiple light pulses of different central wavelengths are involved or when we consider slow electrons and very intense fields, it is possible to observe even multiple electron energy-loss or energy-gain peaks in the resulting electron spectra (see Figure 11b).[217,224] Such an effect is vital for longitudinal modulation of the electron wave function (e.g., for the generation of ultrashort attosecond electron pulses suitable for experiments in which a high temporal resolution is targeted).

### 11.3 Applications of Optical Near-Field Shaped Electron Beams in Electron Microscopy

Shaping the transverse component of the electron wave function via the interaction with optical near fields was demonstrated a few years ago in a proof-of-concept experiment.[194] Illumination of a hole in a metallic thin film with circularly polarized light led to the excitation of in-plane chiral patterns of propagating surface-plasmon polaritons, whose electromagnetic fields coupled to the incident electrons and generated vortex e-beams (Figure 11c). Surface-plasmon polaritons supported by a thin metallic film were also utilized in an alternative experimental setup,[232] where a certain degree of tunability of the transmitted electron wave function was achieved by varying the plasmonic near-field interference patterns using different illumination conditions, such as the direction of light incidence or polarization.

Subsequent theoretical work explored the potential of optical near fields for preparing completely arbitrarily transversely shaped electrons. In particular, by selecting the electron wave function component that has absorbed (or emitted) one photon after interacting with an optical field structured through an SLM, one could straightforwardly synthesize arbitrarily shaped lateral e-beam profiles. Special attention was paid to exploring the possibility of correcting for the spherical aberration produced by a realistic objective lens (see Figure 11d). Proof-of-principle experiments confirmed the feasibility of this theoretical suggestion by demonstrating that Laguerre–Gauss optical beams illuminating a thin film transparent to the electrons and opaque for light can generate the corresponding e-beam profiles.[233] These and related possibilities are discussed in Section 12.

Besides tailoring the transverse electron wave function component, early work associated with pioneering PINEM experiments showed that near-field electron–photon interactions can control the electron wave function longitudinally (i.e., the component along the e-beam direction) and eventually generate trains of attosecond electron pulses,[14,15,169] which can be further compressed through concatenated PINEM interactions.[225] A recent theoretical proposal suggested that by employing multiple PINEM interactions in a parallel arrangement (Figure 11e), it is possible to achieve a well-defined electron focal spot both in space and time (i.e., a combined spatiotemporal compression) within sub-ångström and sub-femtosecond scales.[226] The concept of temporal lensing has recently been invoked to propose electron single-pulse compressed down to the zeptosecond domain in a scheme relatively immune to the typically small degree of electron temporal coherence.[38]

### 11.4 Challenges and Future Directions

Although electron–light interaction is emerging as a promising versatile approach for the precise spatiotemporal control of fast e-beams, numerous challenges should still be overcome to materialize some exciting applications. One of the fundamental limitations is the requirement of relatively high light intensities for efficient modulation. Intense light is typically introduced by employing ultrafast laser pulses, which require the synthesis and synchronization of electron pulses. This strategy is adopted in ultrafast electron microscopes (Section 7), although it suffers from low average electron currents (needed to prepare well-controlled electron pulses) and, thus, long acquisition times. A different solution consists in increasing the interaction time, as recently proposed for the realization of CW longitudinal electron-beam shaping by ponderomotive interaction of co- and counter-propagating light beams relative to the electron.[37] Another challenge is the integration of tailored light into the electron microscope. Proof-of-principle demonstrations in electron microscopes commonly rely on conventional platforms and do not offer much freedom in the placement and physical size of the electron–photon interaction region. Developing a dedicated electron microscope platform that offers more freedom in the alignment of electron and laser pulses might resolve this issue. Some of the theoretically suggested applications could also suffer from limitations on the photon side. For instance, adaptive and single-pixel measurement schemes[214,215,216] require rapid and often complex modulation of the probe. However, when relying on SLMs, the frame rate ($<$100 Hz) restricts the modulation speed. Slow electrons constitute another direction of interest because of the stronger ponderomotive phase (inversely proportional to the electron velocity), so this avenue could be explored in scanning electron microscopes, where more space is available to optically actuate on the electron. Ultimately, electron–light interaction holds the potential to fine-tune the electron wave function in the transverse and longitudinal directions, thus suggesting the development of light-based e-beam pulsers, splitters, and lensing elements.

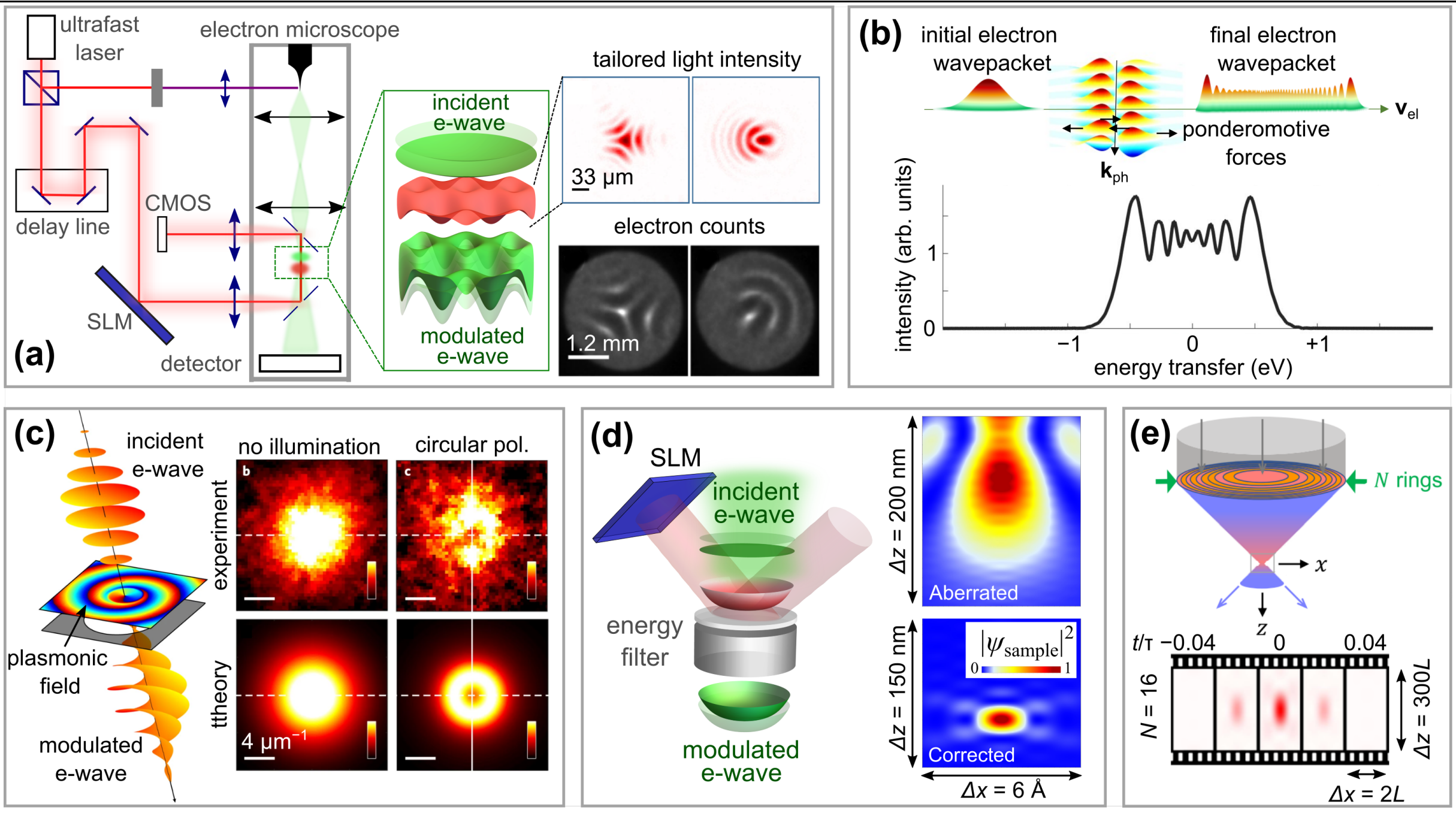

**Figure 11.** Optical e-beam shaping. (a) Experimental demonstration of transverse ponderomotive e-beam shaping. The setup (left) relies on an ultrafast scanning electron microscope, where the photoemitted femtosecond electron pulses are synchronized with spatially tailored light pulses generated by a spatial light modulator (SLM). Two examples of electron intensities obtained at the electron detector for different laser intensity profiles are shown. Adapted with permission from ref 193 (Copyright 2022 American Physical Society). (b) Schematic of inelastic ponderomotive scattering of slow (a few keV) electrons traversing a traveling optical beam and yielding an electron spectrum with multiple energy loss and gain peaks (bottom). (c) Vortex e-beams have been generated through the interaction with a spiraling plasmonic near field. The latter is produced by illuminating a hole in a thin metallic film with circularly polarized light. Adapted with permission from ref 194 (Copyright 2019 Springer Nature) (d) Theoretical proposal of spherical aberration elimination using tailored optical fields. The scheme (left) relies on the illumination of a thin film with tailored light, producing an efficient interaction with electrons, which are subsequently energy-filtered. Such an electron phase plate could substantially improve the focal spot profile created by an aberrated objective lens, as illustrated by the comparison of aberrated versus corrected focal spots (right density plots). Adapted with permission from ref 239 (Copyright 2020 American Physical Society). (e) Spatiotemporal compression of electron pulses. The suggested scheme (top) considers multiple parallel PINEM interactions in areas within concentric rings. For suitably tailored PINEM interactions, the electrons can be focused in a spatially narrow spot with a compressed temporal profile, as shown in several snapshots (bottom, for $N = 16$ PINEM regions) at different times (in units of the optical period $\tau$). Distances are given in units of $L = \lambda_e/\mathrm{NA}$ (typically $< 1$ nm), where $\lambda_e$ is the electron de Broglie wavelength, and NA is the numerical aperture of the microscope. Adapted with permission from ref 226 (Copyright 2023 American Physical Society).

## 12. NOVEL ELECTRON IMAGING METHODS BASED ON LIGHT-MEDIATED COHERENT ELECTRON WAVE FUNCTION SHAPING

**Beatrice Matilde Ferrari, Cameron J. R. Duncan, Maria Giulia Bravi, Irene Ostroman, and Giovanni Maria Vanacore***

### 12.1 State of the Art in Electron-Beam Shaping

In 2010, the field of e-beam shaping took off with two pioneering works demonstrating the generation of e-beams with helical phase fronts carrying orbital angular momentum.[56,227] Additional works then demonstrated the ability to sculpt electron wave functions using nanoscale phase masks, which further advanced the manipulation of e-beams.[228] Motivated by the need to improve the versatility of the electron phase control, researchers began to develop programmable phase plates using slowly varying electrostatic and magnetostatic fields.[229,230] Unlike photon-based ultrafast imaging methods based on, for example, X-rays and EUV light, shaped electrons can offer better performance in terms of scattering cross-section with materials (giving access to 2D materials and nanosystems), atomic spatial resolution, and a large versatility in terms of beam shaping methods, although reaching the required transverse coherence with electrons might be more challenging than with methods that use electromagnetic radiation.

In parallel, the advent of ultrafast electron microscopy introduced dynamic capabilities to e-beam shaping on a femtosecond scale. Following the initial work on PINEM,[13] which demonstrated the quantized inelastic interaction between electrons and nanoconfined light, a similar scheme was adopted to coherently modulate the longitudinal phase of the electron wave function,[14] proposing the creation of attosecond electron pulse trains. Attosecond coherent modulation of a single-electron wave packet was later realized by adopting more versatile inelastic electron–photon interaction methods that rely on the breaking of the translational symmetry of the light field[15,169,231] and on the elastic electron–photon interaction mediated by ponderomotive forces.[199] This progress laid the groundwork for new motivations in beam shaping, especially regarding the possibility of expanding the e-beam modulation also to the transverse phase profile, directly affecting the spatial and momentum coordinates. By means of chiral surface plasmon polaritons (SPPs), researchers demonstrated the generation of pulsed electron vortex beams,[194] which inspired the use of more complex SPP patterns to provide tunable control over the e-beam.[232] Precise modulation of the transverse momentum distribution of the e-beam was also realized.[180,195,231] The increasing interest in light-mediated e-beam shaping stems not only from its ability to induce ultrafast changes on the femtosecond scale or faster[233] but also from the versatility enabled by advanced light modulation technologies. Recent experiments have demonstrated arbitrary modulation of the transverse e-beam profile using a spatial light modulator (SLM) to shape optical fields. This modulation is imprinted on the transverse electron wave function through either inverse transition radiation[193] or the ponderomotive force.[234] These SLM-based shaping methods largely expand the type of patterns that can be transferred on the electron profile and highlight the growing versatility and impact of e-beam shaping techniques for application in advanced imaging (see also Section 11).

### 12.2 First Applications of Light-Induced Beam-Shaping for Phase-Resolved Imaging

Following such basic studies, several groups have started to exploit the ability to imprint an energy modulation on the e-beam to enable phase retrieval of probed electromagnetic waves via interferometry, as demonstrated by three independent works.[16,17,18] When an e-beam with temporal coherence longer than the laser period interacts with a laser pulse, it splits into energy sidebands whose intensities oscillate with the interaction length. A second interaction point can extend these quantum oscillations in the energy domain, provided there is a well-defined phase relationship between the laser pulses at the two locations. This phase coherence can be ensured by deriving both pulses from a common parent laser pulse. By adjusting the optical path-length difference between the two laser pulses—causing constructive or destructive interference in the energy domain—it becomes possible to extract the complex coupling coefficient between the electron probe and the electromagnetic field under investigation. This allows for the reconstruction of both the phase and amplitude of the electromagnetic field. Constructive interference also enhances contrast in energy-filtered imaging. These works show exquisitely phase-resolved and time-resolved images of surface electromagnetic waves, such as propagating modes across a metal tip and a dielectric nanoresonator,[16] SPP modes around a gold nanoprism,[17] and Bessel modes inside a circular resonator.[18] An important aspect of the method is that when the second interaction point is placed at a precise distance downstream of the first interaction so that the initial energy modulation can evolve into a temporal modulation, the experiment benefits of attosecond bunching and time resolution in the sub-femtosecond range, independent of any additional energy modulation of the e-beam at the sample.

### 12.3 Challenges, Future Goals, and Suggested Directions to Meet These Goals

In this section, we explore the most promising novel imaging methods that can be implemented by exploiting the new concepts of light-induced e-beam shaping, with the potential to achieve unprecedented imaging of materials in a TEM.

#### 12.3.1 *Ultrafast Chiral Imaging of Quantum Materials*

In the field of quantum matter, chirality is intrinsically rooted in the electronic, structural, and topological instabilities that govern the behavior of materials. In such a context, a promising way to manipulate the material properties is to control the chiral order of the system using ultrafast light fields as external stimuli,[235] opening new routes to control their macroscopic functionality for unprecedented opportunities in optoelectronics and quantum computing. So far, investigation of chirality is mainly obtained using optical probes, which however exhibit an inherently low sensitivity and a limited spatial resolution. Light-induced vortex e-beam shaping addresses this pressing need and enables the development of new techniques, such as ultrafast electron chiral dichroism. These techniques allow researchers to investigate the role of chirality in governing the nonequilibrium dynamics of low-dimensional quantum systems with greater sensitivity and at previously inaccessible spatiotemporal scales. In particular, one can exploit the ultrafast vortex phase shaping of a free-electron[180,194] to provide chiral selective probing and coherent control on femtosecond and nanometer scales (see Figure 12a). As an example, such a unique approach will provide beyond state-of-the-art visualization of the ultrafast dynamics of chiral phonons and chiral plasmons in 2D materials, as well as topological chiral carriers in Weyl semimetals.

#### 12.3.2 *Ramsey Imaging of Quantum States in Strongly Correlated Materials*

As described above, several works have demonstrated that the phase-controlled interaction of an electron pulse with two independent fields, one serving as reference and the other as unknown, can provide attosecond–nanometer holographic imaging of localized fields coupled to local material excitations, such as plasmon polaritons[16,17] and phonon polaritons.[18] The prospect for the future is to go beyond polaritonic physics, and rather towards the implementation of Ramsey-type holographic imaging for investigating complex quantum states in strongly correlated systems. Collective modes in strongly correlated materials are responsible for several emergent material properties, such as magnetoresistance, multiferroicity, topological protection, and superconductivity, which can be manipulated by light pulses inducing exotic out-of-equilibrium states of matter.[236] Accessing the phase dynamics of a given excitation with nanometer resolution would translate into the dynamical reconstruction of the complete density matrix of the unknown quantum state when monitoring the coherent interaction of the investigated system with a phase-modulated electron wave packet. In Ramsey-holography (see Figure 12b) a first light-based electron modulation stage would split the electron wave function in a quantum coherent superposition of different energy states. Then, the modulated electron packet would interact with the investigated material where a specific many-body state is resonantly excited. Here, the inelastic coupling between the different modes associated with such local excitation and the electron pulse will modulate the electron wave function according to their spatial and temporal evolution. The modulation can be coherently probed by a third interaction point with an additional light pulse, phase-locked with the one that imprints the first phase modulation (homodyne detection), and mapped via energy-filtered imaging in real and reciprocal spaces.

#### 12.3.3 *Correlative Light–Electron Microscopy Via Superradiant Light Emission*

The ability to transfer coherence from a phase-shaped electron wave function to a bound material quantum state is thought to be responsible for the generation of a new type of superradiant light emission,[237,238] especially in the presence of discrete energy states found in low-dimensional systems. The idea is to fiddle with the material degrees of freedom to change their decay probabilities. This aspect is still an open question and so far only theoretical works[237,238] have recently appeared. If experimentally confirmed, the final result would be a single quantized free electron transferring its longitudinal coherence to multiple emitters simultaneously. This concept is general, as it can be applied to any cluster of emitters, and will result in a resonant enhancement of the light emission by such clusters. This can be understood by considering that in conventional CL the radiation flux induced by the electrons is proportional to the electron current and to the density of emitters. In contrast, using phase-shaped electrons, the emitters will result in a coherent state, and thus, the many-body system will emit with a significantly higher rate. The radiation emission by multiple emitters would be then scaling as $N^2$ for a phase-shaped electron pulse versus $N$ for a non-shaped electron wave packet ($N$ is the number of electrons in the pulse). Consequently, the nanoparticles would emit with an enhanced rate (multiplied by the number of emitters), enabling CL superradiance with nanometer resolution. The presence of such enhanced light emission when structured electron packets are used to interrogate materials can open exciting opportunities for imaging weak scatterers (see Figure 12c), especially in the context of correlative light–electron microscopy of biological specimens.

### 12.3.4 *Low-Dose Electron Imaging Via Ultrafast Single-Pixel Reconstruction*

Single-pixel imaging is related to the application of structured-wave illumination for image reconstruction.[214,239] In particular, the method is based on the illumination of a sample using a series of spatially modulated patterns while simultaneously collecting the scattered intensity on a bucket detector. The key aspects are: (1) the spatial modulation of the probe, encoded according to a specific orthogonal basis set; and (2) the characteristic *sparsity* of the acquired images such that compressed-sensing algorithms can be adopted for image reconstruction. In such a case, the number of acquisitions necessary for retrieving an image is generally smaller than the total number of unknown pixels, which directly implies a faster response time, together with a lower radiation dose with respect to conventional methods. Such advantages are extremely interesting in the context of TEM imaging of radiation-sensitive nanostructures in their original environment, for which the electron dose needs to be kept as low as possible and below a critical damage threshold. To implement single-pixel imaging in an electron microscope one has to illuminate the sample using structured e-beams. When performed in combination with time-resolution analysis, efficient and versatile transverse patterning of a free-electron can be achieved through a computer-controlled SLM (see Figure 12d). The latter is used to modulate the incident light field according to the desired spatial pattern, which is then transferred to the transverse electron profile via electron–light coupling. The ability to control both amplitude and phase directly translates into the potential to overcome the Poisson noise of the measurement, which is fundamental for making the single-pixel approach extremely promising for low-dose electron imaging.

### 12.3.5 *Contrast Modulation and Spatial Resolution Enhancements*

Besides the direct imaging methods, light-mediated e-beam shaping is also a very promising tool for improving the performances of the TEM, especially in terms of contrast transfer function and aberration correction. In such a context, transverse modulation of the electron momentum components is key for enabling such potential. Recent theoretical studies[239,240] predict that transverse phase modulation of the electron wave function, mediated by elastic ponderomotive coupling with a phase-modulated light field controlled by an external SLM, can compensate for the spherical aberration introduced by magnetic lenses. This would result in a direct improvement of the spatial resolution in TEM imaging. Similar concepts can be adopted to increase the modulation contrast when imaging weak scatterers. In fact, by introducing additional momentum components within the electron wave packet, the cutoff of the modulation transfer function can be pushed towards higher frequencies, thus improving the contrast in real space at shorter length scales. Strong elastic momentum spread of the electron wave function can also be obtained by adopting terahertz electromagnetic fields as generated via light-induced charged plasmas.[241] In such a configuration, the picosecond-evolving plasma would generate a field configuration that mimics a Laguerre-Gaussian beam able to introduce a lateral phase shift on the electron wave packet that would correspond to a negative spherical aberration coefficient. When properly tuning the parameters of the plasma, one can potentially use such a method to compensate for the positive spherical aberration introduced by the TEM (see Figure 12e).

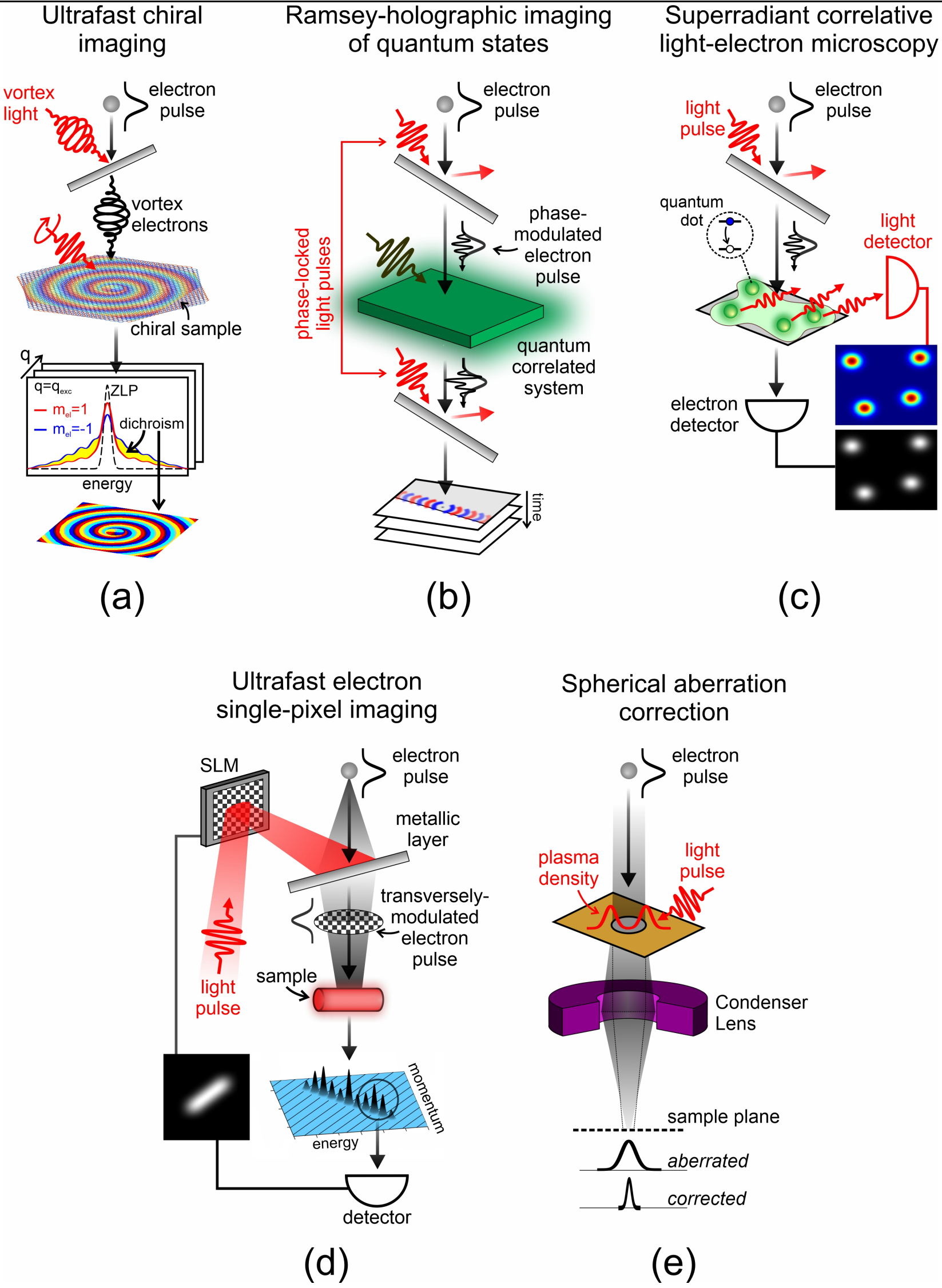

**Figure 12.** Novel electron imaging methods enabled by light-induced e-beam shaping. (a) Schematic picture of the method used to probe chiral ordering in quantum materials via vortex electron pulses and spectral dichroism following energy post-selection. (b) Ramsey-holographic imaging of strongly correlated materials as obtained by two electron-light interaction points (one above and one below the sample) used to prepare and to read the electron state, respectively, following the interaction with the sample quantum state. (c) Superradiant light emission from excited quantum dots owing to the interaction with a phase-modulated light beam becomes key in implementing correlative light-electron microscopy with enhanced sensitivity. (d) Schematic diagram of the single-pixel imaging method with momentum space post-selection as used for image reconstruction with structured illumination patterns. Adapted from ref 214 (Copyright 2023 American Chemical Society). (e) A laser-generated plasma is used to imprint on the electron transverse profile a negative spherical aberration coefficient.

**QUANTUM PHYSICS AND NEW CONCEPTS**

## 13. EXPLORING THE FUNDAMENTALS OF QUANTUM ELECTRODYNAMICS IN TRANSMISSION ELECTRON MICROSCOPES

**Ethan Nussinson, Ron Ruimy, Yuval Adiv, Arthur Niedermayr, and Ido Kaminer***

### 13.1 State of the Art

Quantum electrodynamics (QED) governs the fundamental interactions between electrons and photons. This theoretical framework has provided profound insights into various quantum phenomena and has been instrumental in explaining experimental results, particularly in high-energy physics. In recent years, advances in electron microscopy have opened new avenues for exploring QED processes in unconventional settings: involving measurements of quantum correlations and entanglement in complex electromagnetic environments, a domain captured by macroscopic QED (MQED).[242,243] Electron microscopes provide opportunities for experiments that are more challenging to accomplish in traditional particle colliders, such as coincidence detection, interferometric techniques, and energy–momentum-resolved measurements. These capabilities have inspired a recent surge of interest in exploring quantum processes in electron microscopy, a driving force in the emerging field of free-electron quantum optics.

The theoretical framework of MQED helps to unify and classify these emergent concepts. In all cases, the environment in which electrons and photons interact can significantly influence their properties, fundamentally altering their interactions.[243] The most famous example is the spontaneous emission of photons by free electrons, which is forbidden for free electrons in vacuum in standard QED (Figure 13a). In contrast, free electrons can undergo spontaneous emission in virtually any environment other than a vacuum. The environment, or optical medium, contains a density of photonic modes to which the electron can couple and emit. This effect governs various processes, including Cherenkov, Smith–Purcell, transition, and parametric X-ray radiations, differing by the coupling environment.[5,244] The emitted particles in these processes can be free-space photons or more exotic photonic quasi-particles such as plasmon or phonon polaritons.[243] Interestingly, the first observations of bulk plasmon polaritons were performed in electron microscopy using EELS.[245]

The spontaneous emission by free electrons has been predicted to have quantum recoil corrections when emitting a photonic quasiparticle,[246,247] as observed in the spontaneous emission of X-rays[248] (Figure 13a, middle). The coherence and photonic state of the emitted radiation depends on the electron wave function, exemplifying the quantum nature of the interaction. Achieving strong interactions, with sensitivity to single photons per electron, is currently a primary bottleneck that will enable high controllability of the generated state of the photonic quasiparticles.[249] The first measurement of this strong interaction regime showcased high-order interactions, whereby each electron emitted multiple photonic quasiparticles[250] (Figure 13a, right). Such interactions could be used to herald few-photon Fock states by post-selecting the final energy of the electron.[251]

Stimulated emission and absorption of free electrons are the underlying effects that occur whenever an external field drives electrons.[13] The electron interacts with preoccupied photonic quasiparticle modes that enhance its interaction, enabling it to absorb and emit hundreds of quasiparticles[252] (Figure 13b). This phenomenon governs PINEM experiments.[13,252] When an electron interacts with classical light, its energy spectrum follows an interference pattern, causing the height of the different peaks to change non-monotonically[14] (Figure 13b, middle). When the electron interacts with quantum light,[117] the photon statistics was predicted[31] and demonstrated[253] to be imprinted on the electron energy spectrum after their interaction, which can potentially enable photonic quantum state tomography[254] (Figure 13b, right). Stimulated emission and absorption can be used to shape the electrons.[232,255] Recently, it was shown that such shaping can enable microscopy with sub-cycle attosecond temporal resolution[16,17,18] and enhanced imaging of dynamic electric fields.[18]

### 13.2 Challenges, Future Goals, and Suggested Directions to Meet These Goals

Overall, spontaneous and stimulated electron–photon interactions have been extensively explored both theoretically and experimentally. Pioneering theoretical works in recent years have proposed a range of fundamental interactions between electrons and photons, most of which remain largely unexplored. We expect that the next few years will bring the first experiments of some of these intriguing phenomena. We show how these promising avenues for future experiments can be classified using the language of MQED: including electron self-interactions, electron–electron correlations and entanglement, electron–photon nonlinearities, and coincidence-based creation of quantum light states (Figure 13c-g).

Self-interactions are effects visualized as closed-loop diagrams, where an electron emits and reabsorbs a quasiparticle (Figure 13c). In QED, such self-interactions are the basis for the famous anomalous energy separation between the 2s and 2p levels in the hydrogen atom, also known as the Lamb shift[256] (Figure 13c, middle). The elusive free-electron self-interaction was predicted to be accessible in electron microscopes by a medium that will create a nontrivial electromagnetic vacuum. The self-interaction will then manifest[257] as a phase shift in the electron wave function that could be probed using diffraction experiments (Figure 13c, right).

All the effects mentioned so far were single-electron effects. In general, electron–electron interactions inside electron microscopes are a relatively unexplored field. Such interactions can be split into two types. The first type is parametric interactions (Figure 13d), where the quantum state of the environment remains unchanged, meaning that energy did not transfer from the electrons into photonic quasiparticles or material excitations during the interaction. These effects include direct electron–electron scattering (Møller scattering in the conventional QED description), which was recently explored in electron microscopy by measuring the energy correlations between different electrons emitted in the same pulse[47,258] (Figure 13d, middle). This scattering could potentially be altered and enhanced by introducing electromagnetic structures (Figure 13d, right). This novel phenomenon could go beyond the enhancement of the scattering cross-section and generate entanglement between the electrons, which could be probed by measuring the electrons in different bases after their interaction.

The second type of electron–electron interaction is nonparametric, where the environment changes following the interaction, for example, via the emission of photonic quasiparticles or any material excitation. Measuring the environment after such an interaction could generate entanglement between the electrons. For example, measuring the number of photonic quasiparticles emitted by two electrons either interacting sequentially[117] or simultaneously[50] with a photonic mode, without measuring which electron emitted them can entangle the electrons in the energy basis[117] (Figure 13e, middle). Alternatively, an analogous measurement can entangle electrons in space using an interferometric scheme (such as a *which-path* or a double-slit setup), where only part of each electron interacts with a photonic quasiparticle[259] (Figure 13e, right). In special cases, all the photons emitted by some electrons may, by chance, be absorbed by other electrons, resulting in no net change to the environment. Such events can be viewed either as a special case of non-parametric interactions or as parametric interactions in which the exchanged photons are on-shell rather than virtual.

Substantial recent interest has focused on creating nonclassical light states using free electrons. The creation of nonclassical photonic states always requires some form of nonlinearity. One direction for such effective nonlinearity can arise from the measurement process of the electron by post-selecting the final electron state. The creation of a few-photon Fock state (i.e., a state with a well-defined number of photons) was predicted a decade ago[19] and recently demonstrated[251] (Figure 13f, middle). Recent works have proposed strategies for creating and manipulating more complicated quantum light states that are important for fault-tolerant quantum computation (FTQC). For example, Schrödinger cat states and squeezed states can be generated when the initial electron wave function is modulated before the interaction.[249] More complex states, such as Gottesman–Kitaev–Preskill (GKP) states, can also be generated by applying post-selection on multiple electrons that emit into a shared photonic state[249] (Figure 13f, right).

The second direction for generating quantum light states involves utilizing inherent nonlinearities in the electron-light interaction (Figure 13g). One such nonlinearity is in the recoil of the emitting electron that loses momentum with each emitted photon. By using slow electrons and engineering the dispersion relation of the structure to allow only single-photon emission, one can potentially establish a deterministic single-photon light source[41] (Figure 13g, middle). Another source of nonlinearity is the ponderomotive interaction, which arises from the $A^2$ term in the minimally coupled Hamiltonian, where $\mathbf{A}$ is the vector potential. In QED language, such interactions involve two-photon emission and an intermediate virtual (off-shell) electron. If the electron trajectory is perpendicular to the polarization of the vector potential, such interactions could become dominant. Then, since photons could only be emitted in pairs, they would form a squeezed-vacuum state (SV)[260] (Figure 13g, right).

Many of the phenomena discussed above require either multiple emission or absorption events by the free electron or a strong correlation between the photonic quasiparticle state and the electron state. Both conditions are met when the coupling strength between the free electron and a single photonic mode approaches unity ($\sim$1). Achieving such strong coupling remains a major challenge in the field, motivating ongoing efforts to design optimized cavities and to establish theoretical bounds on attainable coupling strengths.[268,259] Another major challenge is the joint detection of the photonic mode and the electron. Although demonstrated in a few recent studies, such measurements remain limited to specially engineered samples, and no portable or broadly applicable solution has yet emerged.[39,211,212]

In summary, the language of MQED is useful for guiding future investigations of fundamental quantum phenomena within electron microscopes. Unlike conventional QED platforms such as particle accelerators, electron microscopes offer multiple unique opportunities. These include: (1) engineering the electromagnetic environment, as opposed to performing the experiments in vacuum; (2) advanced measurement techniques such as coincidence detection, interferometry, and energy–momentum-resolved measurements; (3) post-selection and heralding capabilities that allow for measurement-induced nonlinearities; and (4) pre-engineering of the initial wave functions, compared to simple plane waves that describe typical collision experiments. These make electron microscopes a versatile platform for exploring otherwise inaccessible QED effects. Going beyond conventional QED, state-of-the-art experiments enabled extensive investigation of spontaneous emission and stimulated emission/absorption—phenomena beyond the scope of standard QED, which MQED naturally describes. Looking to the future of this field, advances in electron–photonic mode coupling and their joint measurement could enable studies of self-interactions, electron–electron interactions, and quantum light generation, opening new paths toward discoveries of new quantum phenomena.

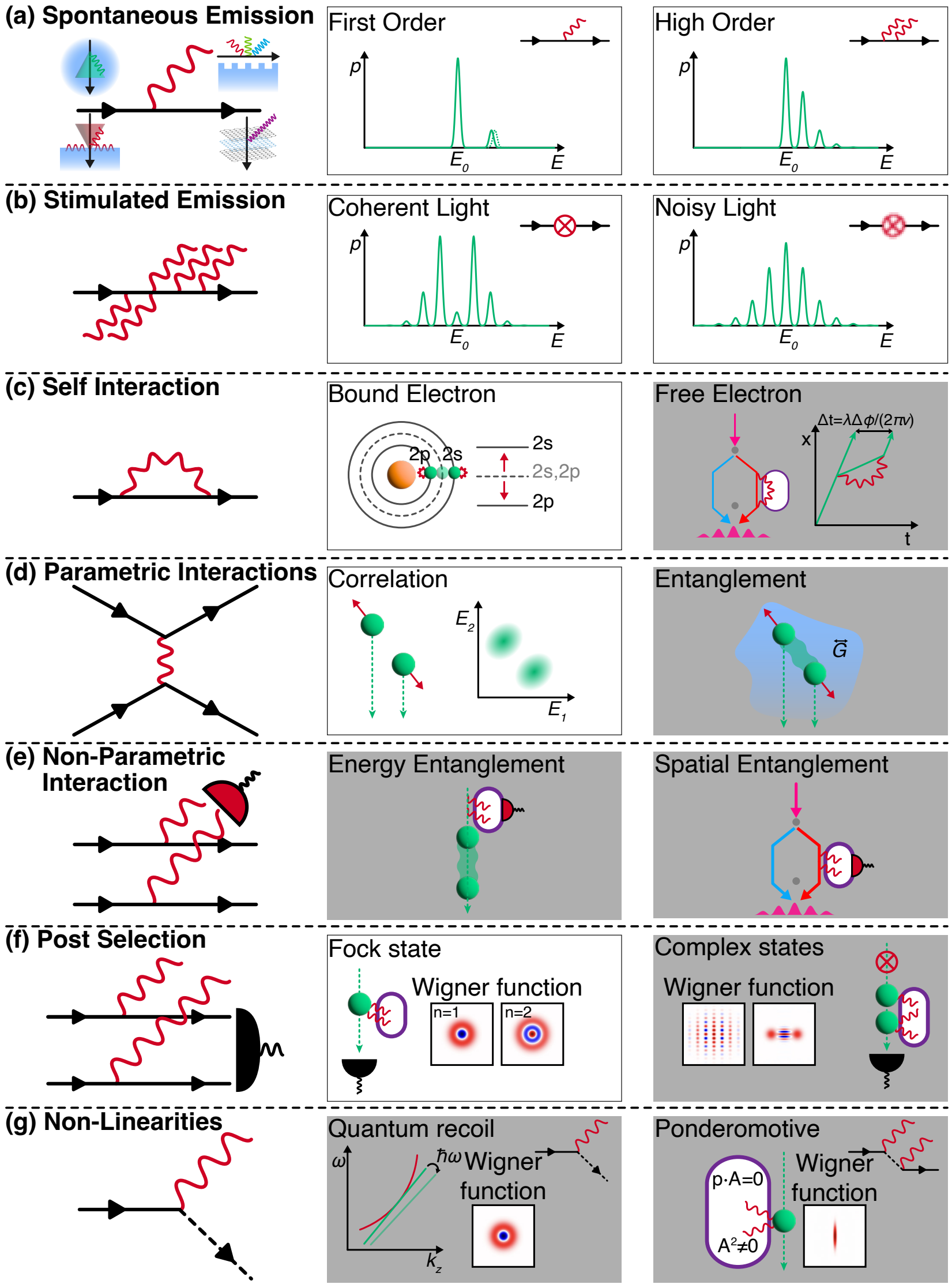

**Figure 13.** Classifying the phenomena of free-electron physics in the language of macroscopic-QED (MQED). Subfigures in grey represent effects that are yet to be realized experimentally. (a) Spontaneous emission of photonic quasiparticles. This emission process is forbidden in QED due to the inability to maintain energy–momentum conservation. However, by altering the electromagnetic environment, spontaneous emission is allowed. Depending on the environment, this effect governs various processes, including Cherenkov, Smith–Purcell, transition, and parametric X-ray radiations. The same tree-level diagram in MQED captures all of these effects. First-order emission process of parametric X-ray radiation, demonstrating the quantum recoil correction (middle).[248] Higher-order spontaneous emission and corresponding electron energy spectrum (right).[250] (b) Stimulated absorption and emission of photonic quasiparticles. These processes are described by the sum of all Feynman diagrams where the electron emits and absorbs photonic quasiparticles. Electron energy spectrum after stimulated interaction with classical (coherent state) light,[14] which is the mechanism behind PINEM (middle).[13] Electrons can also interact with nonclassical states of light,[214] as shown with super-Poissonian photon statistics (noisy or chaotic light) (right).[253] The final energy spectrum depends on the quantum statistics of the photons. (c) QED describes self-interactions of free particles due to quantum fluctuations of the vacuum. In bound-electron systems, such self-interactions are responsible for observing the Lamb shift, the anomalous difference in energy between the 2s and 2p orbitals in hydrogen, celebrated as one of the biggest achievements of QED (middle).[256] For free electrons, alteration of the electromagnetic medium (and the explicit breaking of homogeneity) could lead to such self-interactions, which could be measured through diffraction experiments (right).[257] (d) The direct interaction of multiple electrons could generate entanglement between them. Classical energy correlations mediated by Coulomb interactions have recently been observed (middle).[47,258] By altering the macroscopic electromagnetic environment and its corresponding dyadic Green function $\overleftrightarrow{G}$, these interactions could potentially be enhanced, leading to a strong, controllable generation of entanglement (right). (e) The joint interaction of multiple electrons with a photonic quasiparticle, in combination with post-selective measurement of the photons, could lead to the generation of entanglement between the electrons. Interaction with optical photons could lead to the generation of energy entanglement (middle).[214] Interaction with microwave photons in a path-selective manner could lead to the generation of spatial entanglement (right).[259] (f) A similar interaction, with post-selective measurement on the electrons instead of the photons, can be used to generate nonclassical light states. By using a single electron and analyzing its post-interaction energy loss, Fock states can be generated, as represented by their Wigner functions (middle). When multiple electrons are involved, more intricate photonic states, such as Schrödinger cat and GKP states, can be created. This is achieved by pre-shaping the wave functions of the electrons through stimulated emission. The electrons then emit photons, and their energy is measured post-interaction (right).[249] (g) Nonlinear interactions of the electron with the photonic mode can generate quantum light. For slow electrons, the quantum recoil following the emission of a single photon can detune them from the phase-matching condition, facilitating deterministic single-photon emission[41] (middle). When the electron is coupled to a mode where the vector potential ($\boldsymbol{A}$) is perpendicular to the electron's momentum ($\boldsymbol{p}$), the interaction becomes governed by ponderomotive forces. This interaction, arising from the $A^2$ term in the minimally coupled Hamiltonian, leads to the emission of photon pairs. As a result, squeezed-vacuum light (SV) is generated (middle).[260]

## 14. QUANTUM PHYSICS WITH FREE ELECTRONS

**Valerio Di Giulio, Ofer Kfir, F. Javier García de Abajo, and Claus Ropers***

### 14.1 Introduction and State of the Art

Electron microscopy uses the scattering of free e-beams from materials to study the structure and excitations of an investigated specimen. While phase shifts of the electron wave function yield the atomic-scale structure, inelastic scattering in the form of energy loss and photon emission encodes elemental composition and spectral properties. This information is gained by interrogating elementary interactions, typically considering the resulting change in the electronic or photonic state. In theoretical terms, the full complexity of the microscopic dynamics can be analyzed from the point of view of the joint density matrix $\rho_{e,l,m}$ of the tripartite system composed of electrons, light, and matter. Specifically, electromagnetic coupling mediates the buildup of correlations, transforming an initial separable state $\rho_e^0\rho_l^0\rho_m^0$ into the entangled state $\rho_{e,l,m}$, whose analysis in terms of each of its subsystems corresponds to forming partial traces over all other unobserved degrees of freedom.

To control these three building blocks, rapidly growing experimental and theoretical efforts have been made to act on each subcomponent and their combination, aiming at the underlying quantum features to be used as quantum probes or quantum sources. While the previous section (Section 13) provides a detailed account of electron–light–matter coupling mechanisms in terms of generalized quantum-electrodynamics (QED) processes, in this section, we present and consider recent work and future developments aimed at gaining further insights into the full quantum state $\rho_{e,l,m}$ by focusing on the statistical interrelations among its subcomponents. Figure 14 spans a range of current possibilities and long-term goals, organized around the joint density matrix.

At present, state-of-the-art electron microscopes, equipped with optical systems capable of synchronizing laser and electron pulses at the sample, can readily prepare $\rho_l^0$ in a highly populated coherent state.[14] The resulting inelastic scattering brings the initially quasi-monochromatic electrons into a discrete coherent superposition of energy states spaced by the photon energy, with probabilities following a quantum walk[14] (see Sections 2 and 7). This interaction does not lead to a considerable change in the optical state, thus yielding a separable density matrix $\rho_e\rho_{l,m}$. Because the amplitude and phase of the modulation vary with the scattered optical electric field, nanostructures patterned in the plane perpendicular to the electron trajectory can act as inelastic phase masks, imprinting the spatial distribution of the scattered near field onto the transverse part of $\rho_e$.[194] Longitudinally, the high coherence between different energy components manifests when they mix through free-space electron propagation (because the electron velocity varies with energy), forming trains of attosecond probability-density pulses at specific distances.[14] If taken as the output of a previous phase-locked interaction, the state $\rho_e^0$ can be reconstructed through scanning the phase difference between the two modulations.[169] Moreover, by replacing the laser light with a quantum source, the heights of the peaks in the final electron spectrum can reflect a sub-Poissonian intensity distribution of $\rho_l^0$, a dependence that can be deconvolved to extract information on the photon statistics.[31] An analogous effect has also been observed for super-Poissonian classically fluctuating sources[219] and short laser pulses yielding a varying coupling strength throughout the temporal extension of the electron ensemble.[13]

Further efforts aimed at incorporating optical measurements down to the single-photon level, alongside the capabilities provided by electron microscopes, have also paved the way to perturb, drive, and probe light–matter subsystems by shaping $\rho_e^0$. For instance, the interaction of laterally shaped electrons has been predicted to precisely tune the entanglement between electromagnetic resonances in nanoparticles and the final transverse momentum of the electron, allowing for mode-selective excitation.[22] By employing unshaped electrons, cascaded and single-emitter transitions can be triggered, as demonstrated by the observation of photon bunching[87] and antibunching[261] in CL emission from nitrogen-vacancy centers. Regarding light sources, energy-modulated electrons can produce tailored electromagnetic radiation in the form of polaritonic modes at harmonics of the laser frequency. Interestingly, the generated light state is intimately connected to the quantum phase-space distributions of the electrons, which can be carefully designed by laser shaping and by post-selecting only certain scattering events.[40] Without post-selection, strong time localization of compressed electrons generates coherent photons, once again leading to a separable state[28] $\rho_e\rho_{l,m}$ describing a radiation process that is superradiant and scales quadratically with the electron current.[30] In contrast, conditioning on the final kinetic energies of pre-scattered monoenergetic electrons can herald Fock states from $\rho_{e,l,m}$ when one,[19] two,[210] or more energy loss events are detected, whereas electron energy superpositions hold the promise of generating more complex non-Gaussian light states.[40,249] Finally, multi-photon entangled light has also been predicted to be harnessed in one-dimensional waveguides by leveraging the loss of which-way information after post-selection on undeflected electrons.[262]

### 14.2 Challenges and Future Goals

Considering the panorama presented in the previous section, we proceed to mark several applied capabilities and aspirational goals that would be enabled by deploying the quantum properties of free electrons, as mentioned on the outer rim of the illustration in Figure 14. We organize them by the current challenges they address. A first point is the limited dose that a material can sustain when irradiated by electrons at energies of tens-to-hundreds of keV. To address this issue, future goals require either the suppression of damage per incident electron toward the limit of damage-free electron microscopy, or the extraction of more information per electron in quantum-enhanced spectromicroscopy. Maintaining the instrument's resolution is an additional challenge since a lower dose density is a trivial solution leading to its reduction. A second point is the control of the plethora of secondary radiation emitted from electron–matter interaction. In the visible domain, electron-based quantum light generation raises particular interest due to its potential applications in photonic quantum computations and communications.[249] For energetic photons, X-ray generation by high-quality e-beams may enable quantum sub-Poissonian statistics such as heralded X-rays, or momentum-correlated X-rays with the electrons.[208] A third point is the aspiration to apply quantum tools at the ultimate resolution of an electron microscope, where the figure lists quantum-state tomography and quantum metrology as examples.

**14.3 Possible Directions to Meet These Goals**

To reduce the limitations induced by e-beam damage, we suggest approaches leveraging quantum statistics, aiming at a higher ratio of extracted information per electron. These include sources of number-states as heralded electrons,[263] contrast enhancement by electron holography,[213] or superradiant signal extraction by multi-electron bunches,[47] either from quantum sources[264] or by periodically tailoring on the fly.[265] The prospect of performing useful quantum photonics with free-electron states relies on the near-unity quantum efficiency of electron energy-resolving detectors, as proven by parametric and matter-dependent electron–photon coincidence experiments and measurements of their entanglement (see Section 10 and Figure 10). As these capabilities are still in their infancy, the following basic steps need to be expanded: exploring which useful light quantum states could be achievable,[249] and in parallel combining tools for photonic quantum-state tomography with cutting-edge microscopes. Finally, we expect that quantum sensing at the high spatial resolution of the electron microscope will become a reality soon, following recent demonstrations.[120] Advanced detectors can sort a specific quantum property of a single electron, such as its energy, momentum,[266] or topological charge,[230] coincident with a different output (e.g., single photon detection).

In conclusion, the quantum nature of the electron, embedded in the joint density matrix with other quantum systems (light and material structures), holds a promising perspective in uncovering Å-scale quantum correlations through macroscopic observables. With such rapid progress, several fundamental advancements at the intersection of electron microscopy and optical spectroscopy are expected to emerge soon.

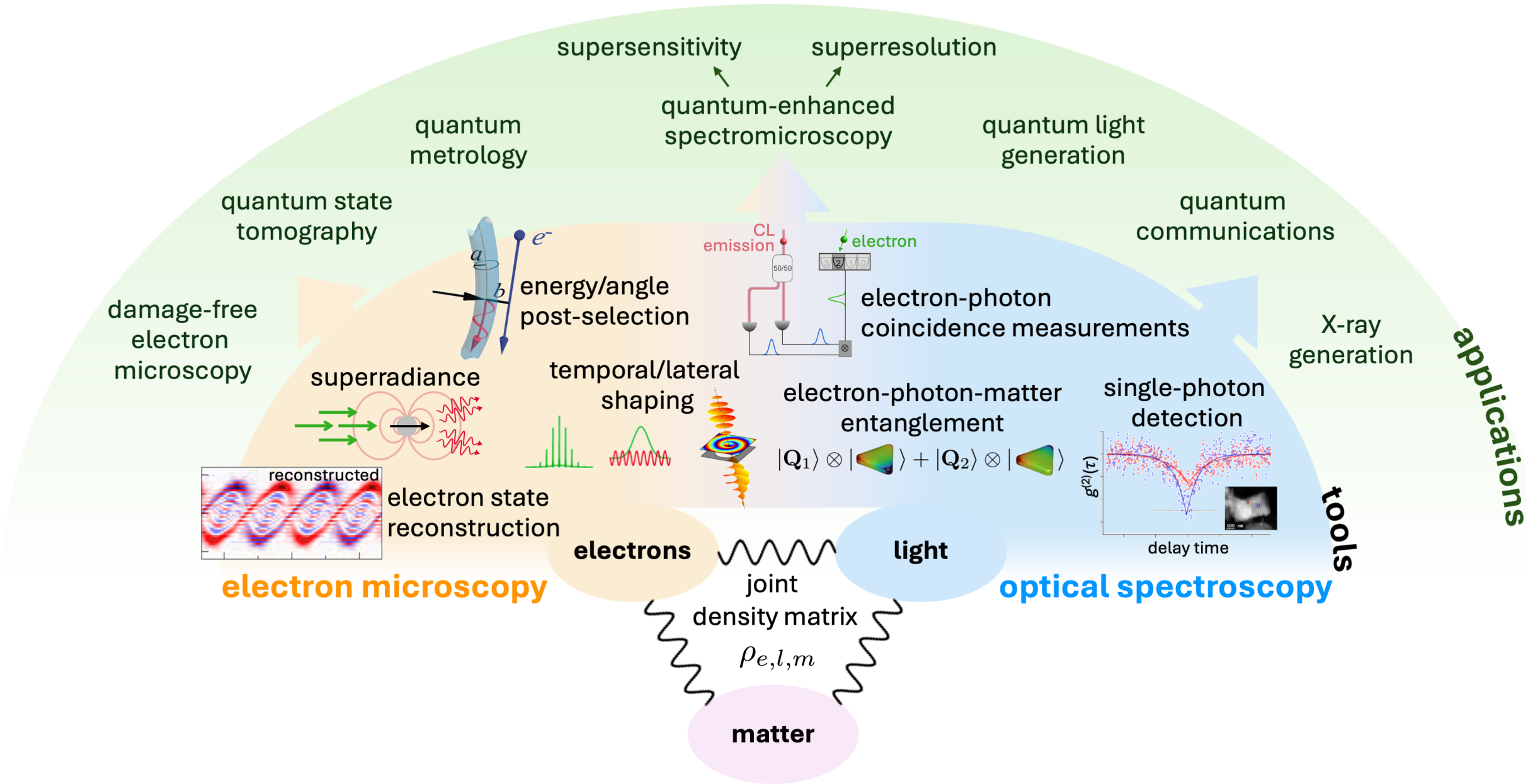

**Figure 14.** Quantum physics and applications at the intersection of electron microscopy and optical spectroscopy. The central elements in this scheme are electrons, light, and matter (lower part), whose characteristics can be described through the joint density matrix $\rho_{e,l,m}$. Several tools have recently been demonstrated or are still under development enabling the spatiotemporal shaping of free electrons through interaction with optical fields, the synthesis of entangled electron–photon and electron–matter states, the projection onto subspaces of interest by electron energy and angle post-selection, the superradiant emission from multiple electrons, and the measurement of correlations. We envision future applications relying on these tools, specifically in the areas of improved electron microscopy (reduction of e-beam damage, superresolution, supersensitivity, etc.), the retrieval of the quantum dynamics and properties of a specimen, quantum metrology, quantum-enhanced spectromicroscopy, and the generation of quantum light, among other feats. More applications are expected to join this list by leveraging the insights gathered within different areas, as described in this Roadmap. Inset images adapted with permission

from refs 19, 22, 169, 194, 210, and 261 (Copyright 2019 and 2017 Springer Nature, 2022 American Association for Advancement of Science, 2013 American Physical Society, and 2011 American Chemical Society).

## 15. NANOPHOTONIC ELECTRON ACCELERATION

**Zhexin Zhao, Roy Shiloh, Yuya Morimoto, Martin Kozák, Peter Hommelhoff***

### 15.1 Introduction

This and Section 16 focus on quantum nanophotonics with low-to-moderate-energy electrons (1–30 keV) inside of an SEM. Substantially smaller kinetic energies of electrons in SEMs compared to typical TEMs (70 to 300 keV) are promising from a theoretical viewpoint because they offer a higher upper limit of both spontaneous and stimulated electron–photon coupling strengths. The different energy ranges will be discussed below in greater detail. At the technological level, SEMs typically come with a much larger sample chamber, allowing us to place conventional optics close to the e-beam, but also to add custom components into the SEM chamber, such as home-built electron spectrometers (Figure 15a). The downside of SEMs is clearly that they are not built for transmission operation, meaning that standard EELS spectrometers cannot be added to commercial SEMs in a standard way. Yet, as we show below, we expect that the experimental flexibility that SEMs offer, in combination with the highly interesting low-energy mode, will open many unforeseen opportunities for electron nanophotonics experiments. So far, our SEMs allowed us to demonstrate multi-interaction zone operation, complex electron phase-space control and particle accelerator on a chip (discussed in this section), as well as longitudinal e-beam shaping by virtue of the Kapitza-Dirac effect (discussed in Section 16). Future developments in the field of quantum interactions between low-energy electrons and photons will mainly aim for optimization of the coupling efficiency to reach its theoretical limits. Combined with the technological advancements it will allow, for example, to utilize free-electron qubits as a platform for quantum information processing or other quantum optical experiments such as measuring electron–photon correlations inside SEMs.

### 15.2 Low-Energy Free Electrons for Optimal Quantum Coupling

15.2.1 *Current State of the Art*

Recent studies of the quantum aspects of the free-electron—light interaction suggested applications in quantum optics, including heralded or deterministic single photon sources, nonclassical photon state generation, and quantum computation (see, for example, ref 184 and Sections 13 and 14 in this Roadmap). All these applications require an efficient coupling between free electrons and photons, where the coupling can be quantitatively described by a unitless coefficient ($g_{Qu}$).[31,267] Therefore, it is important to understand the fundamental limit of the free-electron—photon coupling and search for efficient photonic systems.

Interestingly, studies on the theoretical upper bound of the free-electron—photon coupling show that low-energy electrons can potentially achieve better coupling,[268,269] when the distance between the free electron trajectory and the photonic structure is small. This seemingly surprising argument that a low-energy free electron can maximize the coupling is also demonstrated in the coupling between free electrons and confined optical modes, as discussed in Section 4 as well.[99] Moreover, this argument can be extended to the PINEM interaction, since the electron velocity dependence in the spontaneous (e.g., EELS) and stimulated (e.g., PINEM) free-electron—light interaction shares the same physics. For instance, the optimal free-electron velocity to couple to a confined plasmonic mode of a metallic tip is only a few keV.[191] As the quantum coherent stimulated free-electron—light interaction has been observed in an SEM (Figure 15a inset), it is promising to further explore low-energy free electrons for efficient free-electron—photon coupling.

To achieve strong coupling between free electrons and light, one typical photonic system is dielectric waveguides or closed waveguide resonators (see Sections 6 and 10), where the free electron travels in the vicinity of the waveguide and parallel to the propagation direction of the waveguide mode. In this way, the coupling can be enhanced, where $|g_{Qu}|^2$ typically scales linearly with the interaction length[267] if the phase-matching condition is satisfied (i.e., the phase velocity of the waveguide mode matches the electron velocity). When the normalized free-electron velocity $v/c$ is lower than $1/n$, where $n$ is the refractive index of the waveguide, it is impossible to achieve the phase matching condition with a longitudinally uniform waveguide, resulting in a challenge for low-energy free electrons. Nevertheless, subwavelength gratings (SWGs) can solve this challenge through quasi-phase-matching and achieve coupling strengths comparable to fast free electrons.[268,270] Furthermore, with waveguide dispersion engineering near the (quasi-)phase-matching condition, the scaling of $|g_{Qu}|^2$ with the interaction length can be superlinear.[41,99]

Coupling to plasmonic modes is another promising approach for strong free-electron—light coupling.[99,250] The authors of ref 250 studied the interaction between free electrons and surface plasmon polariton (SPP) modes in the form of 2D Cherenkov radiation, where EELS showed strong coupling features. Furthermore, the theoretical upper bound study showed that the coupling with the SPP modes in simple metallic holes could almost reach the theoretical upper bound.[268]

15.2.2 *Challenges, Future Goals, and Suggested Directions to Meet These Goals*

One fundamental challenge for low-energy free electrons is the fast exponential decay of the optical near field as a function of the separation distance for structures extended along the electron propagation direction. When the waveguide mode satisfies the phase-matching condition, the decay length is $\gamma v\lambda/2\pi c$, where $\gamma = 1/\sqrt{1 - v^2/c^2}$ and $\lambda$ is the optical wavelength. Thus, efficient coupling with low-energy free electrons typically requires a small separation distance (e.g., tens of nanometers for visible and near-IR light). Furthermore, when considering electron sources with the same brightness, the low-energy e-beams used in SEMs have in general a larger geometrical emittance than the high-energy e-beams in TEMs, making it harder to focus them to small spot sizes.

In addition, although low-energy electrons have a higher theoretical upper bound of coupling, the coupling coefficient with simple designs is about one order of magnitude lower than the upper bound, especially with dielectric waveguides.[268,269] Thus, it is crucial to open the design space and numerically optimize the structures to achieve efficient coupling, for instance, by applying inverse design, which has been successfully used to design dielectric laser accelerators and Smith-Purcell radiation generators,[219,271,272] and exploring a broader range of wavelengths and materials.[269]

The plasmonic systems, which can support efficient coupling, generally have non-negligible absorption. Such loss can limit their application in quantum optics. To reduce the influence of absorption loss in plasmonic systems, it is worthwhile to explore and optimize systems with materials exhibiting polariton resonances while maintaining low absorption, such as transparent conducting oxides.

Guiding free electrons with confined transverse dimensions can further increase the interaction length and boost the coupling with photonic structures. Electrostatic electron guiding based on auto-ponderomotive potentials can guide the electrons for tens of centimeters with tens of micrometer beam size,[273] while optical ponderomotive guiding can theoretically confine the beam size to sub-micrometer.[41,193] The combination of guiding and efficient coupling could be a fruitful direction to achieve arbitrarily strong free-electron—light interaction.

**15.3 On-Chip Particle Acceleration with Low-Energy Electrons**

15.3.1 *Current State of the Art*

Dielectric laser accelerators (DLA) are an emerging technology with exciting prospects for research using e-beams.[274] DLAs offer in particular an extensive amount of control over electrons: the generation of attosecond pulses, spatial shaping, and temporal gating, as well as energy shaping and acceleration, and quantum light–matter interactions have been demonstrated(see ref 244 and references therein). Moreover, a key promise of DLA technology is to bring new applications and devices such as widely tunable photon sources to the market and provide a viable solution to small labs wishing to pursue various kinds of research directions with free e-beams.

A key direction in DLAs is their use as on-chip linear electron accelerators with GeV/m-scale acceleration gradients, about 10–100 times higher than conventional radiofrequency-based accelerators. This gradient is enabled and limited by the dielectric properties and damage threshold of the accelerator material at optical frequencies. Driven by femtosecond lasers, these chip-sized nanophotonic accelerators require complex phase-space control to guide electrons through the nanostructure and overcome constraints from the tiny structures and Lorentz force[275,276] (Figure 15d). With precise phase-space manipulation, electron acceleration from sub-relativistic energies is feasible and scalable, as recently demonstrated[277,278] (Figure 15e,f). Scalability makes DLAs a reliable solution for nearly arbitrary final energies and allows their use as intermediate stages in large accelerator facilities.

DLAs have also been shown to generate short, attosecond electron bunch trains and provide sub-optical-cycle gating[198,279,280] (Figure 15b,c). Using DLAs, the incoming electron pulse is subjected to a localized, spatially periodic velocity modulation. Over a short propagation distance, this modulation transforms into a density modulation with a train of experimentally shown attosecond-short bunches down to 270 as. It is noteworthy that the nanophotonic structures for DLA experiments have been developed for efficient coupling of swift electrons to light and require good mastery of cleanroom fabrication processes.[219,281,282]

Owing to the quantum nature of the single electrons most usually employed in sub-relativistic experiments, novel light–matter interaction applications have been proposed and pursued. In this sense, extended DLA structures have been experimentally proven to be suitable for investigations into quantum light–matter interaction. Examples are the imprinting of light statistics onto the electron wave function,[219] various proposals to use free electrons as qubits,[184] and the feasibility of performing such experiments in an SEM.[146]

15.3.2 *Challenges, Future Goals, and Suggested Directions to Meet These Goals*

DLA technology follows much in the footsteps of traditional radiofrequency technology, although a mere adaptation is insufficient and novel solutions are needed. For example, currently, the most notable challenge is realizing the confinement of the e-beam throughout the structure in both transverse directions (so in full 3D) in addition to the existing 2D mechanism.[283] This has to be realized in a nanophotonic structure and driven optically, representing a design and fabrication challenge. Exerting this control will assist in preserving an optimal electron throughput.

Currently, using SEMs with laser-triggered sources, DLAs are limited to currents of 1–10 fA. To increase this, three approaches are clear: parallelizing on-chip nanophotonic channels[284], improving electron sources, and increasing the laser repetition rate. By parallelizing in 2D, we can increase the current proportionally: 1,000 acceleration channels result in a 1 mm structure width and can increase the output current to 1–10 pA. Due to strict input beam requirements (<100 pm×rad normalized emittance), high-current large-emittance flat-cathode electron sources are not viable. Instead, high-coherence multi-tip arrays[285] and custom electron optics for pulsed operation are promising solutions.

Another paramount requirement of DLAs, especially in the context of quantum experiments with low-energy electrons, is the ability to couple light and control electrons at low energies. Practically, the nowadays standard dual-pillar structure design is limited in its unit-cell periodicity by fabrication: a smaller periodicity is required for efficient coupling to lower-energy electrons. For standard near-IR laser wavelengths, this translates to a minimum starting energy of about 5 keV at the highest coupling. DLA experiments with lower starting energies can be done by utilizing less efficient nanophotonic structures or longer laser wavelengths. A full theoretical investigation into the limits of quantum-coupling efficiency is discussed in Section 15.2 above.

With appreciably high coupling, we foresee fully integrated devices, a few cubic centimeters in size, generating pulsed beams of even longitudinally shaped electrons. Next to fundamental electron light coupling experiments, they might find use as a source in diverse settings including ultrafast light sources, scattering experiments, and, more generally, new electron-based imaging devices (see Sections 11 and 18). In the future, this technology might even be applied in high-energy physics: once the electrons are ultrarelativistic, the nanophotonic structure becomes nearly perfectly periodic, leaving mainly the mechanical alignment of consecutive stages or integration into storage rings as a technical hurdle, which should be straightforwardly solvable

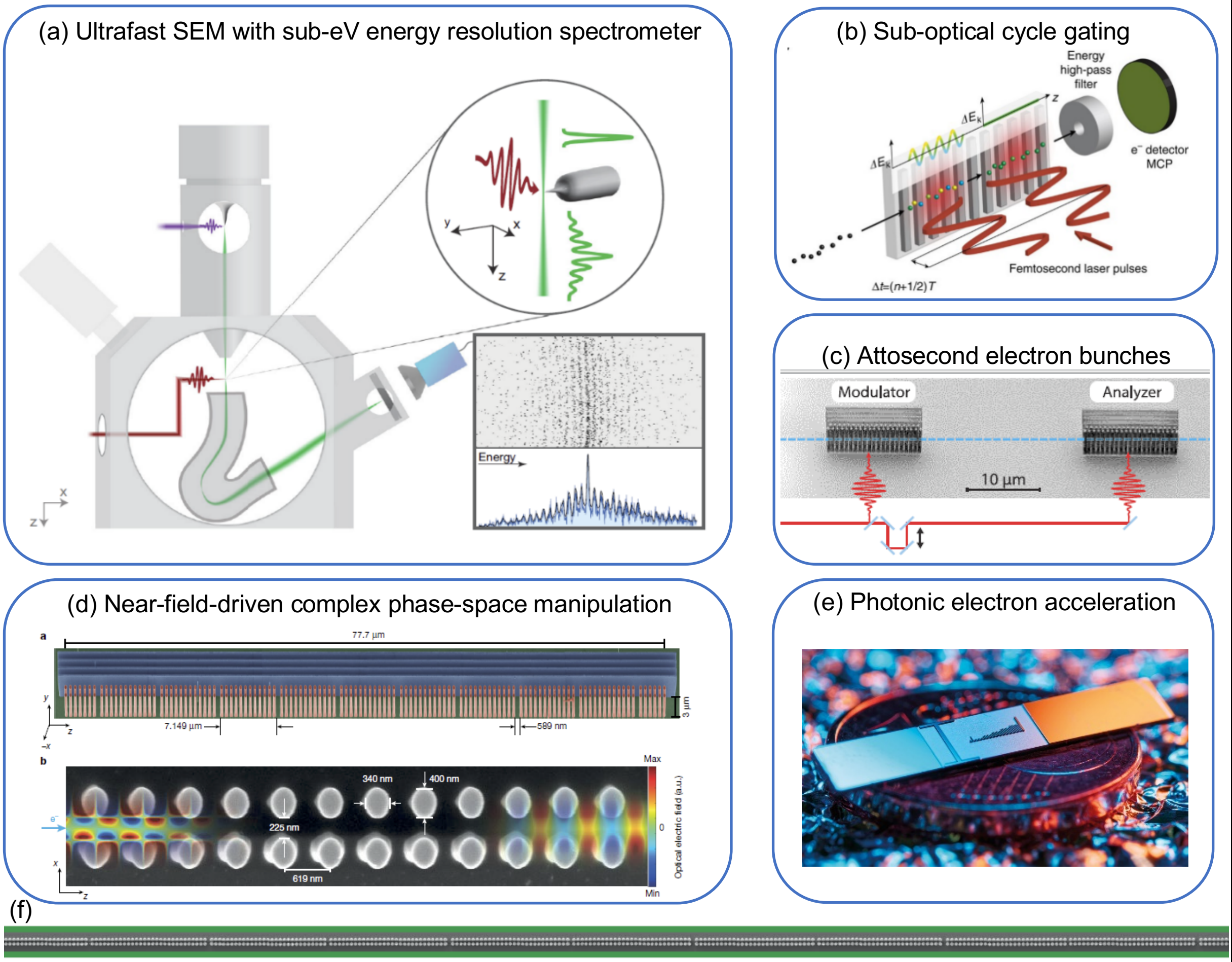

**Figure 15.** Nanophotonics with low-energy electrons in an ultrafast SEM (USEM). (a) Layout of an USEM equipped with a high-resolution electron spectrometer for characterizing electron spectra after inelastic interaction with optical fields. Adapted from ref 146 (Copyright 2022 American Physical Society). (b) Sub-optical cycle gating of electrons with two optical pulses at a grating structure. Adapted from ref 198 (Copyright 2017 Springer Nature). (c) Modulation and bunching of electrons induced and detected by the linear interaction with optical near fields in two subsequent periodic nanostructures.[279,280] Adapted from ref 279 (Copyright 2019 American Physical Society). (d) SEM images of the structure for complex optical phase-space control (guiding and bunching action) of electrons. Adapted from ref. 276 (Copyright 2021 Springer Nature). (e) Picture of a silicon chip hosting five groups of accelerator channels with increasing lengths from 100 to 500 µm. Each group contains eight individual accelerator channels.[277,278] (f) SEM image of roughly ten macrocells of the accelerator on a chip (about 200 out of 500 µm of the longest structure shown in (e). Adapted from ref 277 (Copyright 2023 Springer Nature).

## 16. KAPITZA-DIRAC PHYSICS AND SCATTERING WITH LOW-ENERGY FREE ELECTRONS

**Yuya Morimoto, Zhexin Zhao, Roy Shiloh, Peter Hommelhoff, Martin Kozák***

### 16.1 Kapitza-Dirac-Type Low-Energy Electron Control

16.1.1 *Introduction*

As proposed by Kapitza and Dirac[27] already in 1933 electron matter waves may scatter off the periodic structure made of standing light waves. The interaction between the electron wave function and light in Kapitza–Dirac-type experiments can be understood in two different pictures. When the electron interacts with a coherent state of light, a semiclassical description can be applied to derive a phase modulation of the electron wave, which is proportional to the ponderomotive potential of the light fields (the potential proportional to the local light intensity) integrated over the time of the interaction. In the particle picture, the electron emits and absorbs two photons from the incident light waves in a stimulated manner, which allows for fulfilling momentum and energy conservation laws. In the classical implementation, the electron wave packet scatters an optical standing wave with its wave vector perpendicular to the electron propagation direction[36] (Figure 16a). In the short interaction regime, the electron forms a diffraction pattern corresponding to a coherent superposition of discrete transverse momentum states. In the long interaction regime, the Bragg condition selects one diffracted order.[286] The transverse Kapitza–Dirac effect thus may serve as a coherent e-beam splitter, creating two or multiple beams. Alternatively, the optical ponderomotive potential can be applied as a phase plate in electron microscopy[213] or as a means to shape an e-beam.[287] The interaction of free electrons with an optical ponderomotive potential has recently been generalized to different geometries allowing the control of not only the transverse momentum components of the electron wave packet but also the longitudinal one,[171,217] enabling, for example, compression of electron pulses to attosecond duration[199,230,230] (Figure 16b).

The Kapitza–Dirac-type quantum control of electrons has attracted attention due to its versatility and possibility to modulate e-beams by light in free space. Electron modulation by the linear interaction with the electric field of light requires the presence of a material to slow down the light phase velocity to efficiently couple to the velocity of the electron. Due to the small spatial extent of the interacting optical near-fields, the e-beam has to be focused and propagated in close vicinity of the structure, leading to unavoidable electron scattering. Moreover, the maximum field amplitude is limited by the damage threshold of the material. In contrast, the vacuum interaction mediated by the ponderomotive potential prevents electron scattering of the solid-state structures and it typically extends over spatial regions determined by the envelopes of the interacting pulsed laser beams. Furthermore, this type of control requires relatively high peak electric fields of the light waves on the order of 1 V/nm, which are only achievable with ultrashort laser pulses or by using optical cavities.[213] There are several avenues for future development of the Kapitza–Dirac-type quantum control of electrons, which may bring new functionalities in electron microscopy, spectroscopy, and diffraction experiments.

#### 16.1.2 *Spatiotemporal Control of Electron Wave Packets*

Nowadays, the intensity of optical fields can be shaped in space and time using spatial light modulators and pulse shapers. Due to the direct correspondence between the spatial profile of the light beam/pulse and the phase profile imprinted into the modulated electron wave packet, the spatial and temporal shaping may be combined to generate, for example, superpositions of electron vortex states with a helical density profile applicable as a probe of local chirality of electromagnetic fields[180,288] (Figure 16c). An unexplored direction of electron–photon interactions is the possibility of quantum coherent temporal shaping of the electron wave packets by optical pulses with time-dependent frequency, which can serve for electron monochromatization[289] or, in principle, for close-to-arbitrary manipulation with the time/energy structure of electron wave packets.

#### 16.1.3 *Outlook*

One of the most striking challenges in the research of electron–light interactions is to utilize the quantum nature of light by enhancing the coupling such that the interaction between an electron and a single photon becomes observable. Going in this direction, the Kapitza–Dirac effect can be viewed as an opportunity to apply a similar principle as the one used in homodyne detection of individual photons by utilizing the interference nature of the two optical waves generating the ponderomotive potential. Another quantum aspect of light that can be studied using electrons is the quantum statistics of photons, which has been shown to influence the inelastically scattered electron spectra.[219] Similar effects may be expected for electron diffraction at an optical standing wave formed by a superposition of a coherent beam with a bright squeezed vacuum state of light. When considering the interaction with coherent light, interesting effects are expected beyond the nonrecoil approximation (long/strong interaction regime), where the electron dynamics together with the quantum interference between amplitudes of electron transitions between discrete momentum states enables a rich variety of possibilities for spectral/temporal electron shaping.

### 16.2 Scattering of Shaped Low-Energy Electrons

#### 16.2.1 *Introduction*

The laser-driven control of slow electrons described above and in Sections 11, 11, and 15 can provide a novel opportunity to modulate the electron–matter interaction. The scattering of electrons by atomic targets forms the basis of e-beam imaging and processing. With monoenergetic beams used in ordinal electron microscopes, the electron–matter interaction can be tuned depending on the materials' magnetism or chirality via the modulation of transversal beam profiles and phases. In contrast, light-driven e-beam shaping provides a novel degree of freedom for the control of electron–matter interaction. The light-modulated e-beams have broadband energy spectra and associated temporal (i.e., longitudinal) densities (Figure 16d). It has been predicted that an excitation process induced by electrons passing by can be modulated when the temporal density is shaped into a train of short pulses whose spacing matches with the cycle period of the light resonantly exciting the target.[265] Besides this phenomenon, called free-electron—bound-electron resonant interaction[265] (FEBERI), direct beam profiles on a detector and elastic scattering processes can also be modulated with the longitudinal beam shaping.[290,291]

16.2.2 *Modulation of Electron–Matter Interactions by Electron Wave-Packet Shaping*

We consider the scattering process depicted in Figure 16d. A light-modulated e-beam described by the momentum-space wave function $\phi_e(\mathbf{k}_i)$ with a incident momentum $\mathbf{k}_i$ interacts with a target. The final state of the electron is described by the sum of the unscattered part $\phi_e(\mathbf{k}_f)$ and the scattered part, which is proportional to $i \int f(\mathbf{k}_i, \mathbf{k}_f)\phi_e(\mathbf{k}_i)d\mathbf{k}_i$, where $\mathbf{k}_f$ is the momentum of the electron after the interaction, $i$ is the imaginary unit, and $f(\mathbf{k}_i, \mathbf{k}_f)$ is the scattering amplitude from $\mathbf{k}_i$ to $\mathbf{k}_f$, whose phase depends on the target location. We assume a spatially fixed target, for example atoms in a solid, having a large momentum uncertainty. If there are multiple atoms inside the coherent size of the beam, $f(\mathbf{k}_i, \mathbf{k}_f)$ is given by the sum of the contributions from the atoms. The unscattered and scattered contributions can interfere when the two paths cannot be distinguished by any of the quantities specifying the final state such as the electron's momentum, the target's momentum, or the electronic state.[292] The interference is dominantly taking place at small scattering angles since $\phi_e(\mathbf{k}_f)$ is peaked at the forward direction and, thus, can modulate signals at small angles. For elastic scattering ($|\mathbf{k}_i| = |\mathbf{k}_f|$), it occurs even with monoenergetic electrons and causes a decrease in signals at near-zero angles (i.e., the optical theorem) and an asymmetric angular pattern on a detector with a spatially focused beam, allowing for atomic-resolution differential phase contrast imaging in a STEM. The interference effect is proportional to $i\phi_e^*(\mathbf{k}_f) \int f(\mathbf{k}_i, \mathbf{k}_f)\phi_e(\mathbf{k}_i)d\mathbf{k}_i + \mathrm{c.c.}$ and, thus, depends on the amplitude and phase of the wave function. Therefore, it should be able to control the differential phase contrast with the beam modulation, or inversely, to determine the wave function through the observation of the interference effect. However, interference with broadband e-beams has been scarcely studied so far and future theoretical and experimental investigations are awaited. Unlike monoenergetic beams, the interference might also be induced by inelastic scattering with light-modulated broadband e-beams. The relative phase between the different momentum components related to the temporal shape and coherence can play a role.

At large scattering angles, the scattered part provides the dominant effect. Mathematically, the scattering signal is given by $\left|\int f(\mathbf{k}_i, \mathbf{k}_f)\phi_e(k_i)d\mathbf{k}_i\right|^2 = \left[\int f(\mathbf{k}_i', \mathbf{k}_f)\phi_e(\mathbf{k}_i')d\mathbf{k}_i'\right]^* \left[\int f(\mathbf{k}_i, \mathbf{k}_f)\phi_e(\mathbf{k}_i)d\mathbf{k}_i\right]$, showing that the two paths starting from initial momenta $\mathbf{k}_i$ and $\mathbf{k}_i'$ reaching the same final momentum $\mathbf{k}_f$ are contributing coherently to the signal. Therefore, the relative phase between $f(\mathbf{k}_i', \mathbf{k}_f)\phi_e(\mathbf{k}_i')$ and $f(\mathbf{k}_i, \mathbf{k}_f)\phi_e(\mathbf{k}_i)$ matters. Recently, we have shown numerically that modulations of the total scattering probability as well as the angular profiles of scattered electrons can be achieved by shaping the e-beam wave function $\phi_e(\mathbf{k})$ in both space and time.[290,291] Yet, applications of the light-modulated beams in scattering and collisions are still in their infancy and more opportunities will be suggested by future studies. The longitudinal control of low-energy e-beams by light will provide new opportunities for controlling electron–matter interactions, paving the way for future e-beam applications, including damage-less microscopy and efficient e-beam processing. The possibility to control the quantum state of electrons and to shape the electron wave function in space almost arbitrarily can be utilized in quantum ghost imaging or in imaging based on shaped illumination of the sample combined with single-pixel detection (see Section 12), both of which may significantly reduce the electron dose required for image acquisition.

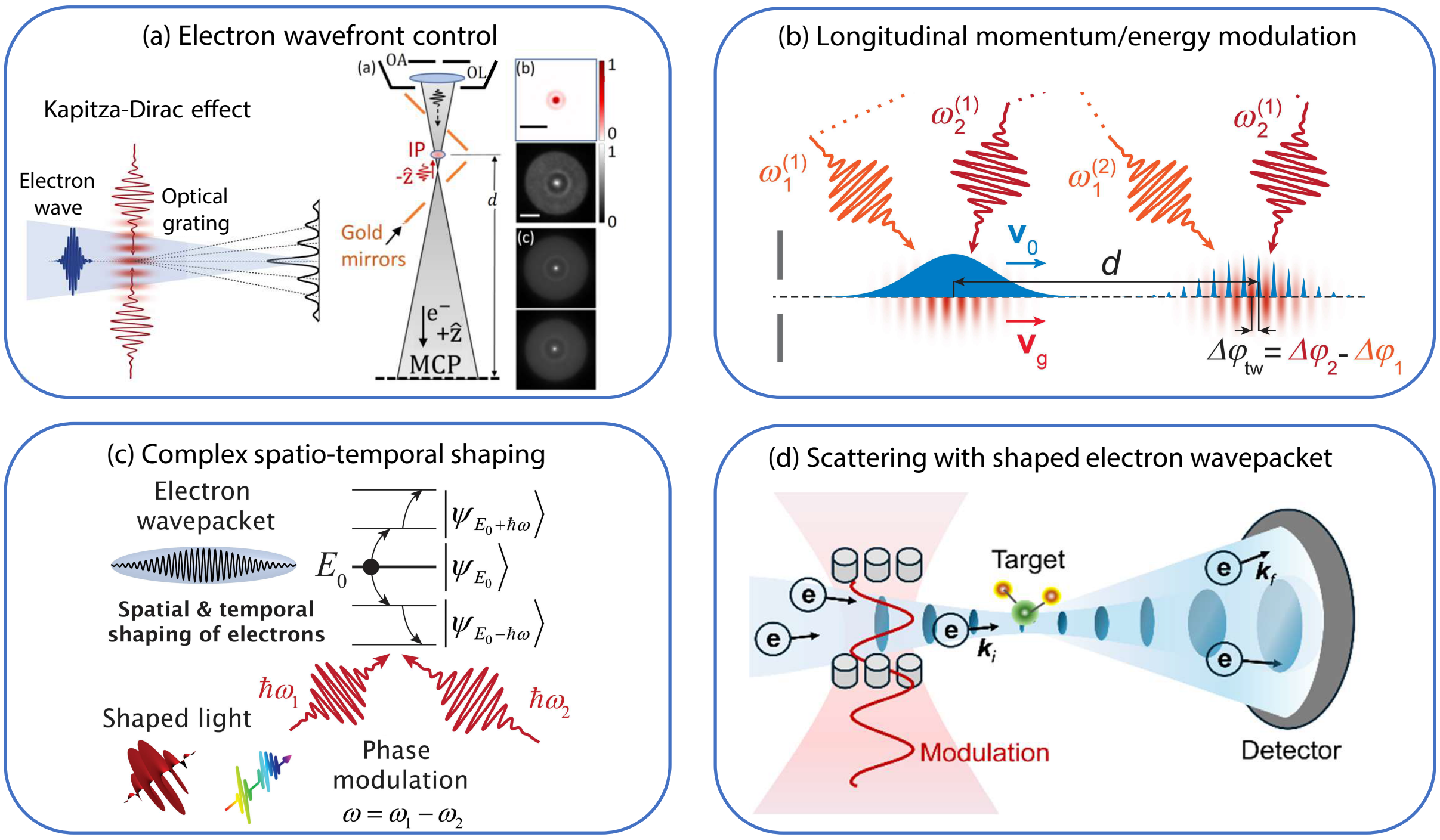

**Figure 16.** Kapitza–Dirac physics and scattering with low-energy free electrons. (a) Electron scattering at a standing wave formed from two intense laser pulses (left). Shaping of an e-beam inside of an SEM with the help of a transversally shaped pulsed laser beam counterpropagating to the e-beam. The optical ponderomotive potential is used to optimize the focusing properties of the e-beam, and more (right). Adapted from ref 193 (Copyright 2022 American Physical Society). (b) Optical traveling waves co-propagating with the e-beam inside of an SEM are excited to imprint an energy modulation to electrons in a first interaction zone. After propagation, the energy modulation translates into a density modulation, generating an electron attosecond pulse train.[171,230,217] Adapted from ref 199 (Copyright 2018 American Physical Society). (c) Quantum-coherent spatiotemporal shaping of electrons by shaped optical fields, including optical vortex beams.[180,288] (d) Control of the electron scattering interaction by shaping the electron wave function.[290,291,292]

## 17. MANY-BODY STATE ENGINEERING IN CORRELATED MATTER VIA SHAPED ULTRAFAST ELECTRON BEAMS

**Francesco Barantani and Fabrizio Carbone***

Periodically driven quantum systems are attracting attention for their potential to realize new exotic states of matter with advanced functionalities and novel properties. Only recently, the advances in ultrafast laser technology have allowed experimental implementation of the driving conditions dictated by the characteristic frequencies of quantum states in materials, finally bridging the gap with theoretical predictions developed before. This line of research is termed *Floquet engineering*, and it encompasses experiments in cold atoms, strongly correlated matter, and van der Waals materials and semiconductors.[293,294,295,296]

Most, if not all, of current studies focus on ultrafast light pulses as the periodic driving mechanism. However, recent progress in the manipulation and control of ultrafast electron pulse technology offers the possibility to engineer temporal distributions as well as spatial profiles of free electron wave functions. Such a possibility has intriguing consequences for Floquet engineering, as electrons provide some significant advantages over light such as atomic-level focusability, transfer of momentum, and very small penetration depth ideal for nanosized and low-dimensional systems.

Typical frequencies involved in Floquet phenomena range from tens to hundreds of terahertz, requiring the preparation of electron pulse sequences separated by intervals on the order of hundreds of attoseconds to tens of femtoseconds. Recent studies[16,169,231] have demonstrated the experimental realization of similar electron pulse trains, while theoretical works[30,265,297] have proposed spectroscopic approaches leveraging periodic electron driving. In the following, we discuss how periodic electron driving can be applied to the investigation of exciton dynamics in strongly correlated systems by exploiting the capabilities of ultrafast TEMs.

Excitons are bound states made of negative (electron) and positive (hole) charges held together by the Coulomb force. Their binding becomes stronger when both charges occupy the same site. Excitons typically form when electrons and holes are excited across a direct band gap in a material and are prevalent in many semiconductors and insulators, whether band gap-driven or Mott–Hubbard, as well as in 2D materials, where the reduced dimensionality further enhances their binding energy. In strongly correlated systems, understanding the interplay between excitons and various degrees of freedom—such as spins, structural excitations, charge ordering, and superconductivity—is crucial yet experimentally challenging. A prominent example is the ongoing effort within the community to investigate excitonic insulators and to confirm their very existence.[298,299,300,301,302,303]

Due to its inherently dynamical nature, an ideal experimental protocol for exciton investigation should map the energy–momentum dispersion as a function of time during the creation, propagation, and decay. To achieve this, one must combine high temporal resolution with simultaneous energy and momentum resolution.[231] This is because excitons are often very sharp features in a material's spectrum and their formation can occur on a sub-femtosecond timescale. Furthermore, to disentangle the interaction with coexisting orders, it is important to investigate the dynamical evolution of the excitons' dispersion across the phase diagram of the material hosting them.

For example, the interplay between excitons and unconventional superconductivity has been a topic of long debate:[304,305] in cuprates, recent high-resolution resonant inelastic X-ray scattering (RIXS) experiments provided evidence of a direct interaction between localized excitons and the spin background surrounding them.[306] A second example is the coupling between low-energy magnetism and excitons in 2D antiferromagnets:[307,308,309] in this context, recent X-ray studies have significantly advanced the understanding of the microscopic origin of these so-called *dark excitons*.[310] From these two examples, it is evident the importance of momentum-resolved information, which is granted by the electron momentum in electron energy-loss studies.

To circumvent the limitations in energy resolution of typical ultrafast EELS experiments (around 0.5 eV), we propose a new experimental protocol based on the coherent control of the excitons by a tailored sequence of electron pulses. One can temporally modulate the e-beam and obtain a train of attosecond electron bunches by a coherent light-driven interaction.[15,169,311] This protocol is illustrated in Figure 17a, where a wavelength-tunable laser pulse scatters off a sharp tip and interacts with the e-beam via its near-field, thereby modulating the electron wave function. By controlling the propagation distance of the phase-modulated electron pulse, the desired bunching effect can be achieved, with a temporal spacing that matches the inverse of the exciton frequency (Figure 17b). In this configuration, excitons are coherently excited by the electrons, similar to the plasmon excitation case described in ref 231. By varying the delay between the attosecond pulses, the excitons can be driven on- and off-resonance, resulting in a modulated EELS probability at their characteristic energy.

Finally, by mapping the EELS response as a function of the time delay, a high-resolution exciton spectrum can be obtained. One can combine this scheme with a slit placed in the back focal plane, following the approach employed in ref 312, and directly extract the momentum-resolved EELS response, eventually enabling the simultaneous determination of both exciton dispersion and lifetime, as sketched in Figure 17c. In such a time-domain protocol, the energy resolution is given by the inverse of the temporal window represented by the largest possible spacing between the attosecond pulses in the train. Therefore, in the currently available configuration, a resolution between 40 and 400 meV is possible.

The advantages of the approach are twofold:

1) Because these experiments are performed in a TEM, energy and momentum-resolved information can be combined with nanometer spatial resolution. This offers a unique playground to look at exciton spatially resolved dispersion and attosecond-resolved lifetime. Among possible candidates, excitons in $Cu_2O$ are known to be very large[313] (up to microns) and ideal candidates that can be mapped by energy-filtered imaging.

2) Our protocol intrinsically offers the possibility to study excitons under out-of-equilibrium conditions. It is sufficient to temporally clock the attosecond electron train with a resonant light pulse to optically drive the exciton and, for example, explore its excited states. Such a protocol will provide additional degrees of freedom for Floquet engineering of the material's excitonic response.

Similar concepts can be applied to any collective excitation that can be probed via EELS—such as phonons, magnons, and plasmons—and harnessed to further investigate mutual coupling, revealing their reciprocal nonlinear interactions. Ultimately, leveraging electrons as coherent excitation channels will serve as a new experimental tool to explore the rich phase diagram of correlated materials, in particular when embedded or microfabricated into nanostructures, where the spatial resolution of transmission electron microscopy will be a crucial advantage.

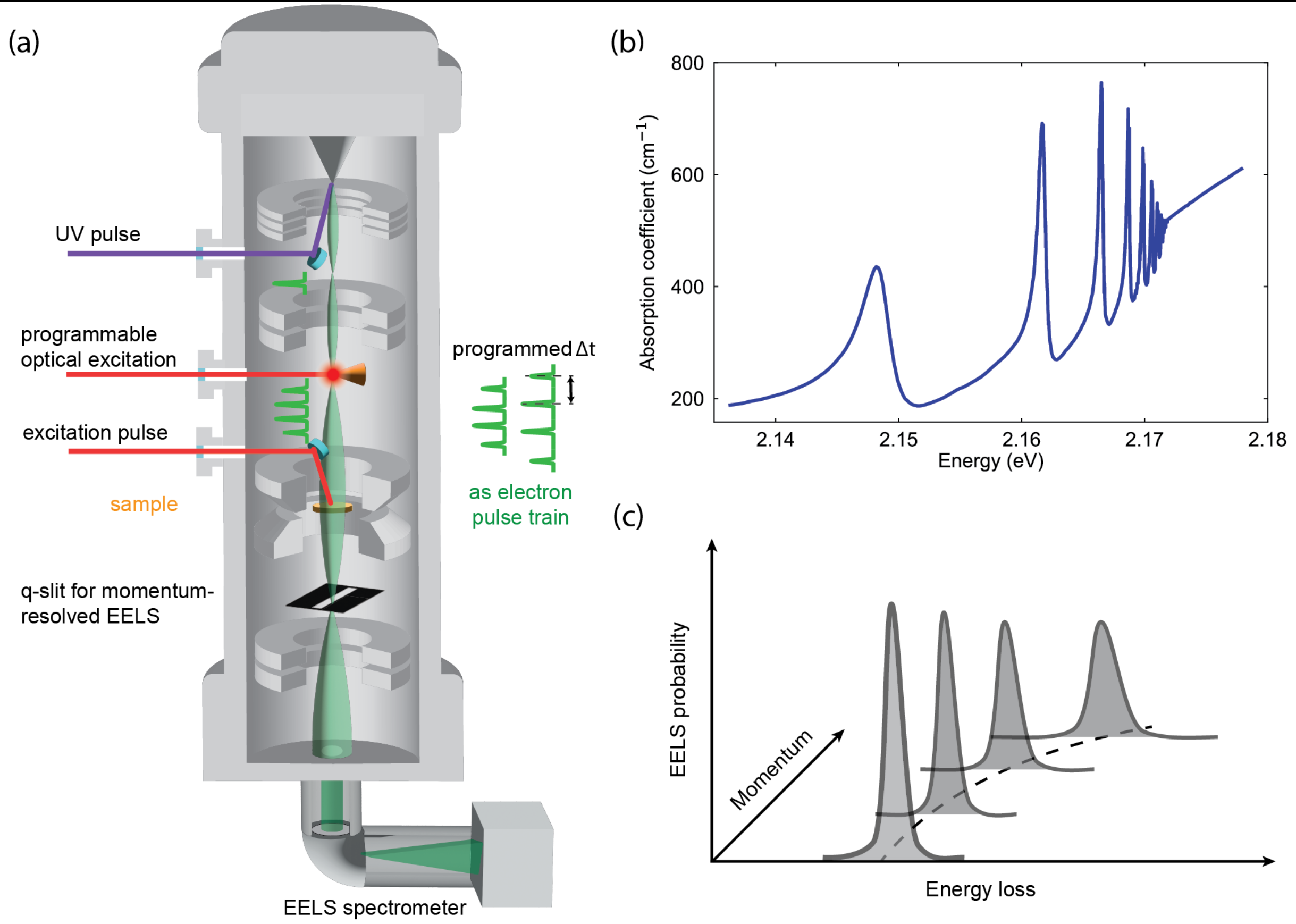

**Figure 17.** Experimental scheme of the proposed Floquet approach. (a) A train of attosecond electron pulses is obtained by interacting photoemitted electrons with the near-field scattering on a sharp tip. (b) The pulse train then excites specific collective modes, namely excitons, in the material. For example, ideal candidates are the excited states of the giant exciton in $Cu_2O$ (data taken from ref 314 ). (c) Depending on the pulse train spacing and duration, one can excite different levels of the exciton and, due to the finite momentum of the electrons, the exciton energy dispersion can be directly measured via momentum-resolved EELS.

**APPLICATIONS IN MATERIALS SCIENCE**

## 18. EXCITONS AND EXCITON-POLARITONS PROBED BY ELECTRON BEAMS

**Fatemeh Chahshouri, Nahid Talebi,* and Mathieu Kociak**

EELS[315] and CL[316] are powerful methods for imaging and characterizing the exciton properties of semiconducting nanomaterials. Correlating electron scattering (elastic or inelastic) with CL[120] or X-ray[209] emission offers deep insights into the chemical and structural properties of nanomaterials as well as their optical behavior. These findings enable precise control over nanoscale light–matter interactions, which is essential for developing next-generation semiconductor devices.

Electron-beam techniques, offering high spatial resolution, have been widely employed in SEMs and STEMs to study excitonic excitations in semiconductors such as transition metal dichalcogenides (TMDs),[315,317,318,319] Perovskites,[320] ZnO,[321] and carbon nanotubes.[322] In the following paragraphs, we summarize some of the reports for characterizing the spectral, spatial, and temporal distributions of the excitonic responses of these materials.

### 18.1 Exciton Physics Explored from Cryogenic Conditions to Room Temperature

TMDs are well known to host room-temperature excitons due to the large binding energy at the K point of the Brillouin zone.[7] Studies on single-layer TMDs[7] demonstrated that the intensity and position of these excitonic peaks vary with temperature. EELS spectroscopy at 150 K revealed well-separated A and B excitons in $MoSe_2$ and $MoS_2$ layers due to the spin-orbit interactions,[1,7] while, by increasing the temperature, the excitonic spectral features shifted to lower energies and broadened, and at 220 K, the two exciton peaks separated by 220 meV were hardly distinguishable.[7] So far, many studies have been conducted at cryogenic temperatures to achieve clearer excitonic responses. However, room-temperature EELS and CL spectroscopy have also successfully detected both A and B excitons in multilayer TMDs.[318] Additionally, temperature-dependent studies on high-quality ZnO microwires (MWs) showed that an increased temperature significantly reduced exciton mobility in the compressive regions of bent MWs, while the exciton lifetime remained unchanged (Figure 18a1,2).[321]

### 18.2 Time-Resolved Spectroscopy of Excitons

Picosecond-time-resolved CL (pTRCL) spectroscopy with high spatial and temporal resolution is an ideal method for studying charge-carrier recombination processes in semiconductors. Using this technique, Corfdir *et al*.[323] measured the decay dynamics of free excitons, donor-bound excitons (D°X), and excitons bound to basal stacking faults (BSF-bound excitons) in GaN. Moreover, the pTRCL has been used to investigate $D^0X_A$ exciton hopping in ZnO microwires,[321] revealing a constant exciton lifetime of 105 ps along straight part of the MWs at various temperatures (Figure 18a3).[321] This technique is now available in STEMs[86,324] to carry out *in situ* EELS–CL measurements as well.[315]

Recently, Talebi *et al*.[325,326,327] have developed a new technique based on an electron-driven photon source (EDPHS) within an electron microscope to perform time-resolved spectroscopy and interferometry with femtosecond temporal and nanoscale spatial resolution. As demonstrated in Figure 18b, this technique involves sequential e-beam interaction with the EDPHS and sample. Using piezo stages, the time delay ($\tau$) between the e-beam and EDPHS radiation arriving at the sample is controlled, enabling Ramsey-type interferometry.[328,329] Moreover, the radiation from the sample is superimposed with coherent EDPHS radiation. The visibility of the Ramsey-like interference fringes is analyzed versus the delay between the EDPHS and the sample, allowing for examining the decoherence time of the generated superposition. Using a $WSe_2$ flake as a sample, spectral interferometry revealed a mutual coherence of 27% between EDPHS and sample radiation. They additionally, mapped the decoherence time of self-hybridized exciton-polaritons in a $WSe_2$ flake to be approximately 90 fs.[101]

### 18.3 Excitonic Response of Vertically Stacked 2D Materials

Adjusting vertically stacked semiconductors of 2D materials at either zero degrees or higher twist angles can influence exciton excitation[330] and tune interlayer coupling.[331] As illustrated in Figure 18c, in twisted bilayers of an hBN-encapsulated $MoSe_2$ monolayer, the intensity and wavelength of the CL excitonic peak vary with the electron probe site.[332] Furthermore, it has been demonstrated that the localized tensile strain, introduced by mechanical stress during the synthesis of $hBN/1L - WSe_2/hBN$ heterostructures, causes a redshift in the CL spectra of excitons.[333] It has been reported that monolayer stacked TMDs can host trions ($X^-$) and lower energy localized exciton (L), in addition to the typical A ($X_A$), B ($X_B$), and C ($X_C$) excitons (Figure 18d).[127] Bonnet *et al.*[127] demonstrated that the absence of residues in hBN encapsulated $WS_2$ monolayers alters the local dielectric environment, increases the free electron density, and leads to trion formation. EELS and CL measurements on the sample revealed localized modulation of trion emission (Figure 18d) when chemical variations on nanoscale dielectric patches change the intensity of $X_A$, and $X^-$. Furthermore, it has been shown that near-field coupling of monolayer and few-layer TMDs with graphene or graphite, with or without hBN encapsulation, can also tune the exciton line shapes and charge states.[66]

## 18.4 Coherence: Exciton-Photon Coupling

Strong interaction between excitons and waveguiding modes in thin films can form exciton polaritons in TMD flakes and result in self-hybridization.[318] It has been shown that the Cherenkov radiation released after electron illumination on the $WSe_2$ flake (60-80 nm) can be trapped inside the sample and couple to the excitons, thereby enhancing the exciton-photon coupling strength.[318] The CL spectrum of these flakes exhibits superbunching and indicates high coherence. By interfering the CL signal generated from exciton polaritons scattered from the edge of the flake with the transition radiation, the coherence level of the radiation is explored.[318] The criterion for constructive or destructive interference depends on the position of the e-beam with respect to the edge, resulting in spatial interference patterns when scanning the flake perpendicular to the edge of the flake (Figure 18e).[318]

Furthermore, by coupling TMDs with plasmonic nanostructures and lattices, the strength of electron–photon interaction can be tuned as well.[334] For instance, Thi Vu *et al.*[335] used a plasmonic nanopyramid array to enhance the luminescence from A and B excitons of $MoS_2$. Moreover, Fiedler *et al.*[336] used a monocrystalline gold nanodisk on top of a $WS_2$/hBN heterostructure to improve synchronization between many exciton emitters excited by the e-beam and to enhance electron–emitter interactions for observing superbunching with a $g^2(0)$ (second-order degree of coherence) up to 2152 ± 236.

Extending polariton studies to hybrid nanoscale systems by combining metal nanoparticles and TMDs also enhances electron–photon interactions. Localized plasmons in this sense act as mediators for shrinking the mode volume and enhancing the electric field intensity. In a hybrid system consisting of a silver truncated nanopyramid (TNP) and few-layer $WS_2$ flakes, Yankovich *et al.*[67] demonstrated plexciton formation due to the overlap between the dipolar localized surface plasmon (LSP) mode of the silver nanoparticles and the A-exciton state of $WS_2$. As shown in Figure 18f, this coupling leads to polariton splitting up to 130 meV in EELS measurement at different corners of the Ag TNP.[67] The strength of the interaction between excitons and plasmons can be further tuned by the thickness of the TMD flakes coupled to a plasmonic Bloch mode when strong exciton-plasmon coupling in 60 nm-thick $WS_2$ flakes form a flat band in the dispersion diagram of the hybrid system.[26] Recent studies have further demonstrated that $WS_2$ nanodisks with a large diameter-to-height aspect ratio can support optical anapoles and anapole-exciton hybrids,[337] which appear as dips in the EELS spectra.

In conclusion, electron probe techniques applied to exciton physics have already granted an enormous understanding of these quasiparticle excitations in different material systems at nanoscale spatial resolution. Further challenges remain to be explored, including correlations among excitons, as well as between excitons and other quasiparticles such as photons and plasmons, which could be directly mapped at the nanoscale. A promising roadmap involves electron–photon coincidence measurements combined with two-photon correlation analyses at the emission wavelengths corresponding to the various quasiparticle excitations. To unravel these correlations, nonlinear optical schemes such as transient absorption spectroscopy and 2D spectroscopy could be integrated with ultrafast electron probes. In such schemes, a series of phase-locked, ultrashort optical pulses could coherently excite the system (possibly with EDPHS structures), while electron beams acting as heterodyne detectors track the spatially resolved evolution. Realizing such a platform will require advancements in the coherent and incoherent excitation of electron beams, including the development of longitudinally and transversely shaped beams with tailored coherence properties.

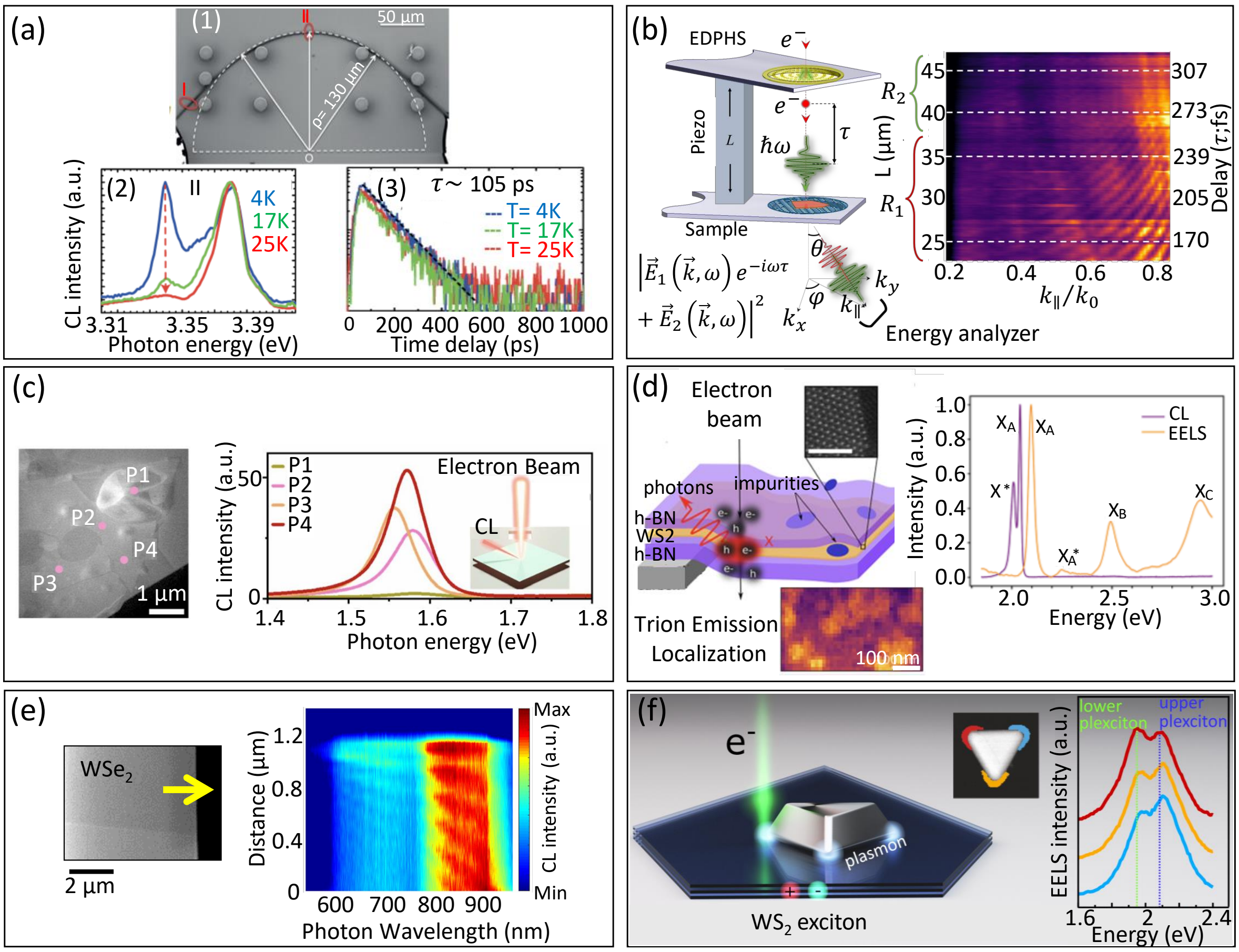

**Figure 18.** Probing exciton excitation, lifetime, coupling, and coherence in 2D semiconductors. (a) 1: SEM image of a probed ZnO MW with a diameter of 2.33 μm. 2: Temperature-dependent CL spectra in the bent region (part II in SEM image). 3: Time-resolved CL for measuring $D^0X_A$ lifetime, as a function of temperature. Adapted from ref 321 (Copyright 2015 AIP Publishing). (b) Schematic of the developed setup for spectral interferometry. The map shows the interference pattern of the EDPHS and sample radiation at the emitted wavelength of 800 nm in the momentum-delay map, where region R1 corresponds to the existence of mutual coherence, whereas region R2 demonstrates the degradation of the visibility of the interference fringes due to exciton-polariton decoherence. Adapted from ref 101 (Copyright 2023 Springer Nature). (c) STEM image of hBN-encapsulated monolayer $MoSe_2$suspended on a TEM grid (left), and CL spectra recorded at the positions indicated by P1, P2, P3, and P4 (right). Adapted from ref 332 (Copyright 2023 Royal Society of Chemistry). (d) Sketch of a sample constructed by a $WS_2$ monolayer (orange) encapsulated by two hBN flakes (purple, 20 and 5 nm thickness) on a holey carbon substrate (gray). Localized Trion emission map and EELS–CL spectrum showing Trion excitation. Adapted from ref 127 (Copyright 2021 American Chemical Society). (e) Schematic of a $WSe_2$ thin film supporting waveguided optical modes that are confined in the transverse direction and can strongly interact with excitons. Measured CL spectra for $WSe_2$ thin films with depicted thicknesses. Adapted from ref 318 (Copyright 2022 Wiley). (f) Schematic of a strongly coupled plexciton system composed of an Ag TNP and few-layer $WS_2$. The red, yellow, and blue electron energy-loss spectra are from each corner of the coupled TNP-layer $WS_2$ system, where the STEM image shows corners. Adapted from ref 67 (Copyright 2019 American Chemical Society).

## 19. PLASMONIC AND QUANTUM NANOMATERIALS PROBED BY ELECTRON BEAMS

**Wiebke Albrecht,* Sergio Rey, Toon Coenen, Erik Kieft, and Johan Verbeeck**

### 19.1 Introduction

The interaction of light and free electrons is crucial for plasmonic and quantum nanomaterials because it helps us understand how local structural features, such as defects or particular morphological details, affect electromagnetic fields, energy and charge transfer at the nanoscale, and quantum phenomena. In this section, we report on this interaction studied by electron-based spectroscopies. Ultimately, materials are used in actual applications and should be analyzed under application-relevant conditions. Here, we include the example of plasmon-mediated photocatalysis, which is expected to benefit greatly from utilizing local light-free electron interactions.

## 19.2 State of the Art

### 19.2.1 *Plasmonic Nanomaterials*

Plasmonic nanomaterials are one of the most obvious systems that benefit from spatial and spectral mapping of the optical response as they confine electromagnetic fields below the diffraction limit in a morphology-dependent manner. Electron-beam probing can offer high-resolution near-field information not accessible in classical far-field optical detection. In addition, unlike optical techniques, e-beams can excite both radiative and nonradiative plasmon modes, providing a more comprehensive understanding of plasmonic behavior. Furthermore, electron-based methods allow for site-specific analysis, enabling the study of individual nanostructures, defects (Figure 19a), and complex plasmonic interactions at the subwavelength scale, thereby revealing local field enhancements and quantum effects.[191,338,339] Combining e-beams with *in situ* modification techniques further allows for the dynamical manipulation of plasmonic systems[340] (Figure 19b).

### 19.2.2 *Plasmonic photocatalysis*

Catalytic materials are essential for producing fuels, plastics, fertilizers, and pharmaceuticals but catalytic processes significantly contribute to climate change. Utilizing light-driven catalysis, such as plasmonic excitation, would be a game changer—not only for climate-friendly technologies but also because ultrafast optical excitations can guide reactants and catalysts through a time-dependent energy landscape, overcoming the classical Sabatier limit.[341] Due to their nanoscale size, electron microscopy is a key tool for characterizing catalysts. Atomic-scale morphological changes in catalysts significantly impact their activity, selectivity, and stability. *In situ* TEM is the only technique that captures these changes dynamically with high spatial resolution under relevant conditions.[342] With recent advances in optically-coupled TEM, catalytic processes can now also be followed under light excitation with atomic resolution[343] (Figure 19c). Electron-based spectroscopies are hereby relevant as they reveal catalyst composition and electronic structure and are now used for local product detection.[342,344] Combined with structural data, they link active sites to selectivity and activity, a key goal in catalysis[345] (Figure 19d), and reveal dynamic operando information.[346]

### 19.2.3 *Quantum Technologies*

Qubits, the fundamental units of quantum technology, are controllable two-level systems used in quantum sensing, computing, and networks. Optically active quantum emitters show great potential in computing and cryptography. Electron-beam-based spectroscopy is a powerful tool for studying these emitters, offering high spatial resolution, broadband excitation, and spectrally resolved data. It helps determine atomic composition[347] (Figure 19e), strain effects,[348] decay lifetimes, and quantum efficiency,[89] while also linking structural properties to changes in emission wavelength, brightness, linewidth, and phonon coupling[349] (Figure 19f), which will be crucial for on-demand quantum light sources.

## 19.3 Challenges and Future Goals

Material research utilizing electron excitations, in particular under relevant application conditions, face several challenges, which are summarized in this section.

### 19.3.1 *Beam-Sensitive and Organic Materials*

Organic and beam-sensitive materials have become important in novel electronic, energy, and quantum materials but are also the basis of catalytic conversion processes. Electron-based spectroscopies require higher electron doses than electron imaging due to lower inelastic cross-sections,[350] posing challenges for beam-sensitive materials that struggle to withstand even a single image. In addition, EELS identifies atomic constituents but lacks in organic molecule selectivity compared to bulk spectroscopies. Advancements in detection, instrumentation, and data processing, addressed, for example, in the European research project EBEAM,[11] are key to overcoming these limitations.

19.3.2 *In Situ/Operando Metrology*

*In situ* and *operando* electron microscopy are crucial for energy materials research, but real conditions are hard to replicate due to space and vacuum constraints. Micro-electromechanical-based devices help but struggle to precisely correlate product formation with the imaged region. Electron-based spectroscopies can overcome this limitation if the following key challenges are overcome: (1) studying e-beam effects, especially with high-dose requirements for spectroscopy; (2) developing fast spectroscopy methods and analysis tools to handle noisy data; and (3) improving environmental cell designs to enhance X-ray energy-dispersive spectroscopy (EDS) and EELS detection. Advancing spectroscopic cells and tailored detectors for *in situ* studies is essential.

19.3.3 *Dynamic Information*

Measuring time-dependent phenomena is a desirable complement to the high spatial resolution of (S)TEM. Depending on the detector, time resolutions from milliseconds to sub-picoseconds are achievable.[351] Improving time resolution further is limited by the need for sources that deliver both high brightness and current while keeping pulse charge intact despite Coulomb interactions. Pump-probe-like techniques can help, but they often have *empty* pulses, with less than one electron per pulse on average, increasing measurement time and only suitable to study reversible processes. In contrast, irreversible processes require single-shot detection, and even with recent nanosecond resolution,[352] electronic switching speeds remain a barrier to capturing faster events.

19.3.4 *Extension to 3D Information*

(S)TEM is inherently a 2D projection technique. To access information along the projection direction, several techniques have been developed. (1) Ptychography-based methods yield 3D information (see Section 20). (2) The tomographic principle applies to many nanomaterials and can be used in spectroscopic techniques when the signal varies monotonically with thickness, as is typical for X-rays, CL, and core-loss EELS.[353] However, caution is required for low-loss EELS, which can represent vector fields (e.g., localized surface plasmons) and limit CL and EELS applications.[354] Of course, beam damage is also a major concern for tomographic acquisition due to longer exposure times to the e-beam.

19.3.5 *Heterogeneous Systems*

Even with atomic resolution, electron microscopy faces challenges due to the heterogeneity of most industrial materials, which are rarely single crystalline or single phase. The same holds true for colloidally synthesized nanomaterials such as plasmonic nanoparticles, which can display significant size and shape heterogeneity. This complexity makes it difficult to link macroscopic properties to microstructure, necessitating statistically relevant sampling through automated data acquisition and analysis—a shift from traditional manual operation by experts.

19.3.6 *Statistics and Reproducibility*

Electron microscopy in materials science is often user-driven, leading to several challenges. Results can be user-dependent, as choices on imaging, sequencing, and instrument settings introduce uncertainty. Additionally, limited image/data acquisition can hinder statistical analysis, causing variability in results and difficulties in reproducing outcomes. To fully exploit electron microscopy and spectroscopy, reliable connections between microscopic data and macroscopic material behavior are crucial, emphasizing the need for improved reproducibility and statistics.

19.3.7 *Sample Preparation*

Accurate measurements demand careful sample preparation. TEM lamellae must be precisely targeted to avoid defects, ion damage, and oxidation. Nanoscopic samples should represent the bulk: nanoparticles need a homogeneous, varied orientation distribution, and colloidal samples must be free from organic residues. Especially for sensitive materials, new cleaning methods are needed to overcome this challenge.[355] To mitigate heat and radiation damage, encapsulation with graphene or similar layers can reduce mass loss through diffusion, and metal layers can act as effective heat sinks.[150] This is of particular interest for time-resolved measurements, where drift correction is essential.

## 19.4 Suggested Directions to Meet Goals

### 19.4.1 *Combination with Complementary Techniques*

Combining complementary techniques on the same sample addresses many of the challenges mentioned above and provides a more complete understanding.[356] Popular alternative characterization techniques include optical spectroscopies, X-ray techniques, scanning probe microscopy (SPM), and mass spectrometry (MS). Photons are less damaging than electrons and enable non-destructive analysis of thicker samples with less preparation. By utilizing the large range of photon energies and the plethora of developed optical techniques, obtainable information ranges from sensitive chemicals to ultrafast dynamics, while diffraction-limited resolution can be addressed by identical location TEM if possible.[357,358] SPM is also a non-damaging complementary alternative to electron microscopy, that can more easily operate under ambient or liquid conditions and correlate local properties not easily accessible in the electron microscope (electric, mechanical, and magnetic) to topography but is limited to surface information. MS delivers detailed chemical and isotopic data with high throughput, even though it is generally destructive and offers lower spatial resolution. However, it can be conveniently paired with *in situ* or *operando* electron microscopy to provide complementary chemical insights.[359]

### 19.4.2 *Technological Developments*

Ongoing technological advancements show promise in overcoming the mentioned challenges. Freely programable phase plates[216] hold the potential for rapidly adjusting imaging setups that reveal the most information per incoming electron[360] and have the potential to impose prior information to further optimize the quantum efficiency of the measurement. Similarly, further improvements in coincidence measurement techniques will enhance information extraction per event by isolating correlated signals, which can be particularly useful when signal-to-noise rather than damage threshold is the limiting factor. Given the high speed of electrons in TEM, the opportunity arises to take snapshots of time-varying phenomena and evanescent fields with frequencies that far surpass optical frequencies and the TEM can act similar to an extreme bandwidth oscilloscope with atomic-scale spatial probing. This requires the development of fast shutters and detectors as well as ultracold photocathode sources.

### 19.4.3 *Automation*

In order to improve the reproducibility and reliability of electron microscopy in materials research, it will be critical to further automate the imaging processes, and to remove the user dependencies as much as possible. The automation is a stepwise process, where initially specific steps of a workflow can be documented in a standard operating procedure and subsequently automated. In the preparation phase, steps such as reliable sample loading/mounting, recipe loading, and alignments/calibrations are of interest. In the acquisition phase, one can consider automatic navigation, possibly guided by specific input and followed by automatic data collection. In the analysis phase, live data processing and visualization, image stitching, and batch processing can aid the automation process. In all of the phases above, proper recording of all relevant parameters in metadata form is essential. Furthermore, it is likely that they can be enhanced/augmented by using state-of-the-art machine-learning/artificial-intelligence (ML/AI) tools.

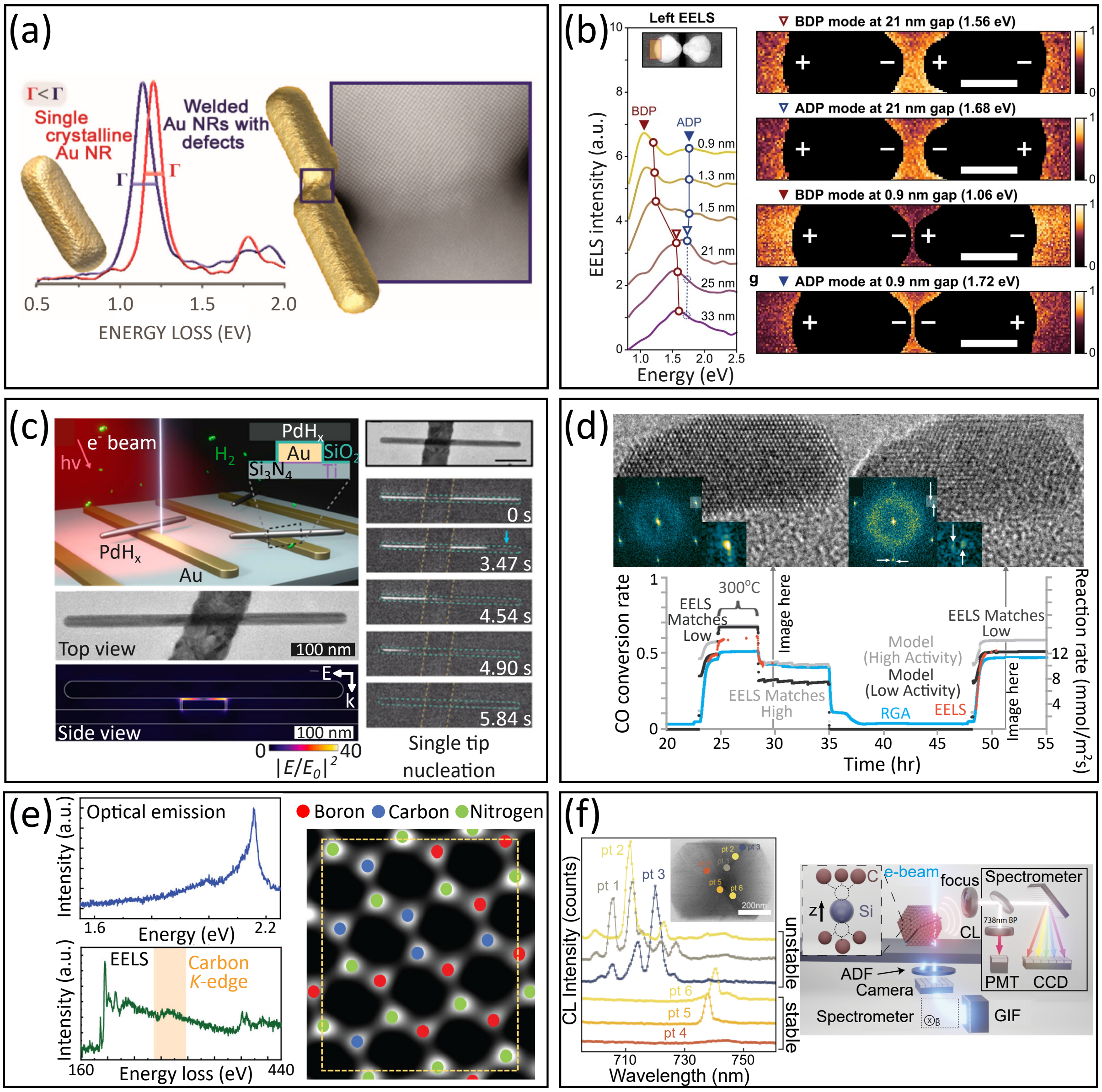

**Figure 19.** Local structure–property correlation in plasmonic, catalytic, and quantum nanomaterials probed by e-beams. (a) EELS and electron tomography are used to demonstrate plasmon broadening induced by a single defect in Au nanorods. Modified with permission from ref 339 (Copyright 2020 American Chemical Society). (b) EELS enables spatial and spectral mapping of bright and dark bonding plasmon modes for a gold nanodisk dimer, for which particle distances were modified *in situ* through a nano-electromechanical system. Modified with permission from ref 340 (Copyright 2021 Springer Nature). (c) Optically coupled TEM allows imaging and control of light-induced chemical transformations in nanoparticles. Modified with permission from ref 343 (Copyright 2021 American Chemical Society). (d) Operando experiment of CO conversion on Ru catalysts correlating structural modifications of the catalyst to changes in conversion efficiency. Modified from ref 345 (Copyright 2024 American Chemical Society). (e) High-resolution STEM plus EELS determination of the chemical environment of quantum emitters in hBN correlated to their optical emission. Modified from ref 347 (Copyright 2024 American Chemical Society). (f) Correlation of local optical properties to structural features of Si-vacancy single-photon emitters in nanodiamonds using cryo-CL. Modified from ref 349 (Copyright 2024 National Academy of Science).

## 20. ELECTRON PTYCHOGRAPHY FOR LOW-DOSE IMAGING

**Hoelen L. Lalandec Robert* and Johan Verbeeck**

### 20.1 Introduction

Ptychography, originally introduced by the work of Hoppe,[361,362,363] denotes a class of computational imaging methods permitting the reconstruction of the spatially dependent scattering power of an object through its illumination by coherent radiation and a subsequent position- or momentum-wise collection of the probing particles. This measurement is repeated under differing conditions, typically while laterally shifting or tilting the illumination, thus allowing for the correlative treatment of the distinct recordings and the retrieval of a specimen-induced phase shift map. Ptychography can therefore be understood as an extension of coherent diffractive imaging[364] paradigm, where a lack of prior knowledge is compensated by the exploitation of information redundancies within the multidimensional scattering dataset. Due to technological limitations, in particular related to detector electronics at the time of its original development, the technique could only be demonstrated in the 1990s.[365,366,367] Nowadays, its wide field of applications encompasses, for example, polymers,[368] zeolites,[369] viruses,[370] and halide perovskites.[371]

Here, we focus on electron-based implementations of ptychography. Note that similar (or even optically identical) experimental setups exist in X-ray and electron microscopies. However, they remain based on different beam–specimen interaction models and are thus distinct in terms of the retrieved physical quantities. Achievable resolutions, largely determined by radiation wavelengths and numerical aperture, are also much higher in the electron-based implementation. This points out that specific probing particles are used for different applications and fields, though practical implementations and computational frameworks may remain comparable.

### 20.2 Implementation

Several geometries exist for the practical implementation of ptychography, including those based on structured illumination[372] or the near-field evolution of scattered fields,[373] employing multiple tilts of the specimen[374] or encompassing spectroscopic information.[375] The most popular is arguably the focused-probe approach, illustrated in Figure 20, where the illuminating radiation is made to converge on a solid-state specimen, while a 2D scattering distribution is acquired in the far-field by a pixelated detector. The probe is furthermore scanned in the two lateral real-space dimensions, with a unique recording performed at each position. Redundancy among the scattering patterns is ensured by keeping the spatial interval of the scan grid small enough so that a significant overlap of illuminated area occurs in between single measurements. Conventionally, the object is represented by a multiplicative transmission function, being then retrieved by the process. From the computational side, a distinction is made between direct analytical solutions[376,377] and iterative algorithms, which encompass, for example, those based on the ptychographic iterative engine[378,379] as well as maximum likelihood estimators.[380]

### 20.3 Application to Low-Dose Electron Microscopy

An important application of electron ptychography is its use for beam-sensitive specimens, where the transfer of energy to the specimen is prevalent, leading to, for example, knock-on displacement of atoms, heating, or radiolysis. In practice, those damage mechanisms impose a critical dose beyond which the structure of the imaged specimen is lost. Depending on the type(s) of damage occurring, this critical dose may range between $10^{-1}$ and $10^{3}$ $e^{-}/Å^{2}$. There, the interest of ptychographic computational imaging lies in its efficient use of all available information, as it aims at directly matching the experimental acquisitions with theoretical expectations. In that context, the precision with which the specimen's electrostatic potential can be retrieved displays a nonlinear relationship with the dose invested, while the accuracy is dependent on the correctness of the assumed interaction model.[381,382] This can be demonstrated by arguments of information theory such as the Cramér-Rao lower bound. This parameter may furthermore be useful to make predictions on the fundamental capacity for frequency transfer[383,384] by the acquisition of the scattering data, as well as on the best resolution achievable while conserving a targeted signal-to-noise ratio in the micrograph. The convergence of a ptychographic reconstruction under conditions where Poisson noise is prevalent has also been investigated empirically.[385,386]

The low-dose application of ptychography implies a need for cameras with high detector quantum efficiency (DQE) and short frame time. Though the required evolution in DQE was fulfilled by hybrid-pixels direct electron detection granting single-electron sensitivity, matching the standard recording speed of conventional electron microscopy techniques was made possible only recently thanks to the introduction of event-driven detectors.[387] Whereas most electron cameras possess a frame-based readout, where stacks of 2D frames are produced through the recording, such event-driven detectors only register a list of single counts, each being identified by the responding pixel and the time of arrival. This recording paradigm is especially well-suited for low-dose measurements where only sparse data is expected and where fast scans are important.

## 20.4 Conclusion

Ptychography constitutes a general framework for phase retrieval based on the correlative use of multiple scattering patterns, in principle requiring no prior information other than the interaction model. The appeal of the focused-probe, electron-based variant can be related to two main factors: its dose efficiency and the capacity for superresolution.[388] From this point, the two most important directions of progress for this emerging method are already clear. The first one consists in continuing efforts aiming at adapting more sophisticated interaction models to ptychography in order to image materials with the highest possible resolution and solve complex material problems, while potentially putting new recording dimensions to use and complexifying the experimental setup. This is, in essence, the high-dose, high-computation, application of this technique. The second direction consists in providing ptychography as a general tool for the live (real-time) imaging[389] of arbitrary specimens and further multidisciplinary work, thus introducing a need for streamlined calculations. Such a development is important for straightforward use in fields where the capacity for fast and reproducible measurements is necessary, and where beam damage is frequent. In particular, the integration in a more extensive experimental process is desirable. An example can be found in biology,[390] where a single-particle analysis procedure may be used to produce a three-dimensional (3D) map of, for example, a virus or a protein, which is based on a high number of pre-acquired 2D micrographs.

This low-dose application of ptychography is also of great interest for nanophotonic materials and semiconductors in general, due to their high susceptibility to radiolysis by the incident electrons. In particular, techniques involving the collection of a CL signal are known to cause unwanted damage due to the need to work with high e-beam doses, rendered necessary by the low emission probability per incident electron. In this context, complementary dose-efficient imaging techniques providing structural information are useful to ensure an accurate correlation with the measured optical properties.

Furthermore, ptychographic and ptychography-like computational methods are of interest for future applications of coherent CL, such as those involving the reconstruction of optical near fields by prior propagation into the far field. Such experimental developments could profit from the already available knowledge on coherent diffractive imaging employing light. In that respect, it is also noteworthy that X-ray scattering is currently dominant in the field, for instance permitting users to overcome the imperfections of focusing optics by refining the assumption made on the illumination. This capacity for lensless imaging is of particular importance, as manufacturing difficulties are typically prevalent in providing such devices to synchrotron facilities. More generally, this approach opens the way for further applications in photonics consisting in the retrieval of 3D wave fields through the prior acquisition of scattering patterns.

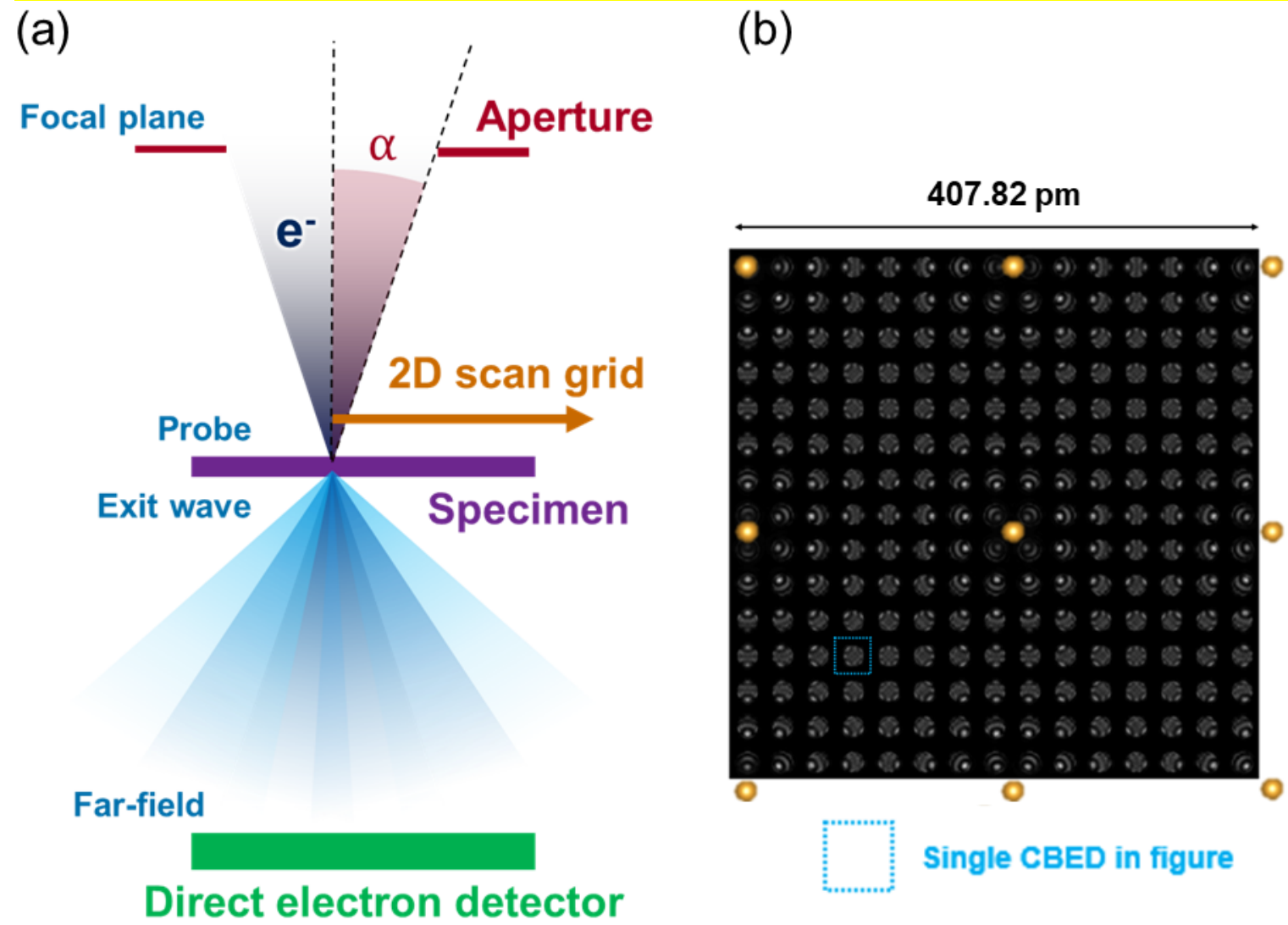

**Figure 20.** Ptychography with a scanning transmission electron microscope (a) Illustration of the focused-probe electron ptychography method. A convergent beam of high-energy electrons is incident on the specimen. A far-field scattering pattern is collected by a direct electron detector. The electron probe is scanned along the two dimensions of space. (b) Simulated examples of a convergent beam electron diffraction pattern obtained for a collection of scan positions. Recordings are done over a [100]-oriented Au unit cell, with atomic positions indicated in the figure.

## 21. INDUSTRIAL PERSPECTIVE

**Toon Coenen,* Frank de Jong, Erik Kieft, and Magdalena Solà-Garcia**

### 21.1. Current State of the Art

Since the invention of the electron microscope (EM) by Ernst Ruska in 1931,[391] there has been a close interaction between the scientists developing the fundamental concepts into working prototypes and high-tech industries which further developed electron microscope products to be used in various fields. AEG and Siemens worked on the first commercial TEM prototypes in Germany, followed by companies like RCA (USA), Philips (NL), and JEOL (Japan), among others. In this dynamic environment, many new improvements and additions were (and still are) introduced, often pioneered in research institutes, and then further developed and commercialized by either larger companies or by focused startups. Applications in materials and life sciences developed, supported by novel sample preparation techniques. The resolving power of the microscopes was improved over the years, making use of sophisticated electron-optical correctors and brighter electron sources, from tens of nanometers originally down to 50 pm in current commercial microscopes. The development of computer-controlled microscopes, as well as improved digital detectors and subsequent extensive computer processing of the resulting data enabled more automation, more complex experiments on many more samples and generally gave an enormous boost to the usefulness of the various EM instruments. Cryo-cooling of the samples enabled more detailed investigations of proteins and other macromolecules down to the atomic level. Scanning techniques were developed which resulted in SEMs allowing surface investigations in many different fields. This also enabled analytical electron microscopy (both for SEM and STEM), in which other signals resulting from the bombardment of the sample with high-voltage electrons are used. X-ray spectroscopy (XRS), EELS and (IR-visible-ultraviolet) light in CL techniques can be used for elemental and chemical analysis of samples. Operando techniques were developed to overcome the limitations of the vacuum environment (needed for the e-beam), and to study dynamic behavior of samples.

One of the most recent and exciting developments is to combine electron microscopy with light which can be done in multiple ways as is also outlined above. Here, we will briefly summarize these developments from an industrial perspective. In a first instance, light that is generated by an e-beam (CL) can be used to understand (optical) materials properties at the nanoscale.[316] Initially, the technique only allowed measuring light intensity and spectral content, but recent technical advancements have boosted the number of possibilities now allowing measurements of many facets of light such as polarization,[392] momentum distribution,[393,394] and dynamics, such as $g^{(2)}$ (second-order autocorrelation[87]), lifetime measurements (with a pulsed e-beam, see below),[81,84,395] and novel pump-probe schemes enabling new research and development in the materials science and semiconductor realms. A separate branch of the combination of light and electrons in microscopy is known as correlative microscopy, which in some cases can be performed in one instrument. Several commercial implementations of modules and complete instruments have been developed. This field is reviewed by Ando *et al*.[396] and is outside the scope of this paper. A comparatively recent development is the introduction of laser light into the EM. Pulsed laser beams are used to trigger sample dynamics in pump-probe EM, where the response of a sample is studied with the e-beam after an external excitation, analogous to all-optical ultrafast science, but now at nanometer scales. Also for ultrafast electron–light–matter interactions, pulsed lasers are often used in order to reach the desired light intensities. The use of pulsed laser beams requires pulsing of the e-beam (to be used as the probe in the experiments) at a commensurate timescale. Developments started as early as the 1970s, with implementation of a variety of beam modulation techniques.[397,398,399] The modern era of time-resolved EM kicked off in the early 2000s with pump-probe experiments using laser pulses both for pumping the sample, and for generating the electron pulses through photoemission. Bostanjoglo[400] pioneered single-shot pulsed TEM imaging at nanosecond timescales, quickly followed by the Dynamic TEM (DTEM) lab at LLNL.[401] Meanwhile, the lab of Zewail broke the picosecond barrier for the study of repeatable sample processes,[166] bringing the concept of femtochemistry to the EM world. Almost in parallel, a similar evolution happened for SEM.[84,402,403] As with the development of EM overall, innovations coming out of universities and research institutes were supported and then further developed by established companies (FEI, currently part of Thermo Fisher) and start-ups (most notably IDES, originating from LLNL). The past decade has seen a modest growth of the market as well as consolidation (IDES being acquired by JEOL).

Parallel to the exciting developments in photoemission sources, there have been significant advancements in a trend toward beam blanking and chopping methods in recent years, which pose a viable alternative for beam pulsing. These approaches offer advantages in terms of integration, fast switching between pulsed/CW operation, stability, and frequency range (up to GHz repetition rates are possible while keeping the baseline microscope performance intact). When using chopping, the total beam current is limited by the current range that can be used in CW operation, whereas with laser-driven photoemission, more intense electron pulses can be generated, which benefits DTEM and certain classes of stroboscopic/analytical experiments. Photoemission microscopes can reach pulse lengths down to 100s of femtoseconds,[141] in contrast to the picosecond and nanosecond timescales of typical beam blanking/chopping techniques. However, new developments, such as the use of a resonant radiofrequency (RF) deflection cavity, are enabling similar time resolutions.[404] Synchronization between RF deflectors and femtosecond pulsed lasers can benefit from a large body of particle accelerator based research, and sub-100 fs jitter levels are readily achievable. Deep sub-picosecond overall time resolution has been demonstrated with the resonant RF cavity chopper.[405] For the nanosecond domain, fast electrostatic beam pulsers are available, which can be used for a range of applications, including blanking during flyback in STEM to prevent unnecessary sample exposure, and further pulse picking of resonantly generated MHz-GHz pulse trains. On the SEM side, integrated ultrafast beam blankers with time resolutions down to 50 ps are now available. These can be used for time-resolved CL imaging in compound semiconductors as well as electrical failure analysis in the development of advanced logic devices to name two examples.

The applications addressed by electron microscopy nowadays are usually divided into three sectors. In the semiconductor industry EMs are extensively used for metrology, defect review, and failure analysis both in-line and in the development labs. In the life sciences EMs are used for subcellular investigations, with a prominent role for EM in structural biology revealing the atomic structure of macromolecules such as proteome complexes and virus particles. The remaining sector, materials science, includes a large variety of academic and industrial users in metallurgy, chemistry, geoscience, automobile and nanotechnology. The renewable energy area (including energy storage, photovoltaics, energy conversion systems, etc.) presents an important and fast-growing subsector. As described in this roadmap, quantum science and nanophotonics are developing into more mature sectors, driven by many of the developments described here.

The EM industry landscape features a limited number of companies offering a broad product range of EMs: Hitachi (HHT) and JEOL in Japan, Tescan (Czech Republic), Thermo Fisher Scientific (Czech Republic, Netherlands and USA) and Zeiss (Germany). Bruker (formerly Nion, USA) offers very specialized STEMs. Some new EM vendors are currently developing in China and South Korea. A few companies concentrate fully on (in-line) tools for the significant semiconductor market such as Applied Materials and KLA (USA based with main facilities in Israel). Other companies are specializing on attachments and modules and include CEOS (correctors), Gatan (cameras, EELS, and CL, now part of Ametek), Bruker (XRS), Oxford Instruments (XRS) and EDAX (XRS, now part of Ametek). Over the years small companies have started often as university spin-outs focusing on specific modules such as Delmic and Attolight (CL); Hummingbird, DENS solutions and Protochips (*in situ* specimen holders); ASI, Advacam, Dectris, Direct Electron, Imascenic, Quantum Detectors and Tietz (direct and hybrid pixelated electron detectors), NenoVision (in-situ AFM), IDES (now part of JEOL), DrX Works, and Euclid Techlabs working on specific modules for time-resolved EM. The global electron microscopy market has been estimated to be $3.9B in 2022, with an expected compound annual growth rate of 8.4% until $7.44 billion by 2030 (see, for example, ref 406). We note that this report excludes semiconductor metrology, inspection, and defect review tools so the actual market is larger.

## 21.2. Challenges, Future Goals and Suggested Directions

### 21.2.1 *Time-Resolved EM*

Ultrafast electron microscopy tools have the potential to democratize ultrafast nanoscale science and to bring these capabilities into individual laboratories, complementary to large-scale synchrotron and free electron laser facilities. However, despite the technical progress that has been made in the last 15–20 years in developing (time-resolved) EM techniques which involve light, they currently remain restricted to specialized laboratories equipped with highly advanced dedicated equipment. While impressive performance levels can already be reached and system complexity has been reduced, the technology is still relatively immature (technology readiness levels <6) in terms of cost and useability, demanding substantial tuning. This makes it suitable for early adopters but not yet for mainstream user groups. Strong connections to larger application sectors mentioned above have yet to be established. Fast beam blanking/chopping techniques have been introduced to the market as outlined above, which minimize the user interaction required for operating a pulsed-beam experiment and reduce the sensitivity of the EM tool to environmental conditions such as vibrations. Thus, the bar for penetrating the general EM market is lowered, but further integration and automation of various additional components (including light injection and collection paths into the microscope column, time-resolved detection, synchronization of signals and specialized sample holders) will be needed to serve more main-stream user groups. Besides these technical advancements, standardization of relevant metrics like temporal resolution and standardization of measurement procedures in terms of reference materials/samples will help users and suppliers to properly benchmark and maintain their equipment. In terms of expanding the market for the technologies covered here, we see potential for use beyond the expert EM labs. In particular, there is potential to link time-resolved CL imaging to compound semiconductor materials analysis in photovoltaics, power electronics, microLEDs, and lasers. Furthermore, the use of pulsed beam technologies in failure analysis of advanced semiconductor devices can be further developed. In these cases, it will be critical to work with advanced semiconductor research labs such as IMEC, CEA-LETI, and/or Fraunhofer institute(s), and industrial customers to bring the technology further and to understand the needs for each application.

### 22.2.2 *Contrast enhancement and damage reduction*

A more general grand challenge in EM is to maximize image contrast for a given electron dose. In particular, beam-sensitive (organic) compounds such as battery materials, perovskites, metal-organic frameworks, and biological samples embedded in vitreous ice (cryo-EM for investigation of proteins, macromolecular particles or subcellular structures) get irreversibly damaged before enough information is collected, hampering the adoption of electron microscopy in such applications. Additionally, many organic materials have a lack of general EM contrast because the materials are composed of a non-crystalline collection of light elements which only impose weak phase shifts on the beam. In terms of mitigating beam damage caused by e-beam there are several parallel tracks that are being pursued. First, cryogenic cooling of samples can stabilize materials and mitigate beam damage.[407] Second, there has been an ever ongoing improvement in electron detection in terms of efficiency and speed in direct-electron and hybrid-pixel TEM camera technology,[408] to the point where, through electron counting, the sensitivity is approaching the inherent shot noise limit. Third, new imaging approaches such as integrated differential phase contrast STEM imaging[409] and ptychography[410] have potential to extract more image information for a given dose. Fourth, optimized illumination conditions and novel non-raster scan approaches[411] provide further avenues to mitigate damage effects. A separate field is developing in which a laser is used to influence the e-beam wavefront directly. One important application is the use of the laser to induce a phase shift in the unscattered e-beam (after the sample) using the ponderomotive effect, similar to the concept of a Zernike phase plate in light optics. This effect can be harnessed by using a strong optical field in a Fabry-Pérot cavity for example.[213] Such a phase shift is essential to enhance the contrast in notoriously low-contrast samples which are also prone to e-beam damage as mentioned above. Methods to enhance phase contrast in TEM have been recently reviewed in ref 412. The phase shifting in the e-beam can also be attained by using electronic phase tuning.[216] This can be extended to arbitrary waveform shaping providing even more control over how materials are imaged (a first commercial wavefront shaper has been developed). We expect that the use of new contrast-enhancing imaging modes and methods to control sample damage will have a profound effect on electron microscopy on beam sensitive samples. Particularly in life science, there is potential to improve cryo-EM imaging for structural biology. A general vision on how the enabling technologies described above could impact the specific key application fields is provided in Figure 21.

| Enabling technologies | Quantum Science & Nanophotonics | Materials Science | Semiconductors | Life science |
|---|---|---|---|---|
| **Contrast enhancement & damage reduction in beam-sensitive samples** | Localize and study single quantum emitter systems | Improve sensitivity and resolution in analysis of batteries, MOF's and perovskites | Increase throughput in (2D) semiconductor device inspection and metrology at the nanoscale | Accelerate high-resolution reconstruction of biomolecules/complexes |
| **Time-resolved (pump-probe) EM with mature pulsing technology** | Explore quantum effects in/with free electrons<br><br>Characterize photonic components with high resolution in space, time, and energy | Analyze in-situ materials dynamics at the nanoscale for batteries, catalysis and PV applications under optical/electrical stimuli | Characterize carrier dynamics and defects in semiconductor devices and packages (in operando) | Enhance particle/vesicle detection |

**Figure 21.** Vision on how key enabling technologies in contrast enhancement/damage reduction and time-resolved electron microscopy could impact the main application sectors.

## 22. CONCLUSION

**F. Javier García de Abajo***

This Roadmap emphasizes the wide range of ideas generated by teaming up free electrons and light to develop and implement new forms of microscopy with improved resolution and access to previously unobservable phenomena (Sections 5, 6, 7, and 10). In addition, leveraging the impressive control exerted over the energy and lateral spatial profile of e-beams in electron microscopes, integration with ultrafast optics has opened the doors to an increase in temporal resolution that is quickly entering the attosecond scale while retaining nanometer spatial precision (Sections 7 and 8).

Beyond their uses in microscopy (Sections 3, 4, and 18–21), free electrons are emerging as excellent tools to synthesize, characterize, and manipulate the quantum states of photons, optical excitations in nanostructures, and the electrons themselves. We anticipate a continuous growth in the use of free electrons to explore and exploit quantum physics with capabilities that are on par with those of photons but featuring distinct appealing attributes. A new era of quantum free electronics is thus emerging, where the area of *free electrons for quantum nanophotonics*—the title of this article—is only the beginning. Compared to photons, free electrons offer the following advantages:

- *Robustness*. The spatiotemporal manipulation of the free-electron wave function through electron optics and optical-field interactions surpasses what can be achieved with single photons using classical, nonlinear, and quantum optics (Sections 11, 11, 15, and 16). Upon interaction and entanglement with materials and their excitations, free electrons retain their single-particle properties in ways that photons cannot. Additionally, parallel detection of a large number of single free electrons is possible with a low background and high efficiency (Section 3).

- *Post-Selection for the Synthesis of Quantum-States*. In a way analogous to how single-photon detection events can populate an interference pattern in a double-slit experiment, the detection of an individual electron projects its quantum state onto the measured energy, position, and angle, thus making a selection of the excitations it has left behind in the interaction with light and materials (Section 2 and 17). From this perspective, the collection of EELS spectra in an electron-by-electron detection fashion is a manifestation of the quantum nature of free electrons and their interactions. Still in its infancy, the application of this idea, which has recently been leveraged to create single- and few-photon states,[21] holds the potential to synthesize engineered quantum states in, for example, confined optical excitations at designated positions and times (Sections 9, 13, and 14).
- *Strong Coupling to Individual Quantum Excitations*. It is a challenging task to make photons interact deterministically with individual quantum excitations, generally requiring elaborate setups.[413] In contrast, free electrons can undergo much larger coupling to optical excitations, reaching near-unity probabilities at low kinetic energies[414] (down to the few-eV range where chemistry occurs). At high energies, the quest for strong coupling to individual quantum modes is attracting much attention and we anticipate innovative solutions to this problem, although it currently remains as a challenge.
- *Strong Electron-Electron Interactions*. The Coulombic interaction among electrons opens a vast range of possibilities for producing quantum superpositions, as well as for leveraging concepts such as superradiance and novel forms of pump-probe spectromicroscopy, recently unlocked through the observation and characterization of multi-electron beams.[47,48]
- *Nonlinear Evolution and Recoil*. While the manipulation of light waves in nonlinear nanophotonics is limited by the weak anharmonic response of known materials, electron waves can be strongly influenced by nanoscale features such as defects or free charges. Additionally, the electron energy combs produced by IELS (e.g., in PINEM) exhibit anharmonic-ladder characteristics due to recoil when the photon energy is comparable to the electron kinetic energy. From this perspective, low-energy free electrons present an opportunity to realize a strongly nonlinear response (encoded, for example, in their momentum degrees of freedom), combined with strong coupling to external stimuli—the properties of a still unrealized, strongly scattering two-level atom.
- *Sensitivity to Fluctuations*. From thermal optical fields to noise emerging as a distribution of vibrational excitations in a material, the free-electron quantum state undergoes decoherence that can potentially be traced, providing information about the electromagnetic and material environment.[25] An extension of this idea could be applied to disruptive forms of quantum sensing and metrology in directions that are complementary (and inaccessible) to those offered by quantum optics.

These properties open up a unique set of possibilities waiting for further investigation and exploration in new directions. However, several significant obstacles remain, most prominently: (1) the low excitation probabilities typically observed in EELS and CL experiments at high kinetic energies (as discussed in the *strong coupling* entry in the list above); and (2) the low temporal coherence of electrons produced by currently available sources (discussed in more detail in Section 8). We hope that research in these areas will be appealing, as they requires ingenious solutions in the best tradition of scientific discovery.

The field is ripe for breakthroughs based on the endeavors and prospects summarized in this work, and we thank our co-authors for preparing a cohesive set of sections and presenting their insightful perspectives, along with concise formulations of the state of the art, goals, opportunities, and ways to achieve them. Beyond the materialization of this collective effort in new concepts and methods, we anticipate unsuspected advances in fundamental science and, ultimately, significant benefits to society at large.

## ▪ LIST OF SELECTED ACRONYMS

| | |
|---|---|
| 2D, 3D | two-, three-dimensional |
| CL | cathodoluminescence |
| CW | continuous-wave |
| EEGS | electron energy-gain spectroscopy |
| EELS | electron energy-loss spectroscopy |
| IELS | inelastic electron–light scattering |
| IR | infrared |
| PINEM | photon-induced near-field electron microscopy |
| SEM | scanning electron microscope/microscopy |
| SLM | spatial light modulator |
| STEM | scanning transmission electron microscope/microscopy |
| TEM | transmission electron microscope/microscopy |

## ▪ AUTHOR CONTRIBUTIONS

Authors listed at the beginning of each section are responsible for the content of that section. The corresponding author in each section should be the person to be contacted if questions arise regarding the content of that section.

## ▪ ACKNOWLEDGMENTS

This work has been supported in part by the European Commission (EC) under Grants No. 101017720 FET-Proactive EBEAM (Sections 1, 2, 3, 4, 6, 7, 9, 10, 11, 14, 18, 19, 20, 21, and 22) and No. 964591 SMART-electron (Sections 1, 2, 11, 11, 13, and 17) as well as by the funding sources listed next. Section 2: the European Research Council (ERC) (101141220 QUEFES); the Spanish MCINN (CEX2019-000910-S); the Catalan CERCA Program; Fundaciós Cellex and Mir-Puig. Section 4: the Dutch Research Council (NWO); ERC (10101932 QEWS); EC (ANR-17-EURE-0009 NanoX and ANR-23-CE09-0018 LUTEM). Section 5: JSPS KAKENHI JP21K18195 (T.S), JP22H01963 (K.A.), JP22H05032 (K.A.), JP24H00400(T.S.), and JST FOREST JPMJFR213J (T.S.). Section 6: the French National Agency for Research (ANR) (ANR-10-EQPX-50 TEMPOS-CHROMATEM and QUENOT (ANR-20-CE30-0033); ERC (101141220 QUEFES). Section 7 and 9: the German Research Foundation (DFG) (Gottfried Wilhelm Leibniz program RO 3936/4-1). Section 10: ANR (ANR-10-EQPX-50 TEMPOS-CHROMATEM, ANR-20-CE42-0020 JCJC grant SpinE); the Belgian Research Foundation - Flanders (FWO) (G042920N). Section 11: A.K.: the Czech Science Foundation GACR (Junior Star grant 23-05119M); N.T.: ERC (101170341 UltraSpecT) and A. v. Humboldt Foundation (AvHF); F.J.G.A.: ERC (101141220 QUEFES). Section 13: R.R.: the Adams fellowship of the Israeli Academy of Science and Humanities; Y.A.: Scholars Program of the Clore Israel Foundation; I.K.: the Gordon and Betty Moore Foundation (GBMF) (GBMF11473) and EC (851780). Section 14: O.K.: the Israel Science Foundation (ISF) (2992/24 and 1021/22) and the National Quantum Science and Technology program of the Israeli Planning and Budgeting Committee. Sections 15 and 16: P.H.: ERC (AccelOnChip), GBMF (11473), DFG (HO 4543/7-1, HO 4543/8-1, and SFB-TR 306 QuCoLiMa); M.K.: Czech Science Foundation (22-13001K), ERC (101039339 eWaveShaper), and OP JAK (TERAFIT No. CZ.02.01.01/00/22 008/0004594); Y.M.: JST FOREST JPMJFR2228 and MEXT/JSPS KAKENHI JP21K21344; R.S.: DFG (TA 1694/5-1) and ISF (1025/24, 1027/24); Z.Z.: AvHF Fellowship. Section 17: F.C.: the Swiss NSF (200331) and the Airforce Office for Scientific Research (SCR0833393); F.B.: NSF fellowship P500PT_214437. Section 18: N.T.: Volkswagen Foundation (Momentum Grant) and ERC (101170341 UltraSpecT). The authors of Section 5 would like to thank Dr. N. Yamamoto, S. Yanagimoto, Dr. H. Saito, Dr. T. Yuge, and Dr. R. Okamoto for valuable discussions. C.R., A.F., M.S., and J.G. acknowledge the continued support from the Göttingen UTEM team and collaborators. F.J.G.A. is indebted to Archie Howie for many enjoyable and stimulating discussions.

### Notes

Views, findings, conclusions, and recommendations expressed in this Roadmap are those of the authors and do not necessarily reflect the views of the funding agencies.

Each section has been independently authored by different research groups. This Roadmap is a collection of opinions from these independent groups, and does not represent a collaborative effort.

## ▪ CONFLICT OF INTEREST

Section 4: The authors declare the following competing financial interest(s): A.P. is cofounder and co-owner of Delmic B.V., a company that builds instrumentation for CL microscopy in SEMs, which are used for research described in this section. Section 6: M.K. has licensed know-how and patents to Attolight, which is manufacturing the Mönch system used for some research presented in this section. Section 21: T.C. is employed by Delmic B.V. (see above). F.d.J. and E.K. are employed by Thermo Fisher Scientific, a company that builds S(T)EMs and dual beam systems amongst a large range of products in materials science, life science, and semiconductor applications.

## ▪ REFERENCES